\documentclass[aps,groupedaddress,floatfix,nofootinbib,multirow,twocolumn]{revtex4}

\input epsf

\begin{document}
\title{Dynamical Dark Matter:~
     I.  Theoretical Overview}
\author{Keith R. Dienes$^{1,2,3}$\footnote{E-mail address:  {\tt dienes@physics.arizona.edu}},
        Brooks Thomas$^{4}$\footnote{E-mail address:  {\tt thomasbd@phys.hawaii.edu}}}
\affiliation{
     $^1$ Physics Division, National Science Foundation, Arlington, VA  22230  USA\\
     $^2$ Department of Physics, University of Maryland, College Park, MD  20742  USA\\
     $^3$ Department of Physics, University of Arizona, Tucson, AZ  85721  USA\\
     $^4$ Department of Physics, University of Hawaii, Honolulu, HI  96822  USA}
%  \date{\today}
\begin{abstract}
   In this paper, we propose a new framework for dark-matter physics.
   Rather than focus on one or more stable dark-matter particles,
   we instead consider a multi-component framework in which the dark matter of the universe
   comprises a vast ensemble of interacting fields with a variety of different masses, mixings,
   and abundances.  
   Moreover, rather than impose stability for each field 
   individually, we ensure the phenomenological viability of such a scenario
   by requiring that those states with larger masses and Standard-Model
   decay widths have correspondingly smaller relic abundances, and vice versa.  
   In other words, dark-matter stability is not an absolute requirement in such 
   a framework, but is balanced against abundance.
   This leads to a highly dynamical scenario in which cosmological quantities 
   such as $\Omega_{\rm CDM}$ experience non-trivial time-dependences beyond those
   associated with the expansion of the universe.    
   Although 
   it may seem difficult to arrange an ensemble of states 
   which have the required decay widths and relic abundances, 
               we present one particular example in which this balancing act occurs naturally:
   an infinite tower of Kaluza-Klein (KK) states living in the bulk of large extra
   spacetime dimensions.  Remarkably, this remains true even if the stability 
   of the KK tower itself is entirely unprotected.
   Thus theories with large extra dimensions --- and by extension, certain limits of string theory ---
   naturally give rise to dynamical dark matter.
   Such scenarios also generically give rise to a rich set of collider
   and astrophysical phenomena which transcend those usually associated with dark matter.
\end{abstract}
%  \pacs{95.35.+d,98.80.Cq,14.80.Rt,11.25.Wx,11.25.Mj}
\maketitle

%========================================================================
%          KEYSROKE-SAVING MACROS, nothing complicated 
%========================================================================
\newcommand{\newc}{\newcommand}
\newc{\gsim}{\lower.7ex\hbox{$\;\stackrel{\textstyle>}{\sim}\;$}}
\newc{\lsim}{\lower.7ex\hbox{$\;\stackrel{\textstyle<}{\sim}\;$}}

\def\beq{\begin{equation}}
\def\eeq{\end{equation}}
\def\beqn{\begin{eqnarray}}
\def\eeqn{\end{eqnarray}}
\def\calM{{\cal M}}
\def\calV{{\cal V}}
\def\calF{{\cal F}}
\def\half{{\textstyle{1\over 2}}}
\def\quarter{{\textstyle{1\over 4}}}
\def\threehalf{{\textstyle{3\over 2}}}
\def\ie{{\it i.e.}\/}
\def\eg{{\it e.g.}\/}
\def\etc{{\it etc}.\/}
\def\Nhex{N_{\mathrm{hex}}}

%     The following macros are to create the "blackboard bold"
%     characters for "R" (set of real numbers),
%     "C" (set of complex numbers), and "Q" (set of rational numbers).

\def\ket#1{{ |{#1}\rangle}}
\def\inbar{\,\vrule height1.5ex width.4pt depth0pt}
\def\IR{\relax{\rm I\kern-.18em R}}
 \font\cmss=cmss10 \font\cmsss=cmss10 at 7pt
\def\IQ{\relax{\rm I\kern-.18em Q}}
\def\IZ{\relax\ifmmode\mathchoice
 {\hbox{\cmss Z\kern-.4em Z}}{\hbox{\cmss Z\kern-.4em Z}}
 {\lower.9pt\hbox{\cmsss Z\kern-.4em Z}}
 {\lower1.2pt\hbox{\cmsss Z\kern-.4em Z}}\else{\cmss Z\kern-.4em Z}\fi}
%========================================================================

\input epsf

%========================================================================
%========================================================================
%               MAIN TEXT BEGINS HERE
%========================================================================

%========================================================================

\section{Introduction, motivation, and summary}

Situated at the nexus of particle physics, astrophysics, and cosmology
lies one of the most compelling mysteries that faces physics today:
that of unravelling the identity and properties of dark matter~\cite{reviews}.  
 From measurements of galactic rotation curves and velocity dispersions to 
observations of the gravitational lensing of galaxy clusters 
and the detection of specific acoustic peaks of the cosmic microwave background (CMB), ample
circumstantial evidence suggests that most of the matter in the universe  
does not interact strongly or electromagnetically.  
Such matter is therefore electrically neutral (dark) and presumed non-relativistic (cold).
Beyond these properties, however, very little is known about the nature of dark matter.
Fortunately, the current generation of dark-matter experiments have 
unparalleled sensitivities, and new data concerning the possible 
direct and indirect detection of dark matter can be expected soon.
This data will therefore go a long way towards not only resolving
this pressing cosmological mystery, but also constraining the possibilities for
physics beyond the Standard Model (SM).

Many theoretical proposals for physics beyond the Standard Model give rise
to suitable dark-matter candidates.  However, most of these dark-matter 
candidates consist of a single particle (or a small collection of
particles) which are stable on cosmological time scales as the result of
a discrete symmetry.
Examples include the lightest supersymmetric particle (LSP) in supersymmetric
theories, 
and the lightest Kaluza-Klein particle (LKP) in certain higher-dimensional theories  
in which the Standard Model propagates in the bulk~\cite{KKdarkmatter}.
In the first case, the LSP is stabilized by the assumption of an $R$-parity 
symmetry, while in the second case the stabilizing symmetry is a so-called
``KK parity''.  However, in all cases, the ability of these particles 
to serve as dark-matter candidates rests squarely on 
their stability.  
Indeed, any particle which decays into Standard-Model
states too rapidly is likely to upset traditional big-bang nucleosynthesis (BBN)  
and its successful predictions
of light-element abundances.  Such decays can also leave unacceptable imprints in the 
cosmic microwave background and diffuse X-ray and gamma-ray backgrounds. 
For this reason, 
stability is often the very first criterion required for the phenomenological
success of a hypothetical dark-matter
candidate.

There is, of course, one important exception to this argument:  a given dark-matter
particle need not be stable if its abundance at the time of its decay 
is sufficiently small.
A sufficiently small abundance ensures that the disruptive effects of the decay of
such a particle will be minimal, and that all constraints from BBN, the CMB, {\it etc}\/., 
will continue to be satisfied.

In this paper, we shall consider a new 
framework for dark-matter physics which takes advantage of this possibility.
Specifically, we shall consider a multi-component framework 
in which the dark matter of the universe
   comprises a vast ensemble of interacting fields with a variety of different
masses, mixings,
   and abundances.  
   Rather than impose stability for each field 
   individually (or even for the ensemble of fields as a whole), 
   we shall ensure the phenomenological viability of such a scenario
   by requiring that those states with larger masses and larger decay widths
   into Standard-Model fields
   have correspondingly smaller relic abundances, and vice versa.  
   In other words, dark-matter stability is not an absolute requirement in such 
   a framework, but is balanced against abundance.
   As we shall demonstrate, this leads to a highly dynamical scenario 
in which cosmological quantities 
   such as $\Omega_{\rm CDM}$ experience non-trivial time-dependences beyond those
   associated with the expansion of the universe.    
We shall therefore refer to such a scenario as ``dynamical dark matter''.

In general, it might seem difficult (or at best fine-tuned) to have
an ensemble of states which are not only suitable candidates for dark matter
but in which the abundances and decay widths are precisely
balanced in this manner.  
However, it turns out that 
theories with large extra spacetime dimensions 
not only naturally provide such ensembles of states, but 
do so in a manner which is virtually intrinsic to their construction. 
If the Standard Model is restricted to a brane floating in a higher-dimensional space,
it then immediately follows that any field propagating in the bulk of this space 
must be neutral under all Standard-Model
symmetries.  As a consequence, such bulk fields can have at most gravitational 
interactions with the physics on the brane, and will therefore appear as dark matter
from the perspective of an observer on the brane.
Moreover, from the perspective of this four-dimensional observer, such bulk fields will appear as an 
infinite tower of individual Kaluza-Klein (KK) modes.  
This, then, would constitute our dark-matter ensemble.

At first glance, such a scenario for dark matter would appear to face a major 
phenomenological hurdle:  in the absence of additional symmetries
or {\it ad-hoc}\/ assumptions, an entire Kaluza-Klein tower of bulk states will generally
be unstable:  the heavy KK states in such a tower will generically decay into 
not only lighter KK bulk states but also Standard-Model brane states,
and the lighter KK states will also decay into Standard-Model brane states.
Even the stability of the lightest modes of the bulk field is not guaranteed.
This instability of the Kaluza-Klein tower therefore appears to pose 
a serious threat for the survival of big-bang
nucleosynthesis  in its traditional form, and can similarly
disturb the X-ray and gamma-ray backgrounds.

Fortunately, there are two 
critical features of Kaluza-Klein towers 
which can play off against each other in order 
to render such a scenario phenomenologically viable.
As one goes higher and higher in a generic KK tower,
it is true that the decay width of the KK states generally increases with the KK mass.
However, it is also true that the cosmological abundance associated with such states
can often {\it decrease}\/ with the KK mass.
This is particularly true if we imagine that these states are cosmologically produced through
misalignment production, as turns out to be particularly appropriate for such scenarios.
As a result, it might be possible that all KK states
which decay before or during BBN have such small abundances that the destructive effects of
their decays are insignificant, while at the same time a significant fraction 
of the KK tower survives to the present day and thereby contributes to the observed 
total dark-matter abundance.
Thus, the surviving dark matter at the present
day would consist of not merely one or two states, but a significant fraction 
of an entire interacting KK tower.
Through the existence of such ``dark towers'',
theories of large extra spacetime dimensions therefore provide an 
ideal realization of our general dynamical dark-matter scenario.

In this paper, we shall lay out the general properties of such a scenario and 
explore the extent to which such a scenario is viable.
Moreover, we shall attempt to do so in a completely model-independent way, without making
any assumption concerning the nature of the bulk field in question.
However, it is important to recognize that this entire approach represents a 
somewhat unorthodox approach to dark-matter physics.  
By balancing the stability of the different dark-matter components against their 
abundances across a large or even infinite ensemble,
% an infinite tower of KK modes,
the dark matter in this scenario is 
intrinsically {\it dynamical}\/ ---
its different components 
continue to experience non-trivial mixings and decays  
throughout their cosmological evolution, with such dynamical behavior continuing 
until, during, and  beyond the current epoch.
Moreover, because the dark matter in our scenario has multiple components,
its phenomenology cannot be characterized in terms of 
a single mass or annihilation cross-section.
This can therefore lead to an entirely new dark-matter phenomenology,
profoundly changing the way in which such dark matter might be observed and constrained
through collider experiments and astrophysical observations.
Indeed, 
within the specific context of large extra dimensions,
we shall see that
one important new phenomenon that emerges for such dark matter
is the possibility of ``decoherence'' --- \ie, the phenomenon  in which  only a single linear
combination of KK modes couples to brane physics at one instant before decohering 
and becoming essentially invisible to the brane at all subsequent times.

It is also important that this framework not be 
confused with recent proposals concerning so-called ``Kaluza-Klein dark matter''~\cite{KKdarkmatter}.
Theories of KK dark matter require that the entire Standard Model propagate
in the large extra dimensions~\cite{Antoniadis,DDGLargeED,UED}, and that the lowest
excited KK mode of a Standard-Model field (such as the lowest-excited
KK photon or neutrino) be stable as the result
of an internal geometric symmetry such as KK parity~\cite{UED,KKdarkmatter}.
Such theories of KK dark matter are therefore similar to theories of supersymmetric
dark matter ---
they are theories of a single, stable, dark-matter particle.
While phenomenologically consistent, such a point 
of view is diametrically opposed to what we are suggesting here.
Moreover, in doing away with the infinite tower of KK states and focusing
exclusively on the single lightest KK mode,
such ``KK dark matter'' theories 
also do away with that part of the physics which is intrinsically higher-dimensional.
The resulting scenarios are therefore insensitive to the rich physics that can emerge from
an entire tower of Kaluza-Klein states acting in unison, with 
non-trivial masses and mixings governing their dynamics.
Indeed, it is precisely such behavior that would give a window into the nature
of the extra dimensions from which such states emerge.

By contrast, because our scenario balances a spectrum of decay lifetimes against 
a spectrum of relic abundances,
our framework is sensitive to the physics of the entire tower of KK states
and thereby makes use of the full higher-dimensional ``bulk'' 
in a fully higher-dimensional way. 
Moreover,
since Type~I string theories naturally give rise
to closed-string states (such as the graviton, various moduli, and axions) which live in 
more spacetime dimensions than the Standard-Model open-string states 
which are restricted to live on D-branes,
the scenario we shall be investigating is also extremely natural --- and indeed
almost unavoidable --- in certain limits of string theory. 
Our work can therefore be seen as providing a test of the extent to which such string theories
remain cosmologically viable as a function of the volume of the extra dimensions
transverse to the Standard-Model brane.
In other words,
by studying dynamical dark matter and its phenomenological
viability, we are not only exploring a new candidate for dark matter
but also providing new phenomenological constraints on large extra
dimensions and certain limits of string theory.

This paper is organized as follows.
In Sect.~II, we shall introduce our dynamical 
dark-matter framework in its most general form,
without making reference to the specific example of 
large extra dimensions or KK towers.
We shall discuss how lifetimes and abundances can play
off against each other in such a scenario, and sketch the resulting contributions to the
total dark-matter abundance as they evolve in time.
In Sect.~III, we shall then focus on the example of a generic tower of Kaluza-Klein
states emerging from the bulk of large extra dimensions, and 
show that such a tower has all the required properties to be a dynamical
dark-matter candidate, with abundances and lifetimes that satisfy
unique mathematical inverse relations.   
We shall then proceed, in Sect.~IV, to discuss several laboratory
and astrophysical signatures of such a scenario, focusing on
those new features which transcend typical dark-matter signatures
and which might explain why such dark matter has not yet been observed.
Throughout this paper, our analysis shall be as general
as possible without specifying the
precise nature of the bulk field in question.  
We shall then conclude in Sect.~V with a discussion of extensions and
possible generalizations of our dark-matter framework.

This paper is the first in a two-part series.
The primary purpose of the present paper is merely to provide a general theoretical overview
of the dynamical dark-matter framework.  
As a result, we will
not choose a particular species of dark-matter field,  
neither restricting ourselves
to specific numbers nor subjecting ourselves to specific phenomenological bounds.
Instead,
our discussion here will focus on the full range of theoretical
possibilities afforded by this new scenario. 
However, in a companion paper~\cite{paper2} we will provide
a detailed ``proof of concept'' by focusing on the 
particular case that the KK bulk field in question is an axion.
In particular, in Ref.~\cite{paper2} we will demonstrate that a bulk axion field
can satisfy all theoretical and numerical constraints needed to serve
as a dynamical dark-matter candidate, and moreover we will
demonstrate that 
this candidate also satisfies all known cosmological,
astrophysical, and collider bounds on dark matter.
Indeed, in making this assertion, Ref.~\cite{paper2} will borrow heavily from the results
of a third paper~\cite{paper3}, a detailed forthcoming phenomenological study
of axions and axion-like particles in higher dimensions.
Thus, taken together, these papers will demonstrate 
that our dynamical dark-matter scenario remains a very real
possibility for explaining the dark matter of the universe.

\section{Dynamical Dark Matter:  ~General Scenario}

In this section we shall begin by discussing
our dynamical dark-matter
scenario in its most general form, without reference
to specific examples such as those involving extra spacetime dimensions or 
KK towers of bulk fields.

\subsection{$\Gamma$ versus $\Omega$:  ~Balancing lifetimes against abundances}

Broadly speaking, upon positing any new scenario for dark matter, one faces 
certain immediate constraints which must be satisfied.
These constraints ultimately restrict either the 
abundance of the dark matter,
the lifetime of the dark matter, or the relation between the two. 

Let us begin by considering the case of a single dark-matter
particle $\chi$.  
Since this dark-matter particle is presumed unique,
it alone must carry the entire observed dark-matter abundance:
$\Omega_\chi=\Omega_{\rm CDM}\approx 0.23$~\cite{WMAP}.  
However, given this large abundance, consistency with constraints
coming from big-bang nucleosynthesis,
the cosmic microwave background, 
and diffuse X-ray and gamma-ray backgrounds
together require that   $\chi$ have a lifetime 
which meets or exceeds the current age of the universe.
Otherwise, decays of $\chi$ 
run the risk of disturbing BBN and 
its successful predictions for light-element abundances.
Such early decays also have the potential to distort the cosmic microwave background
as well as the X-ray and gamma-ray backgrounds.  

However, because of the quantum-mechanical nature of the decay
process,  not all dark matter will decay at once.
As a result, the lifetime of $\chi$ must actually {\it exceed}\/ the 
age of the universe
by at least one or two orders of magnitude in order to ensure
that $\chi$ has a negligible chance of having already decayed in the recent past.
This likewise implies that such a particle also has a negligible chance of decaying 
either today or tomorrow.
Such a particle $\chi$ is therefore ``hyper-stable''.
Indeed, this is the case for most if not all known single-particle dark-matter candidates.

Hyper-stability is the only way in which a single-particle dark-matter
candidate can satisfy the competing constraints of having a significant abundance 
$\Omega_\chi\sim \Omega_{\rm CDM}$
while simultaneously avoiding the dangerous effects of decaying into Standard-Model particles.
This then results in a dark-matter scenario which is ``frozen'' in time,
with cosmological quantities such as $\Omega_\chi$ evolving only because of 
the expansion of the universe.

However, the primary purpose of this paper is to propose that there is 
another way --- a ``dynamical'' way --- 
to satisfy these competing constraints.
First, we recognize that in some sense, it is somewhat unnatural to consider the 
dark matter of the universe as consisting of only a single particle.
After all, the {\it visible}\/ matter of the universe constitutes only a small fraction of the energy
density attributed to dark matter, and yet is teeming 
with a diversity and complexity, known as the Standard Model, in which
a complex network of elementary particles
is organized according to their own internal principles.    
It therefore seems natural to consider the new opportunities that are open to us 
by taking the dark matter to consist of multiple particles as well.

In this paper, we shall therefore 
imagine that the dark matter consists of a vast ensemble of particles $\chi_i$, 
$i=1,2,...,N$ with $N\gg 1$.
The first observation that follows from this assumption is that none of these particles individually needs to have
a significant abundance, since they may still {\it collectively}\/ yield the 
correct total abundance $\Omega_{\rm CDM}$.
As a special case, for example, we might imagine that each particle $\chi_i$ shares a common
abundance $\Omega_i = \Omega_{\rm CDM}/N$.
However, if these particles were also to have equal lifetimes, then 
this would not solve our second constraint --- that of protecting the successes of BBN and
minimally impacting the CMB and other diffuse backgrounds --- unless each 
of these particles is not only stable but hyper-stable. 
This is because the net effect of the nearly simultaneous decays of each of these $N$ particles would 
be no different
than that of the decay of a single particle carrying the 
full abundance $\Omega_{\rm CDM}$. 

However, the fact that we have multiple particles furnishes us with
an alternate way to satisfy these constraints:
we can imagine that each of these particles has a significantly {\it different}\/ lifetime.
In general, these particles can also have different individual abundances. 
As long as those particles which have relatively short lifetimes also have correspondingly small
abundances, and as long as those particles which have relatively large abundances also have relatively
long lifetimes, we can reproduce the correct total dark-matter abundance $\Omega_{\rm CDM}$
while simultaneously avoiding any damaging effects on BBN, the CMB, etc.
In this way, the existence of a vast ensemble of dark-matter particles $\chi_i$ opens up
the possibility of balancing abundances against decay widths in a non-trivial way across a multitude of states.

This, then, is the essence of our dynamical dark-matter proposal.
The fact that we have distributed the total required dark-matter abundance  
across many states means that no particular state is forced to carry a significant abundance on its own.
We thus have the room to give these states a whole spectrum of lifetimes (or decay widths) without 
running afoul of cosmological constraints.

Note that the usual scenario of a single hyper-stable dark-matter particle
is nothing but a special case of this more general framework:
even though our scenario has $N\gg 1$, it still remains possible that
one particle (or just a few particles) could carry
the bulk of the abundance $\Omega_{\rm CDM}$ 
at the present time.
Following the same logic that applied in the single-particle case, this small subset
of states would then be required to be hyper-stable, and all of the remaining states would have 
abundances that are far too insignificant to be of consequence.
However, the novel features of our scenario emerge in the opposite limit, when we imagine
that {\it none}\/ of our dark-matter states individually carries the bulk of the total relic abundance.
Some fraction of these states then no longer need to be hyper-stable,
leading to a dynamical scenario in which spontaneous dark-matter decays are occurring prior to, 
during, and beyond the current epoch.
As a result of this behavior, cosmological quantities such as $\Omega_{\rm CDM}$ will experience time-variations 
which transcend those due to the ordinary expansion of the universe.

\subsection{Dynamical dark matter:  ~Time-evolution of individual components}

It is possible to outline the salient features of this scenario somewhat
more quantitatively without loss of generality.
For this purpose,
we shall describe the history of the universe as progressing
through four distinct eras respectively associated with inflation (vacuum domination),  
reheating (RH), radiation domination (RD), and matter domination (MD).
Note that the reheating era is itself essentially matter-dominated,
with the matter in this case consisting of the oscillating inflaton.
Likewise, note that even though the current epoch is technically a $\Lambda {\rm CDM}$ universe,
approximating this epoch as matter-dominated is
nevertheless a fairly good approximation.
For the purposes of our general discussion, we shall not specify
the particular energy or temperature scales associated with the transitions between
these eras;  such scales, especially as they relate to inflation
and reheating, are likely to be highly model-dependent.
However, regardless of these energy scales, 
a quantity whose energy density $\rho$ and pressure $p$ are related
through a single-parameter equation of state of the form $p = w\rho$
will have a
relative relic abundance $\Omega\equiv \rho/\rho_{\rm crit}$
that scales as a function of cosmological time $t$ according to\footnote{
       The results in Eq.~(\ref{wbehav}) follow from the general facts that
       $\rho\sim R^{-3(1+w)}$ and $\rho_{\rm crit}\sim H^2$,
      where $R$ and $H$ are respectively the scale factor and Hubble parameter.
      Recall that these latter quantities have the scaling behaviors
    $(R,H)\sim (t^{1/2}, t^{-1})$ in an RD era;
    $(R,H)\sim (t^{2/3}, t^{-1})$ in RH or MD eras;  and
    $(R,H)\sim (e^{Ht}, {\rm constant})$ in an inflationary era.}
\beq
         \Omega ~\sim~ \cases{  
           t^{(1-3w)/2}  &  RD era \cr
           t^{-2w}  &  RH/MD eras\cr
           \exp[ -3 H(1+w)t]  &  inflationary era~.\cr}
\label{wbehav}
\eeq
Recall that $w=0$ for matter, while $w= -1$ for vacuum energy
(cosmological constant) and $w=\pm 1/3$ for radiation and curvature respectively.

For concreteness, we shall assume that the individual components
of the eventual dark matter in our scenario are described by scalar
fields $\phi_i$ with corresponding masses $m_i$ and
widths $\Gamma_i$ describing their decays into Standard-Model states.
For simplicity we shall also assume that these widths $\Gamma_i$ correspond to
processes in which a given dark-matter component $\phi_i$ decays {\it directly}\/
into SM states (\ie, $\phi_i\to {\rm SM}$)
without passing through any other dark-matter components
as intermediate states.  In other words, we shall assume that 
 {\it extra}\/-ensemble decays $\phi_i\to {\rm SM}$ (with combined total
width $\Gamma_i$) dominate over all
possible {\it intra}\/-ensemble decays $\phi_i\to \phi_j+...$.
It turns out that this is a valid assumption for many realistic
scenarios to be discussed in this paper and in Ref.~\cite{paper2}, 
and in the Appendix we shall discuss what happens when this assumption is relaxed.

In a Friedmann-Robertson-Walker (FRW) cosmology parametrized by a Hubble parameter
$H(t)\sim t^{-1}$ in which we assume that 
our dark-matter component fields $\phi_i$ have negligible spatial variations
as well as negligible self-interactions,
these fields will time-evolve according to the differential equation
\beq
       \ddot \phi_i + [3H(t) + \Gamma_i] \dot\phi_i + m_i^2 \phi_i ~=~0~.
\label{harmosc}
\eeq
This is the equation for a damped harmonic oscillator, with critical damping
occurring for $3H(t)+\Gamma_i =2m_i$;  
note that
the single-derivative ``friction'' term in this equation
receives two separate contributions,
one arising from the cosmological Hubble expansion
and the other arising from the intrinsic decay of $\phi_i$.    
As a result, 
at early times for which 
$3H(t)+\Gamma_i > 2 m_i$,
the field $\phi_i$ is overdamped:  it does not oscillate,
and consequently  its energy density behaves like vacuum energy (with $w= -1$).
By contrast, at later times for which
$3H(t)+\Gamma_i < 2 m_i$,
the field is underdamped:  it therefore oscillates, and consequently 
its energy density scales appropriately for matter (with $w=0$).  
The condition
$3H(t)+\Gamma_i = 2 m_i$
thus describes the ``turn-on'' time at which oscillatory behavior
begins and the field begins to act as true matter.

Given these observations, we can now sketch how each of the abundances $\Omega_i$
for each component $\phi_i$ will behave in our scenario.
For concreteness,
we shall assume that
these abundances are all initially established at a common time $t_0$.
Moreover, we shall assume that each component has an  
initial abundance which decreases as a function of its mass.  
While not all production mechanisms have this property,
we shall see in Sect.~III that misalignment production in particular
can accomplish this task.

Immediately upon establishment of these abundances, the states in our
ensemble can be separated into two groups.  Those heavier states with
masses $3H(t_0)+\Gamma_i < 2 m_i$
will all begin oscillating simultaneously.  In other words,
they experience a simultaneous, instantaneous turn-on.  By contrast,
the lighter states with 
$3H(t_0)+\Gamma_i > 2 m_i$
will experience a step-by-step ``staggered'' turn-on,
with lighter and lighter states crossing the
turn-on threshold at later and later times.
Indeed, if we approximate $H(t)\approx \kappa/3t$ where $\kappa$ is a constant within each era,
then a given mode with mass $m_i$ will turn on at a time $t_i\approx \kappa/2m_i$.
Thus the lightest states are necessarily the last to turn on.
Indeed, $\kappa=2$  for the RH and MD eras, while $\kappa =3/2$ for the RD era.

Finally, once these states are all ``turned on'' and behave as matter with $w=0$,
their abundances $\Omega_i$ 
will evolve as discussed above until such times $t\sim \tau_i\equiv \Gamma_i^{-1}$
as these states decay.
Specifically, the abundance $\Omega_i\sim \rho_i/H^2$ associated with each component $\phi_i$ will evolve
according to the differential equation
\beq
   \dot\Omega_i + \left( 3 H + 2{\dot H\over H} + \Gamma_i\right)\,\Omega_i~=~0~.
\label{diffeq1}
\eeq
[This equation follows directly from Eq.~(\ref{harmosc}) upon
use of the general expression $\rho_i= \half(m_i^2\phi_i^2 + \dot \phi_i^2)$
and the virial theorem $m_i^2\phi_i^2 = \dot \phi_i^2$, the latter 
holding for oscillations whose frequencies are large compared with $\Gamma_i$.]
Note that if we ignore the decays of these particles (\ie, if we set $\Gamma_i\to 0$),
the solutions to this differential equation are nothing but the
results given in Eq.~(\ref{wbehav}) for $w=0$.
Upon decay, however, the corresponding abundance $\Omega_i$ drops rapidly to zero;
this occurs when $t\sim \tau_i$, and indeed 
$\Gamma_i \gsim 3 H + 2{\dot H/H}$ at $t\sim \tau_i$.
For simplicity, 
in the rest of this paper 
we shall approximate such decays as occurring
promptly and completely at $t=\tau_i$.
However this approximation will not be critical for any of our conclusions, and can
easily be discarded if needed.

%================== FIGURE ============================================
\begin{figure*}[thb!]
\centerline{
   \epsfxsize 7.1 truein \epsfbox{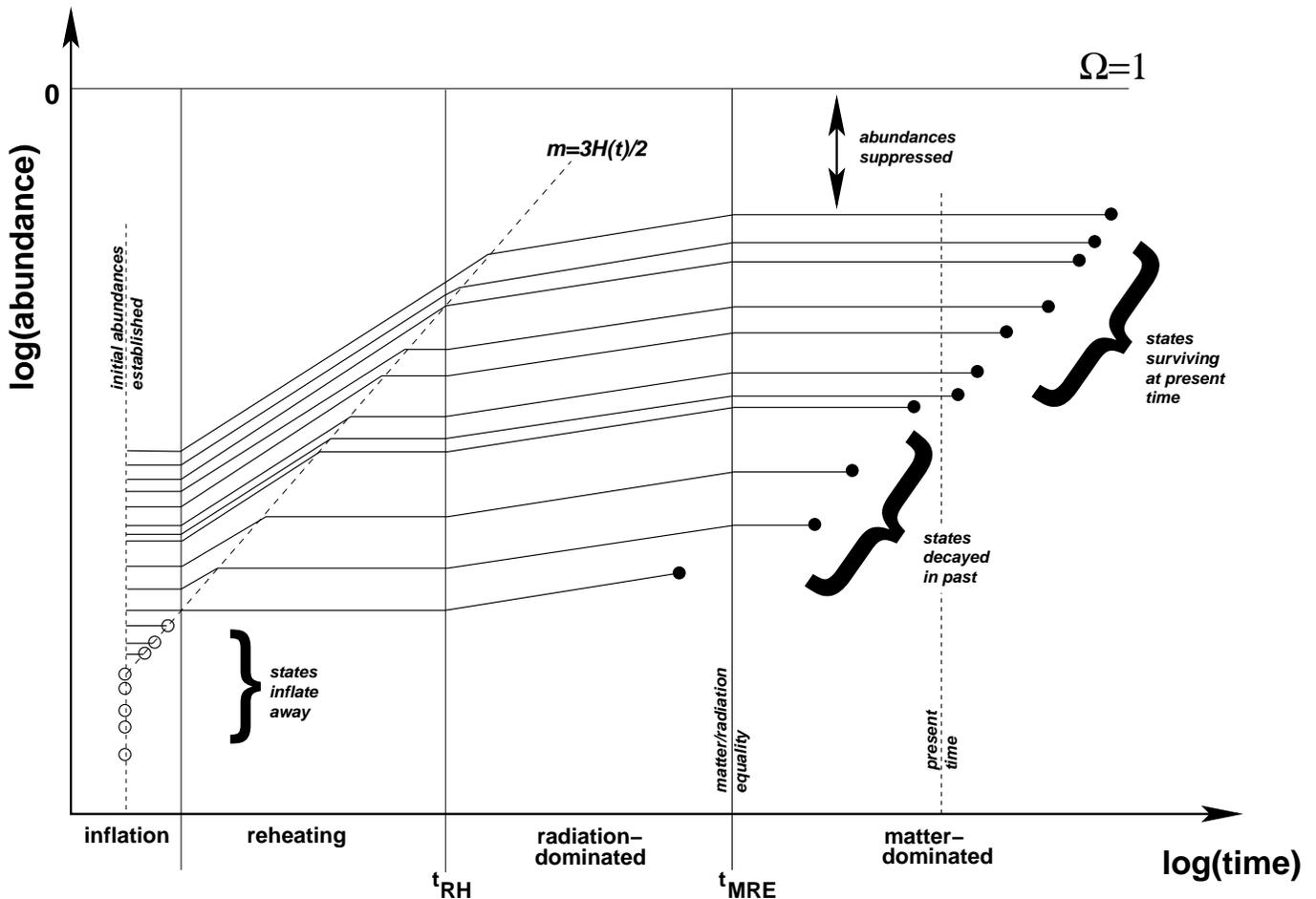} 
 }
%  \vskip -0.2 truein
\caption{A sketch of our dynamical dark-matter scenario in which 
the dark matter of the universe comprises a vast ensemble  
of individual components with different masses, abundances, and lifetimes.
This plot illustrates the  evolution
of the abundance of each dark-matter component as a function of time,
assuming that all abundances 
are initially established at a common
time (chosen here to be prior to or during the inflationary era), with values that decrease as a function
of the component mass.
For all subsequent times, these abundances scale as vacuum energy until 
$3H(t)=2 m_i$, after which point they scale as matter.
Open circles indicate states which inflate away,
while closed circles indicate states which decay into SM particles
with associated lifetimes that decrease with increasing mass.
In our scenario, the lifetimes and abundances are balanced against each other in
such a way that there continue to exist a plethora of dark-matter states 
which survive at the present time:
although each such state has an extremely small abundance $\Omega_i\ll 1$, 
they collectively reproduce $\Omega_{\rm CDM}$.
Nevertheless, because of their extremely small abundances,
states which have already decayed into SM particles leave negligible imprints 
on the CMB and other observable astrophysical and cosmological backgrounds.
An important feature of this scenario is that it is fully {\it dynamical}\/, with
the composition and properties of the dark matter continuing to 
experience a non-trivial time evolution before, during, and even 
after the current epoch.  }
\label{sketch}
\end{figure*}  
%========================================================================

Combining all these features,
we can then sketch the salient features of our scenario
as in Fig.~\ref{sketch}.
In this plot, we have taken the time at which the initial
abundances are established to be during the inflationary era,
but other time intervals are also possible.
We have also assumed that $\Gamma_i\ll 3H(t_i)$ 
when $3H(t_i)=2 m_i$, so that the decay widths $\Gamma_i$ affect the final
decay times $\tau_i$ but not the staggered turn-on times $t_i$. 

Several features which are clear from this sketch help to 
define and characterize this scenario.
First, we see that at the present time, there continue to be
a plethora of dark-matter states.  Although each of these
has an extremely small abundance (exponentially suppressed on
the plot in Fig.~\ref{sketch}), they can collectively produce
a sizable, ${\cal O}(1)$ abundance 
which we choose to identify 
with the observed dark-matter abundance $\Omega_{\rm CDM}\approx 0.23$~\cite{WMAP}.

Second, we observe that our dark-matter states are not governed 
by the notion of stability:  while some are indeed more stable
than others, decays of dark-matter states can occur {\it throughout}\/
the evolution of the universe.  This is not in conflict with observational
constraints because of the extremely small abundances of those states
which decay at critical epochs during the evolution of the universe.
In other words, as discussed above, 
lifetimes and abundances are balanced against each
other in this scenario.

Third, as a consequence of these features, we see that 
nothing is particularly
special about the present time
in this framework.  Dark-matter states need not be held stable
until any special moment, and the current age of the universe
plays no special role in this scenario.
Indeed, as evident from the sketch in Fig.~\ref{sketch},
dark-matter states decay prior to, during, and after the present
era.  What results, then, is a scenario in which the dark matter is 
not frozen in time at the present era, but continues to act as a
highly dynamical component 
of an ever-evolving universe.

%=============================================================

\subsection{Characterizing a given dark-matter configuration:  
    ~$\Omega_{\rm tot}$, $\eta$, and $w_{\rm eff}$}

In general, there are three quantities which we can use
in order to characterize the configuration of our dark-matter ensemble
at any instant in time, and to track its subsequent time-evolution.  
We shall begin by introducing these three quantities and discussing
the relations between them.  Then, we shall discuss several qualitative
aspects of their overall time-evolutions.

\subsubsection{Fundamental definitions and relations}

The first quantity we shall define in order to characterize the
configuration of our dark-matter ensemble is 
the total dark-matter abundance:
\beq
            \Omega_{\rm tot}~\equiv~ \sum_i\Omega_i~.
\label{omtotdef}
\eeq
Note that we should include in this
sum the contributions from only those components in our ensemble which have 
already ``turned on'' (\ie, have begun oscillating) and which are therefore already behaving as true matter.
Restricting such contributions in this way ensures that
$\Omega_{\rm tot}$ can truly 
be associated with a total dark-matter abundance. 
Of course, all of the components in our ensemble will eventually ``turn on'' 
and behave as dark matter after enough time has passed.
It is for this reason that we shall continue to refer to each component
of our ensemble as a dark-matter component, regardless of its particular
turn-on time.

The quantity $\Omega_{\rm tot}$ describes the total dark-matter
abundance at any instant of time.
However, 
we may also define a complementary quantity $\eta$ which
describes how this total abundance is {\it distributed}\/ across
the different components:
\beq
         \eta ~\equiv~ 1-{\Omega_0\over \Omega_{\rm tot}}~.
\label{etadef}
\eeq
Here $\Omega_0\equiv {\rm max}_i \lbrace \Omega_i\rbrace$ is defined
to be the largest
individual dark-matter abundance from amongst our ensemble of dark-matter states.
Thus, $\eta$ measures what fraction of the total abundance $\Omega_{\rm tot}$ is {\it not}\/
carried by a single dominant component.
We see from its definition that $\eta$ varies within the range $0\leq \eta\leq 1$: 
values of $\eta$ near zero signify the traditional situation
in which the total dark-matter abundance is predominantly carried by
a single state, while
larger values of $\eta$ signify departures
from this traditional configuration.
In this sense, then, $\eta$ can also be viewed as quantifying the degree to which
our scenario deviates from the more traditional dark-matter framework at
any instant in time.

Recall that we have assumed for this discussion that 
the more massive components of our ensemble have smaller
initial abundances, and vice versa.
Indeed, this assumption is already reflected 
in the sketch in Fig.~\ref{sketch}.
It then follows that the largest abundance
$\Omega_0$ in Eq.~(\ref{etadef}) will correspond to the lightest component
in our ensemble.
However, 
in the event of a ``staggered'' turn-on, 
the lightest components in the ensemble are necessarily
the last to turn on.
Indeed, prior to their turn-on times,
the abundances 
of these lightest states
contribute
to the total dark-{\it energy}\/ abundance rather than to the total dark-matter abundance.
We must therefore be careful to identify
$\Omega_0$ as the abundance associated 
with the lightest of those components which have already turned on. 
As a result, even the identity of the component whose abundance is to be
identified with $\Omega_0$ can occasionally be time-dependent.

While both $\Omega_{\rm tot}$ and $\eta$ 
characterize the configuration of our dark-matter ensemble
at a given instant in time, one of the critical features
of our dynamical dark-matter scenario is precisely that it
is {\it dynamical}\/ --- \ie, that these quantities have non-trivial 
time-evolutions.
Of course, some of this time-evolution is common to all dark-matter scenarios,
arising due to the Hubble expansion of the universe during its reheating, 
radiation-dominated, or matter-dominated eras.  
There are, however, additional time-dependent effects 
which are unique to our dynamical dark-matter scenario.
For example,
one such effect dominates the physics of the final, matter-dominated era,
and arises because 
the total dark-matter abundance in our scenario has been distributed across 
many individual dark-matter components,
each with a potentially different lifetime.
This phenomenon leads to a slowly falling $\Omega_{\rm tot}$ at late times.
Clearly the time-evolution of $\Omega_{\rm tot}$
during this period is extremely sensitive to  
not only the distribution of the abundances $\Omega_i$ across
the different dark-matter components within the ensemble, but also the
decay widths $\Gamma_i$ which govern the times at which these different
components decay.  

For this reason,
it will be useful to define a quantity
which can meaningfully characterize the aggregate time-evolution of our ensemble.
Moreover, we would like this quantity to characterize this time-evolution
regardless of the particular cosmological era under study,
and likewise to quantify the extent to which this time-evolution 
intrinsically differs from that normally associated 
with the cosmological expansion of the universe.

%================== FIGURE ============================================
\begin{figure*}[thb!]
\centerline{
   \epsfxsize 5.8 truein \epsfbox{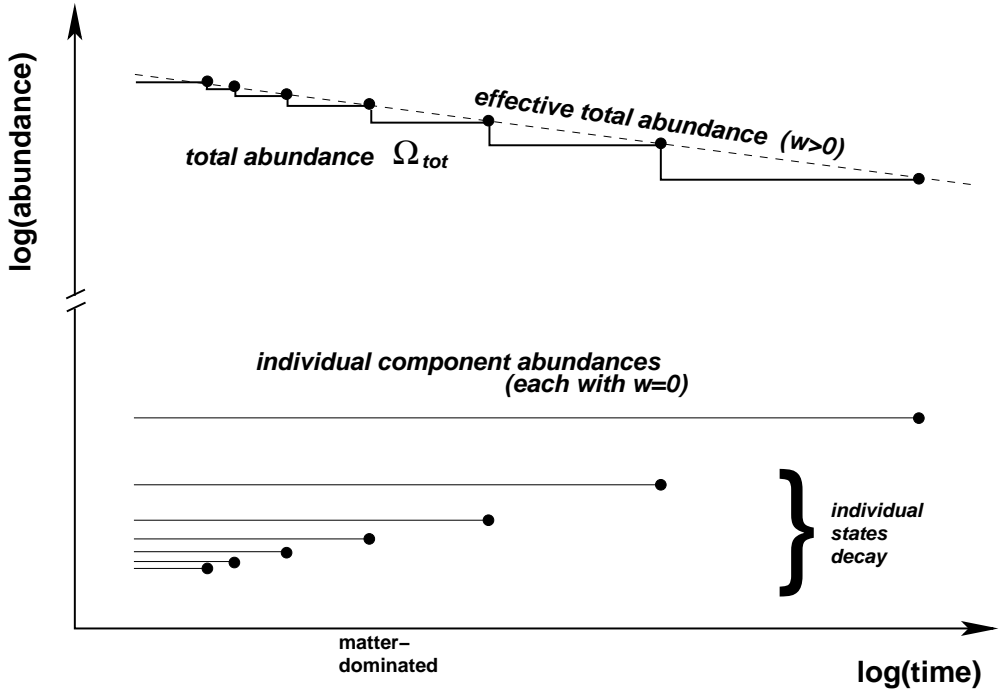}
 }
\vskip -0.2 truein
\caption{A sketch of the total dark-matter abundance in our scenario
during the final, matter-dominated era.
Even though each dark-matter component individually has $w=0$, the
spectrum of lifetimes and abundances of these components conspire 
to produce a time-dependent total dark-matter 
abundance $\Omega_{\rm tot}$ which corresponds to an 
effective equation of state with $w>0$.}  
\label{sketch2}
\end{figure*}
%========================================================================

We have already seen in Eq.~(\ref{wbehav}) 
that the time-dependence of the abundance $\Omega_i$ associated 
with a {\it single}\/ 
dark-matter component can be parametrized 
by its equation-of-state parameter $w$.
It is therefore natural to ask what ``effective'' equation-of-state parameter
$w_{\rm eff}$ might collectively describe our entire dynamical ensemble of 
dark-matter components.
For example, 
even though each individual dark-matter component
behaves as matter (with $w=0$), the decays of these components
at late times
(or even their staggered turn-ons at early times)
might conspire to produce an {\it effective}\/ $w$-value for the entire ensemble which is non-zero.
Such behavior is illustrated in Fig.~\ref{sketch2} for decays that occur during the final
matter-dominated era. 
In all cases and in all cosmological eras, the presence of an effective $w_{\rm eff}$ which differs
(however slightly) from zero would then signify a departure from the traditional dark-matter scenarios.

We can also understand this at a mathematical level.
The fact that each individual dark-matter component has an abundance
which follows the behavior in Eq.~(\ref{wbehav})
with $w=0$ does not guarantee that their {\it sum}\/ $\Omega_{\rm tot}$
must follow the same behavior.
Indeed, the two effects which can alter the time-evolution of the sum $\Omega_{\rm tot}$
in our scenario
are a possible staggered turn-on at early times, and the individually decaying dark-matter 
components at late times.
Thus the time-dependence of $\Omega_{\rm tot}$ need not necessarily follow
Eq.~(\ref{wbehav}) with $w=0$.

One possibility, of course, is that $\Omega_{\rm tot}$ will continue to follow Eq.~(\ref{wbehav}),
but with some other effective value $w_{\rm eff}$.
However, even this outcome 
requires that our individual dark-matter components 
exhibit certain relationships between their abundances and lifetimes 
which need not actually hold for our dark-matter ensemble.
Therefore, in general, we expect that $\Omega_{\rm tot}$ might exhibit
a time-dependence which does not resemble that given in Eq.~(\ref{wbehav}) for
 {\it any}\/ constant $w_{\rm eff}$.
Or, to phrase this somewhat differently, we expect that
in general, 
our effective equation-of-state parameter $w_{\rm eff}$
might itself be time-dependent.
We therefore seek to define a {\it function}\/ $w_{\rm eff}(t)$
which parametrizes a time-dependent equation of state for our dynamical dark-matter
ensemble as a whole.

In order to define such an effective function $w_{\rm eff}(t)$,
let us first recall that the traditional parameter $w$ is fundamentally defined 
through the relation $p=w\rho$
where $p$ and $\rho$ are respectively the pressure and energy density of the ``fluid''
in question.  Of course, in an FRW universe with radius $R$, the conservation law for
energy density $dE= -pdV$ yields the relation
$d(R^3 \rho) = - p d(R^3)$, from which it immediately follows that
$(p+\rho) dR^3/R^3 = -d\rho$ or
$3 (p+\rho) d\log R= -d\rho$.
Recognizing $p+\rho=(1+w)\rho$ and $d\log R= H dt$ where $H$ is the Hubble
parameter, we thus have
\beq
              3H(1+w) ~=~ -{d\log \rho\over dt}~.
\eeq
This is a completely general relation which makes no assumptions
about the time-(in)dependence of $w$.
We may therefore take this to be our fundamental definition for
$w_{\rm eff}(t)$ --- \ie,
\beqn
       w_{\rm eff}(t) & \equiv &  -\left( {1\over 3H} {d\log \rho_{\rm tot}\over dt} + 1\right)~\nonumber\\
                           & = &   \cases{
           - \displaystyle{1\over 2} \left({d\log \Omega_{\rm tot}\over d\log t}\right) & for RH/MD eras \cr
~&~\cr
           - \displaystyle{2\over 3} \left({d\log \Omega_{\rm tot}\over d\log t}\right) + {1\over 3}
                   & for RD era ~.\cr}\nonumber\\
\label{weffdef}
\eeqn
Note that while our derivation has thus far been completely general,
we have specialized to specific cosmological eras in passing to the final
expressions in Eq.~(\ref{weffdef}).  Specifically, we have
written $\rho_{\rm tot}=\Omega_{\rm tot}\rho_{\rm crit}$ and
taken $3H\sim \kappa/t$ where $\kappa=2$ (RH/MD),
$\kappa =3/2$ (RD).

The final expressions in Eq.~(\ref{weffdef}) are easy to interpret physically,
since the double-logarithmic derivatives which appear in these expressions
are nothing but the slopes in the sketches in Figs.~\ref{sketch} and \ref{sketch2}.
However, the important point of this derivation has been to demonstrate that
$w_{\rm eff}$ defined as in Eq.~(\ref{weffdef}) continues to have a direct
interpretation as a true equation-of-state parameter relating energy density and pressure,
even when $w_{\rm eff}$ is time-dependent.
No other definition of $w_{\rm eff}$ would have had this property.

The results in Eq.~(\ref{weffdef}) provide a relation between $w_{\rm eff}$ and $\Omega_{\rm tot}(t)$.
However, it is also possible to derive a similar relation between 
$w_{\rm eff}$ and $\eta$.
Assuming that we restrict our attention to periods of time {\it after}\/ all staggered turn-ons
are complete (so that the identity of the dark-matter component associated with $\Omega_0$ is fixed),
it trivially follows from the definition of $\eta$ in Eq.~(\ref{etadef})
that
\beq
        {d\log (1-\eta)\over d\log t} ~=~ \cases{
             - \displaystyle \left({d\log \Omega_{\rm tot}\over d\log t}\right) & RH/MD eras\cr
             - \displaystyle \left({d\log \Omega_{\rm tot}\over d\log t}\right) + {1\over 2} & RD era~.\cr}
\eeq
Using the results in Eq.~(\ref{weffdef}), we therefore find that
\beq
       w_{\rm eff}(t) ~=~   \cases{
            \displaystyle{1\over 2} \left[{d\log (1-\eta)\over d\log t}\right] & RH/MD eras \cr
                 ~&~\cr
            \displaystyle{2\over 3} \left[{d\log (1-\eta)\over d\log t}\right] & RD era~.\cr}\nonumber\\
\label{weffdefeta}
\eeq
It therefore follows that decreasing $\eta$ corresponds to positive $w_{\rm eff}$, and
vice versa.
As a self-consistency check, we also observe that the standard paradigm --- which has $\eta=0$ for
all times ---  has $w_{\rm eff}=0$ for all times as well.

\subsubsection{ $\Omega_{\rm tot}$, $\eta$, and $w_{\rm eff}$:~
Qualitative time-dependent behaviors}

Having defined the three quantities
$\lbrace \Omega_{\rm tot}, \eta,w_{\rm eff}\rbrace$
that we will use in order to characterize
a general configuration of our dark-matter ensemble,
we now seek to understand the time-dependence that will
be generically exhibited by these quantities
across the different cosmological eras.
As we shall see,  
several qualitative observations can be made even without further assumptions
concerning the individual abundances and decay widths within our ensemble.

Given the sketch in Fig.~\ref{sketch}, it is perhaps easiest
to understand the qualitative behavior of $\Omega_{\rm tot}$ as a function of time.
There is, of course, the time-dependence for $\Omega_{\rm tot}$ which can be associated
with the regular Hubble expansion of the universe and which causes
$\Omega_{\rm tot}$ to increase
during the radiation-dominated era.  This aspect of the time-dependence 
is generally common to all dark-matter scenarios. 
However, as discussed above, there are two additional 
effects which are specifically associated with our dynamical dark-matter scenario
and which cause $\Omega_{\rm tot}$ to experience a further time-dependence.
The first of these is the possibility of a ``staggered'' turn-on across the different
dark-matter components in our ensemble --- \ie, the possibility that some components will
remain longer than others in a state in which 
their abundances $\Omega_i$
grow rapidly as $\sim t^2$ but 
contribute to the total {\it dark-energy}\/ abundance rather than to $\Omega_{\rm tot}$.
This feature, when present, will thus generally cause $\Omega_{\rm tot}$ to experience
a more gradual growth 
(but also a greater eventual maximum value) 
than would otherwise occur in a more traditional dark-matter scenario.
However, at later stages of the cosmological evolution (particularly during the final,
matter-dominated era), we see that our different dark-matter components have
a broad spectrum of lifetimes and decay widths.  
This causes $\Omega_{\rm tot}$ to experience a slow step-wise decline before
finally reaching zero upon the decay of the last-surviving dark-matter component within the ensemble.

Similar qualitative arguments also apply to the time-evolution of $\eta$. 
An initial value of $\eta$ is implicitly determined 
once the abundances for each of the dark-matter components are established.
Of course, if this occurs during an inflationary period, it is possible
that certain more massive components will have inflated away by the time the
inflationary period ends.  If this is the case, then we may regard
the ``initial'' value of $\eta$ to be the value of $\eta$ at the
end of the inflationary period.
  
In general, after that point, the evolution of $\eta$ 
can experience as many as three distinct phases.
Let $t_1$ denote the time at which the last (lightest) dark-matter
component has turned on, and let $t_2$ denote the time at which
the most massive dark-matter component decays.
Assuming $t_2>t_1$, there are therefore three distinct time
intervals which become relevant.

During $t_1\leq t \leq t_2$,
each of the individual dark-matter abundances experiences a common
overall time-dependent scaling behavior
as the universe evolves.  As a result, 
the ratio between the abundances of these components
remains fixed. In other words, $\eta$ remains frozen
during this period (even though $\Omega_{\rm tot}$ may continue to vary).

For $t>t_2$,
by contrast,
it is clear that the decays of the more massive dark-matter components have the cumulative
effect of decreasing $\Omega_{\rm tot}$ without altering $\Omega_0$. 
What results, then, is 
a step-by-step, threshold-by-threshold decline in the value of $\eta$.  
This process continues until only
a single dark-matter component survives and $\eta$ reaches zero.
By assumption, in our scenario this will not happen until the distant future,
at $t=\Gamma_1^{-1}\gg t_{\rm now}$ where $\Gamma_1$ is the decay width of
the second-lightest dark-matter component.

However, during the period $t<t_1$, a ``staggered'' turn-on for the individual
dark-matter components can also generally induce a non-trivial time-evolution for $\eta$.
Indeed, each time a new lightest dark-matter component turns on,
its abundance $\Omega'_0$ suddenly contributes to the total dark-matter
abundance $\Omega_{\rm tot}$.  This abundance
$\Omega'_0$ also displaces the previous largest individual abundance $\Omega_0$.
We therefore find that with each such successive turn-on, $\eta$   
experiences a shift in its value:
\beq
    \eta \equiv 1- {\Omega_0\over \Omega_{\rm tot}} ~~~\longrightarrow~~~
    \eta' \equiv 1- {\Omega'_0\over \Omega_{\rm tot} + \Omega'_0}~.
\label{etashift}
\eeq
By assumption, $\Omega'_0>\Omega_0$.
However, it is easy to see from Eq.~(\ref{etashift}) that $\eta'<\eta$ only
if $\Omega'_0>\Omega_0/\eta$.
Since $0\leq \eta\leq 1$, we see that this condition is guaranteed to be satisfied only if $\eta=1$,
and guaranteed {\it not}\/ to be satisfied only if $\eta= 0$. 
In all other cases,
this condition may or may not be satisfied, and this will cause $\eta$ to either decrease or increase,
respectively.
We also observe that in a very rough sense, $\eta$ tends to stabilize and avoid either
the $\eta=1$ or the $\eta=0$ extremes:  as $\eta \to 1$, it becomes easier and easier
to satisfy the constraint that drives $\eta$ lower, while as $\eta\to 0$, it becomes
easier and easier to satisfy the constraint that pushes $\eta$ higher. 
Indeed, if we imagine that each new abundance $\Omega'_0$ somehow has a random value 
greater than the  previous $\Omega_0$, we can envision an ``oscillatory''
behavior in which $\eta$ varies between its two limits.
Unfortunately, we cannot be more specific about this behavior 
without knowing something further about the 
individual abundances that exist during such a staggered turn-on phase.

We therefore conclude that $\eta$ will take an initial value once the abundances
are established, and that this value can then undergo a non-trivial time dependence
if there is an initial period during which a staggered turn-on occurs.
After the staggered turn-on is complete, $\eta$ will remain frozen until late times
when our individual dark-matter components begin to decay.
This will then cause $\eta$ to fall monotonically, ultimately vanishing when only
the single longest-lived dark-matter component remains.
However, it is regarded to be a fundamental property of our scenario
that $\eta$ is nevertheless presumed to be significantly different from 
zero at the present time.

Finally, we turn to the behavior of $w_{\rm eff}$ as a function of time.
However, given the relations in Eq.~(\ref{weffdefeta}),
it is relatively straightforward to map out the rough time-dependence of $w_{\rm eff}$.
During the staggered turn-on phase (\ie, $t<t_1$),
we have seen that $\eta$ may either increase or decrease;
this implies that $w_{\rm eff}$ may be either negative or positive.
Moreover, the fact that $\eta$ tends to stabilize during this period, avoiding its extreme
values at $\eta=0$ or $\eta=1$, implies that  
$w_{\rm eff}$ will likewise tend to stabilize around zero,
with positive values of $w_{\rm eff}$ ultimately followed by negative values,
and vice versa.   
As indicated above, however, this
assumes that each new abundance $\Omega'_0$ that turns on has a value 
which is greater than the  previous $\Omega_0$ but is otherwise somewhat random.

During the period $t_1\leq t \leq t_2$, by contrast, 
the behavior of $w_{\rm eff}$ is far simpler to describe:  we simply
have $w_{\rm eff}=0$.  
This is completely in accord with our observation
that $\eta$ stays frozen during this period, and that our dynamical ensemble
is behaving as ordinary dark-matter during this period (except with a non-zero
value of $\eta$).
Finally, for the period $t>t_2$ after our individual dark-matter
components have begun to decay, we have argued that $\eta$ is monotonically
decreasing.  This implies that $w_{\rm eff}$ is strictly positive during this period,
a feature which is illustrated in Fig.~\ref{sketch2} and which again serves
as a cosmological ``smoking gun'' for our dynamical dark-matter scenario.

\subsection{A signature of dynamical dark matter:  ~Time-evolution of $\Omega_{\rm tot}$,
$\eta$, and $w_{\rm eff}$ during the final matter-dominated epoch}

As just discussed,
one of the most important signatures of our dynamical dark-matter framework
is the fact that {\it the total dark-matter abundance 
$\Omega_{\rm tot}$ is a time-evolving quantity --- even during
the current matter-dominated epoch.} 
Within such a framework, it is therefore only to be regarded as an accident 
that this quantity happens to reproduce a specific observed value 
$\Omega_{\rm CDM}\approx 0.23$ at the present time.

With only a few additional assumptions, 
it turns out that we can explicitly calculate
the time-evolution of the 
total dark-matter abundance $\Omega_{\rm tot}$ 
during this epoch.
We can also explicitly calculate the time-dependence of $\eta$,
and the resulting equation of state $w_{\rm eff}(t)$.
In the rest of this section, we shall therefore concentrate
on the final matter-dominated epoch.  Indeed, this is the epoch
during which
a non-trivial time-evolution for $\Omega_{\rm tot}$ 
arises only because of the decays of the individual dark-matter components
within our ensemble.

Within this era, each dark matter component $\phi_i$ has a relative abundance $\Omega_i$
which remains constant until it decays at a time $t\sim \tau_i\equiv \Gamma_i^{-1}$.
Taking this decay to be nearly instantaneous, we can thus write
\beq
         \Omega_i(t) ~=~ \Omega_i \,\Theta(\tau_i -t)~,
\eeq
whereupon we see that
\beq
         {d \Omega_{\rm tot}(t)\over dt} ~=~ 
         \sum_i \Omega_i {d\over dt} \Theta(\tau_i - t) ~=~
          -\sum_i \Omega_i \delta(\tau_i -t)~
\label{step1}
\eeq
where we have defined $\Omega_{\rm tot}(t)\equiv \sum_i \Omega_i(t)$
and used the relation $d\Theta(x)/dx= \delta(x)$ where
$\delta (x)$ is the Dirac $\delta$-function.
In the limit that we truly have a large number of dark-matter
states, we can imagine that the spectra of decay widths $\Gamma_i$
and decay times $\tau_i\equiv \Gamma_i^{-1}$
are nearly continuous, with continuous variables $\Gamma$ and $\tau$.
With this approximation, we can
view $\Omega_i$ as a continuous function $\Omega(\tau)$
and convert the sum over states to an integral, \ie,
\beq
           \sum_i ~\Longrightarrow~ \int d\tau \, n_\tau(\tau)
\eeq
where $n_\tau(\tau)$ is the density of dark-matter states per unit of $\tau$,
expressed as a function of $\tau$.
Eq.~(\ref{step1}) then becomes 
\beqn
         {d \Omega_{\rm tot}(t)\over dt} &=& 
           -\int d\tau\, \Omega(\tau) \, n_\tau(\tau) \,  \delta (\tau-t) \nonumber\\
          &=& -\Omega(t) n_\tau(t)~.
\label{taut}
\eeqn

In general, the quantities $n(\tau)$ and $\Omega(\tau)$ are unspecified,
their properties depending on the particular dark-matter scenario
under study and the specific features of our dark-matter ensemble.
However, it will prove convenient to parametrize these quantities
in terms of their scaling behaviors as functions of $\Gamma$:
\beq
         \Omega(\Gamma) ~\approx ~ A \Gamma^\alpha~,~~~~~~~~~
         n_\Gamma(\Gamma) ~\approx ~ B \Gamma^\beta~
\label{step2}
\eeq
with overall (generally dimensionful) coefficients $(A,B)$ and scaling exponents $(\alpha,\beta)$.
Since the abundances of states in our scenario generally have
an inverse relation to their decay widths, we expect that $\alpha <0$.
Note that $n_\Gamma$ in Eq.~(\ref{step2}) is the density of states
per unit of $\Gamma$, whereupon it follows that
\beq
     n_\tau ~=~ n_\Gamma 
                 \left| {d\Gamma\over d\tau} \right| 
     ~=~ \Gamma^2 \,n_\Gamma~.
\label{step3}
\eeq
We thus find that $\Omega(\Gamma) n_\tau(\Gamma)\sim AB \Gamma^{\alpha+\beta+2}$, or
equivalently $\Omega(\tau) n_\tau(\tau)\sim AB \tau^{-\alpha-\beta-2}$. 
Use of Eq.~(\ref{taut}) then leads to the result
\beq
         {d \Omega_{\rm tot}(t)\over dt}  ~=~  - AB \, t^{-\alpha-\beta-2}~.
\label{preresult}
\eeq
Imposing the condition that $\Omega_{\rm tot}=\Omega_{\rm CDM}$ at the present time $t=t_{\rm now}$
and assuming that $\alpha+\beta\not= -1$
then leads to the solution
\beq
     \Omega_{\rm tot}(t) ~=~ \Omega_{\rm CDM} + {A B\over \alpha+\beta+1} \left(
           t^{-\alpha-\beta-1} - t_{\rm now}^{-\alpha-\beta-1}\right)~.
\label{result}
\eeq
For $\alpha+\beta= -1$, by contrast, we have
the solution
\beq
     \Omega_{\rm tot}(t) ~=~ \Omega_{\rm CDM} - AB \log\left( {t\over t_{\rm now}}\right)~.
\label{logresult}
\eeq

Under the assumptions in Eq.~(\ref{step2}), the results in Eqs.~(\ref{result})
and (\ref{logresult})
are completely general in a matter-dominated era.  
Moreover, it is clear from Eqs.~(\ref{result}) and (\ref{logresult}) 
that in all cases, $\Omega_{\rm tot}$ decreases with time.
This is precisely as expected, since all of the time dependence of $\Omega_{\rm tot}$ in
a matter-dominated era arises due to the decays of the individual
dark-matter components within the ensemble.
Notice that some of these functional forms for $\Omega_{\rm tot}$ actually predict that $\Omega_{\rm tot}(t)$ 
will eventually hit zero;  this is also not unexpected, since this corresponds to the final 
decay of the last remaining dark-matter component in the ensemble.
Needless to say, we should not consider any of these $\Omega_{\rm tot}(t)$ functions 
beyond the times when they might hit zero.
Nevertheless, as long as our dark-matter ensemble obeys the scaling laws in
Eq.~(\ref{step2}), the functions given in
Eqs.~(\ref{result}) and (\ref{logresult}) 
correctly describe the behavior of the corresponding
total dark-matter abundances $\Omega_{\rm tot}(t)$.

Given the results in Eqs.~(\ref{result}) and (\ref{logresult}) 
as well as the definition in Eq.~(\ref{weffdef}) for a matter-dominated era, we
can also obtain a solution for the time-dependent equation-of-state parameter $w_{\rm eff}(t)$
associated with our ensemble of decaying dark-matter states.
For $x\equiv \alpha+\beta\not = -1$, we find
\beq
        w_{\rm eff}(t) ~=~ {(1+x) w_\ast \over  2w_\ast + (1+x-2w_\ast) (t/t_{\rm now})^{1+x}}~
\label{weffresult}
\eeq
where 
\beq
         w_\ast ~\equiv~  w_{\rm eff}(t_{\rm now}) ~=~ {AB\over 2\Omega_{\rm CDM} t_{\rm now}^{1+x}}~.
\label{wnow}
\eeq
Note that for $w_\ast \ll 1$, this result is fairly well-approximated by
\beq
        w_{\rm eff}(t) ~\approx~  w_\ast \left({t\over t_{\rm now}}\right)^{-x-1}~.
\label{weffresult2}
\eeq
By contrast, for $x= -1$, we instead obtain
\beq
      w_{\rm eff}(t) ~=~ {w_\ast\over 1-2w_\ast \log(t/t_{\rm now})}
\label{weffresultlog}
\eeq
where 
\beq
         w_\ast ~\equiv~  w_{\rm eff}(t_{\rm now}) ~=~ {AB\over 2\Omega_{\rm CDM}}~.
\label{wnowlog}
\eeq

The behavior of the results in Eqs.~(\ref{weffresult}) and (\ref{weffresultlog}) 
depends critically on the relationship between $x$ and $w_\ast$.
For $1+x<2w_\ast$, we find that $w_{\rm eff}$ always {\it increases}\/ monotonically
as a function of $t$ before reaching $w_\ast$ at $t=t_{\rm now}$.
By contrast, for $1+x>2w_\ast$, this function {\it decreases}\/ monotonically
before reaching $w_\ast$ at $t=t_{\rm now}$.
Finally, for $1+x=2w_\ast$, we have the exact result that $w_{\rm eff}(t)=w_\ast$
for all $t$.  This (admittedly fine-tuned) case illustrates that it {\it is}\/ possible
to achieve a time-independent equation-of-state parameter $w_{\rm eff}=w_\ast$ under the
assumptions in Eq.~(\ref{step2}), and moreover that this value of $w_\ast$ can be tuned to
any positive value desired.
This is indeed the situation illustrated in Fig.~\ref{sketch2}, which is plotted for
$\alpha<0$ and $\beta>0$.  

The above qualitative descriptions indicate the history of $w_{\rm eff}(t)$ prior to the 
present day.  However, in general, this same increasing or decreasing behavior
continues for $t>t_{\rm now}$ (\ie, through and beyond the current epoch), 
with one important caveat:  for $1+x<2w_\ast$,
we see that $w_{\rm eff}(t)$ not only continues to increase, but eventually hits a pole.
However, such poles represent the locations at which 
the corresponding $\Omega_{\rm tot}$-functions have zeroes.  
These poles are therefore unphysical, signalling the decay of the last component within
our dynamical dark-matter ensemble, and we can restrict our analysis of these functions 
to times preceding these critical values.
   
If our dynamical dark-matter scenario is to be in rough agreement with cosmological
observations, we expect that $w_\ast$ today   
should be fairly small (since traditional
dark ``matter'' has $w=0$).  We also expect that the function $w_{\rm eff}(t)$ should not have experienced
strong variations within the recent past.  This suggests that
situations with $x< -1$ are likely to be phenomenologically preferred over those with $x\geq -1$,  
since having $x<-1$
ensures that $0\leq w_{\rm eff}(t)\leq w_\ast$ for all $t<t_{\rm now}$.
Indeed, the more negative $x$ becomes, the closer to vanishing $w_{\rm eff}(t)$ remains
before finally reaching $w_\ast$ at $t=t_{\rm now}$.
However, depending on the detailed properties of the particular
realization of our dynamical dark-matter scenario under study,
values of $x$ near $-1$ or slightly above may also be phenomenologically
acceptable.

Finally, we may also use these results to solve for $\eta$ as a function
of time.
For $x\equiv \alpha+\beta\not = -1$, we find
\beq
        \eta(t) ~=~ {
          2w_\ast + [\eta_\ast (1+x)-2w_\ast] (t/t_{\rm now})^{1+x}~
            \over  
          2w_\ast + [1+x            -2w_\ast] (t/t_{\rm now})^{1+x}}~
\label{etaresult}
\eeq
where $w_\ast$ is given in Eq.~(\ref{wnow}) and where 
\beq
         \eta_\ast ~\equiv~  \eta(t_{\rm now}) ~=~ 1- {\Omega_0\over \Omega_{\rm CDM}}~.
\label{etanow}
\eeq
Likewise, for $x= -1$, we have
\beq
    \eta(t) ~=~ 
       {\eta_\ast -2w_\ast \log(t/t_{\rm now})
        \over
             1    -2w_\ast \log(t/t_{\rm now}) }
\label{etaresultlog}
\eeq
where $w_\ast$ is given in Eq.~(\ref{wnowlog}).
It is not surprising that $\eta$, unlike $w_{\rm eff}$, depends on {\it two}\/
independent dimensionless quantities $w_\ast$ and $\eta_\ast$, since the very definition
for $\eta$ introduces a new quantity $\Omega_0$ which had not previously
appeared.

Note that all time-dependence for $\eta(t)$ cancels, with $\eta(t)\approx \eta_\ast$ for all $x$, if
either $w_\ast \to 0$ or $\eta_\ast\to 1$. 
This makes sense, since in the first case
$\Omega_{\rm tot}$ does not change
while in the second case $\Omega_0\to 0$.
These are the only two ways in which $\eta$ can remain constant.
In all other cases, however, $\eta(t)$ is always a decreasing function of time, as expected.
We also see from Eqs.~(\ref{etaresult}) and (\ref{etaresultlog}) that
$\eta(t)\to 1$ for all $x\geq -1$
as $t/t_{\rm now}\to 0$. 
Indeed, this holds regardless of the values of 
$w_\ast$ or $\eta_\ast$.

While these characteristics successfully conform to our expectations concerning the behavior of $\eta(t)$,
there are some features that the functional forms in Eqs.~(\ref{etaresult}) and (\ref{etaresultlog})
do not accurately capture.
For example, if $x+1 < 2w_\ast/\eta_\ast$, these functions predict that $\eta(t)$ will eventually become
negative beyond a certain late time.
Moreover, while these functions resemble those for $w_{\rm eff}$ in that they properly capture 
the pole that results when $\Omega_{\rm tot}\to 0$, they do not necessarily approach $\eta\to 0$
before hitting this pole.

The reason behind these failures is easy to understand.
Unlike $\Omega_{\rm tot}$ and $w_{\rm eff}$, the quantity $\eta$ has a special characteristic
not shared by the others:  it is sensitive not only to $\Omega_{\rm tot}$, but also to $\Omega_0$.
Features such as having $\eta$ reach zero but not become negative are extremely sensitive to
the value of $\Omega_0$ and the fact that $\Omega_{\rm tot}$ must exactly hit $\Omega_0$
after all but the lightest dark-matter component have decayed.
Indeed, these features are extremely sensitive to the fine-tuned and ultimately discrete nature of the lightest
dark-matter components, and this is precisely the sort of information that our scaling assumptions
in Eq.~(\ref{step2}) are incapable of modelling.
Thus, while we may view the functions in Eqs.~(\ref{etaresult}) and (\ref{etaresultlog}) as
being reasonably accurate for most portions of the cosmological evolution in the matter-dominated
era, we should not maintain this expectation beyond a certain time when all but a few dark-matter
components have decayed.

To summarize, then, in this section
we have presented a dynamical multi-component dark-matter scenario 
in which individual component abundances and lifetimes are balanced and distributed 
across the components in such a way that constraints from BBN and other backgrounds
are potentially satisfied.
An important part of this scenario is the proposition that both $\eta$ and $w_{\rm eff}$ are 
different from zero at the present time, the former significantly so,
and that components of the dark matter are actively decaying prior to, during, and beyond
the current epoch.  As a result, cosmological quantities such as $\Omega_{\rm tot}$ experience
a time-evolution which transcends that due to the ordinary expansion of the universe.

%============================================================================
\section{Dynamical Dark Matter Meets the Incredible Bulk}

Thus far, we have done little more 
than present a new framework for dark-matter physics.
In particular, we have not yet demonstrated that an ensemble of dark-matter states can
easily be assembled in which the individual component abundances are naturally
balanced against lifetimes in a well-motivated way. 
In this section, however, we shall demonstrate that an infinite tower of Kaluza-Klein states
propagating in the bulk of large extra spacetime dimensions
naturally constitutes an ensemble of states with the desired properties.
As we shall see,
this occurs because KK towers 
obey a special ``balancing'' constraint which relates the lifetimes of individual
KK modes to their abundances. 
Specifically, we shall demonstrate that the KK modes within a generic KK tower 
exhibit abundances $\Omega_i$ and SM decay widths $\Gamma_i$ which obey
an inverse relation of the form anticipated in Eq.~(\ref{step2}), \ie,
\beq
      \Omega_i \Gamma_i^{-\alpha}~\sim~ {\rm constant}
\label{ab}
\eeq
for some $\alpha<0$.
This constraint ultimately emerges as a 
consequence of the non-trivial interplay between physics in the bulk and physics on the brane.

\subsection{General setup}

For simplicity, we shall consider our spacetime to take the form ${\cal M}_4\times S_1/\IZ_2$,
where ${\cal M}_4$ denotes ordinary four-dimensional Minkowski spacetime and $S_1/\IZ_2$ denotes 
a line segment which is realized as a $\IZ_2$ orbifold of a circle of radius $R$.
We shall take $z^M\equiv (x^\mu,y)$ to denote the coordinates on this spacetime, with the $\IZ_2$ orbifold action
identified as $y\to -y$,
and imagine that the Standard Model is restricted to a brane at the fixed point $y=0$.  We are therefore
considering a ``toy'' ADD-like scenario~\cite{ADD} with a single flat extra dimension.
Despite the simplicity of this toy model,
we are making no assumptions at this stage about relevant mass scales or the full
number of extra spacetime dimensions that might actually exist in a more fully realized scenario.  
Indeed, we believe that most of the desired properties that emerge from this scenario are likely to be 
retained if we imagine that our spacetime contains additional extra dimensions, or is 
warped~\cite{RS} rather than flat.

In such a scenario, all fields that propagate in the ``bulk'' are necessarily singlets with
respect to all Standard-Model gauge forces.  As a result, such fields can have at most highly
suppressed (\eg, gravitational) interactions 
with the Standard-Model fields,
and thus appear as dark-matter candidates.
Such fields might include the graviton, axion, and other moduli fields.
For simplicity, we shall consider the case in which the bulk field 
is a five-dimensional scalar $\Phi$,
but we shall make no further assumptions about its properties.

Neglecting gravity,  
and with $\psi_i$ generically denoting the Standard-Model fields, 
we see that such a scenario therefore has an action of the form
\beq
S ~=~ \int d^4 x \, dy \, \left[
           {\cal L}_{\rm bulk}(\Phi) + \delta(y)  {\cal L}_{\rm brane}(\psi_i,\Phi)  \right]~.
\label{genform}
\eeq
In general, we may assume that our five-dimensional bulk action 
takes the form
\beq
  {\cal L}_{\rm bulk} ~=~ \half \,\partial_M \Phi^\ast  \partial^M \Phi - \half M^2 |\Phi|^2~
\label{Lbulk}
\eeq
where $\partial_M$ denotes a five-dimensional derivative and where
$M$ is an unspecified bulk mass.  In certain cases, specific symmetries 
may restrict us to the case with $M=0$, but we shall leave $M$ general 
until further notice.

Likewise, the brane action will generically consist of two contributions ---
the usual Standard-Model action ${\cal L}_{\rm SM}$, and an action ${\cal L}_{\rm int}$ 
which arises due to
the interactions between $\Phi$ and the Standard-Model fields:  
\beq
             {\cal L}_{\rm brane} ~=~ {\cal L}_{\rm SM} + {\cal L}_{\rm int}~.
\label{branelag}
\eeq
In general, there are two types of interactions which will concern us.
The first class of interactions result in explicit couplings between $\Phi$ and the Standard-Model fields,
and will ultimately be responsible for allowing $\Phi$ to decay into Standard-Model states.  
We shall discuss such interactions in Sect.~III.B.~
There is, however, another possible type of interaction term which can also
appear within ${\cal L}_{\rm int}$:  this is a possible ``brane mass'' for $\Phi$ itself,
\ie,
\beq     
     {\cal L}_{\rm int} ~\supset~  - \half m^2 |\Phi|^2~.
\label{branemass}
\eeq
Such a brane mass can emerge as an effective operator arising due to  
perturbative or non-perturbative dynamics wholly restricted to the brane.
Note that this brane-mass term must not be confused with 
the primordial bulk mass that appears in Eq.~(\ref{Lbulk});  rather, this term 
has its origins within the physics on the brane itself, and 
appears as part of ${\cal L}_{\rm brane}$
within Eq.~(\ref{genform}) rather than ${\cal L}_{\rm bulk}$.

These minimal assumptions are already sufficient to permit us to 
understand the nature of the resulting Kaluza-Klein spectrum for $\Phi$.
Indeed, the following results are similar to those previously obtained in Ref.~\cite{DDGAxions}. 
As appropriate for compactification on the line segment $S^1/\IZ_2$,
we can decompose our five-dimensional field $\Phi$ in terms of
an infinite tower of four-dimensional modes $\phi_k$,
\beq
     \Phi(x^\mu,y)~=~ {1\over \sqrt{2\pi R}}\, 
            \sum_{k=0}^\infty  r_k \, \phi_k(x^\mu) \cos\left({ky\over R}\right)~;
\label{KKdecomp}
\eeq
the normalization factors
\beq
           r_k~\equiv~ \cases{ 1 & for $k=0$\cr  
                               \sqrt{2} & for $k>0$ \cr}
\eeq
are designed to ensure that each mode $\phi_k$ has a canonically
normalized kinetic term in the resulting four-dimensional theory. 
We then find that
\beqn S &=& \int d^4 x \, dy \, \left\lbrack
  \half \,\partial_M \Phi^\ast  \partial^M \Phi - \half M^2 |\Phi|^2
       -\half \delta(y) m^2 \Phi^2\right\rbrack \nonumber\\
     &=& \int d^4 x \left(\half \sum_{k=0}^\infty \partial_\mu \phi_k^\ast \partial^\mu \phi_k - 
         \half \sum_{k,\ell=0}^\infty {\cal M}^2_{k\ell} \, \phi_k \phi_\ell^\ast \right)~
\eeqn
where the Kaluza-Klein (mass)$^2$ matrix is given by 
\beq
       {\cal M}^2_{k\ell} ~=~ \left( {k\ell\over R^2} + M^2 \right) \delta_{k\ell} + r_k r_\ell \, m^2~.
\label{massmatrix}
\eeq

Given these results, we 
see that this mass matrix would have been diagonal
were it not for the brane mass term. 
This in turn implies that the KK mass eigenstates $\phi_\lambda$
necessarily differ from the KK momentum eigenstates $\phi_k$ --- \ie, there 
is a non-trivial {\it mixing}\/ that is induced as a result of the KK mass.
This mixing turns out to be critical for our analysis.
In general, we may characterize the degree of mixing in terms of the dimensionless parameter
\beq
              y~\equiv~ {1\over mR}~.
\eeq
For $y\gg 1$ the mass matrix is essentially diagonal;
this is what trivially occurs, for example, in the four-dimensional $R\to 0$ limit in which 
the excited KK modes decouple.
By contrast, in the opposite limit $y\ll 1$, the mixing is essentially maximal
across all of the eigenmodes.

It is possible to describe the solutions for the eigenvalues $\lambda^2$ of the (mass)$^2$ matrix 
in closed form, and thereby obtain explicit expressions for the corresponding mass eigenstates.
The eigenvalues turn out to be the solutions to the transcendental equation
\beq
           \pi m^2 R \cot\left( \pi R \sqrt{ \lambda^2-M^2 } \right) ~=~ \sqrt{\lambda^2-M^2}~.
\label{coteq}
\eeq
If $m$ were zero (\ie, no brane mass), the solutions to this equation would be nothing but the
expected eigenvalues
\beq
           \lambda_n^2 ~=~ M^2 + {n^2\over R^2}~,~~~~~~ n\in \IZ~,
\label{spect1}
\eeq
and more generally this remains approximately true when $m\ll 1/R$, \ie, when $y\gg 1$.
What is perhaps surprising, however, is that the presence of a non-zero brane mass does not result
in a further additive shift in this mass spectrum for the KK tower (as does the bulk mass term), 
but instead distorts the
lower mass eigenstates in the tower 
so that they approximately follow the alternate spectrum
\beq
           \lambda_n^2 ~=~ M^2 + {(n+\half)^2\over R^2}~,~~~~~~ n\in \IZ~.
\label{spect2}
\eeq
Remarkably, this is precisely the spectrum which we would normally associate with a 
five-dimensional field $\Phi$ which is taken to
be {\it anti-periodic}\/ (rather than periodic) around the extra-dimensional circle prior to orbifolding!
Indeed, for general values of $y$, the solutions $\lambda_n$ of Eq.~(\ref{coteq}) tend to follow 
the spectrum in Eq.~(\ref{spect2}) for $n\ll \pi/y^2$, 
while they follow the spectrum in Eq.~(\ref{spect1}) for $n\gg \pi/y^2$ 
and smoothly transition between the two spectra for intermediate values $n\sim \pi/y^2$. 
As we have discussed above, this unusual behavior is the consequence of the non-trivial interplay
between brane and bulk physics, and may have applications beyond its appearance here.

For each mass eigenvalue $\lambda$, we can also solve for the 
corresponding mass eigenstate $\ket{\phi_\lambda}$ as a linear combination 
of the KK-momentum eigenstates $\ket{\phi_k}$.
We find the exact result~\cite{DDGAxions}
\beq
            \ket{\phi_\lambda} ~=~  A_\lambda \, \sum_{k=0}^\infty 
            \, {r_k \tilde \lambda^2 \over \tilde \lambda^2 - k^2 y^2} \, \ket{\phi_k}~
\label{mixed}
\eeq
where we have defined the dimensionless eigenvalues 
\beq
           \tilde \lambda ~\equiv~ \sqrt{\lambda^2 - M^2}/m
\eeq
and where
\beq
               A_\lambda~\equiv~ {\sqrt{2}\over \tilde \lambda} 
          \, {1\over \sqrt{ 1 + \pi^2/y^2 + \tilde \lambda^2}}~.
\label{Alamdef}
\eeq  
Given these results, it is straightforward to convert between the mass-eigenstate basis $\ket{\phi_\lambda}$ and
the KK-momentum basis $\ket{\phi_k}$. 
It turns out that there are two specific groups of matrix elements involved in this conversion
which will be of particular interest to us. 
The first involves the KK zero-mode $\phi_{k=0}$, for which we have
the matrix elements
\beq
       \langle \phi_\lambda | \phi_{k=0}\rangle ~=~ A_\lambda~.
\label{zeromodematrixelement}
\eeq
However, the second concerns 
the {\it projection}\/ of the five-dimensional bulk field $\Phi(y)$ 
onto the Standard-Model brane at $y=0$, \ie,
\beq
          \phi'~\equiv~ \Phi(y)\bigl|_{y=0} ~=~ \sum_{k=0}^\infty r_k \phi_k~.
\label{phiprime}
\eeq
For this projection field,
we likewise have the matrix elements
\beqn
         \langle \phi_\lambda | \phi' \rangle  &=& 
           A_\lambda \, \sum_{k=0}^\infty {r_k^2 \tilde \lambda^2 \over \tilde \lambda^2 - k^2y^2} \nonumber\\
       &=&  
            {\pi \tilde \lambda^2 \over y} \cot\left( {\pi \tilde \lambda \over y}\right) A_\lambda ~=~
             \tilde \lambda^2 A_\lambda~
\label{phiprimematrixelement}
\eeqn
where we have made use of Eq.~(\ref{coteq}) in the final equality.

%================== FIGURE ============================================
\begin{figure}[t!]
\centerline{
   \epsfxsize 3.7 truein \epsfbox{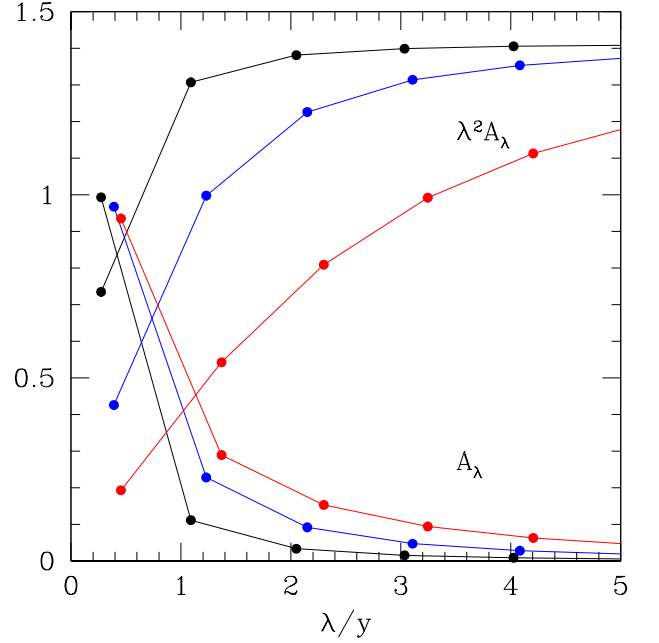} 
 }
\vskip -0.2 truein
\caption{Values of $A_\lambda$ (falling curves) 
and $\tilde \lambda^2 A_\lambda$ (rising curves), plotted as functions
of the mass eigenvalues $\tilde \lambda/y= \lambda R$ for $y=\pi$ (black),  
$y=\sqrt{\pi}$ (blue), and $y=1$ (red). 
For each $y$, there are only a discrete set of corresponding 
allowed eigenvalues $\tilde\lambda$ (indicated with solid dots);  note that the quantity 
$\tilde \lambda/y = \sqrt{\lambda^2-M^2} R$ takes values closer to $\IZ+1/2$ near the bottom
of each tower and shifts to values closer to $\IZ$ as $\lambda$ increases.  In each case, we 
see that $A_\lambda$ falls with increasing $\tilde\lambda$,  while $\tilde \lambda^2 A_\lambda$ 
increases with increasing $\tilde \lambda$ and ultimately
reaches an asymptote $\tilde \lambda^2 A_\lambda\to \sqrt{2}$ as $\tilde \lambda \to \infty$.}  
\label{plot1}
\end{figure}  
%========================================================================

In Fig.~\ref{plot1}, we plot
the values of $A_\lambda$ and $\tilde \lambda^2 A_\lambda$ 
as functions of $\tilde \lambda$ and $y$.
We see that $A_\lambda$ falls with increasing $\tilde\lambda$,  while $\tilde \lambda^2 A_\lambda$ 
increases with increasing $\tilde \lambda$ and ultimately
reaches an asymptote $\tilde \lambda^2 A_\lambda\to \sqrt{2}$ as $\tilde \lambda \to \infty$.  
Moreover, we see that larger and larger values of $\lambda$ are needed to reach this asymptote 
as $y$ decreases.

\subsection{Balancing $\Gamma$ versus $\Omega$}

We now address the central feature underpinning dynamical dark matter:
the balance between the SM decay widths $\Gamma_\lambda$ associated
with each KK mass eigenstate and
the corresponding cosmological abundances $\Omega_\lambda$.
As we shall show, an inverse relation of the form anticipated
in Eq.~(\ref{ab}) naturally emerges across the entire KK tower.

\subsubsection{Abundances $\Omega_\lambda$}

We begin by focusing on
the different cosmological mode abundances $\Omega_\lambda$ that can arise
in such a scenario.

During the course of the evolution of the universe,
there are many production mechanisms through which the different KK states might come
to be populated and thereby acquire non-zero abundances.
One such method, for example, is thermal production;
another relies on purely geometric effects (\eg, 
topological defects such as cosmic strings and domain walls) and the decays associated with
them.
However, there is also a third production mechanism which
exists in cases where the bulk mass $M$ happens to vanish:  this
is so-called ``misalignment production''.

In many string-theoretic contexts, bulk fields often
do have vanishing bulk masses.  Such fields often include gravitational
and/or geometric moduli fields;  they also include various axion-like fields.
Moreover, as we shall demonstrate, the predictions of 
misalignment production are rather straightforward to 
calculate, and are fairly generic for bulk fields as a whole.
We shall therefore take $M=0$ in what follows, and restrict our
attention to abundances established through misalignment production.

It is easy to understand the physical underpinnings of misalignment production
within the framework of dynamical dark matter.
Prior to the brane dynamics that establishes the brane mass $m$, the fact that
$M=0$ implies that our theory
exhibits a five-dimensional shift symmetry $\Phi\to\Phi+c$, where $c$ is a constant.
As a result, any value for $\langle \Phi\rangle$ is equally likely to occur:
\beq
         \langle \Phi \rangle ~=~ \theta f_\Phi^{3/2}~
\label{prefdef}
\eeq
where $\theta$ is a random ${\cal O}(1)$ dimensionless coefficient and 
where $f_\Phi$ is a mass scale (or decay constant) 
associated with the five-dimensional $\Phi$ field in the bulk.
Decomposed into KK eigenstates via Eq.~(\ref{KKdecomp}), 
this non-zero vev for the five-dimensional field $\Phi$
implies a non-zero vev for the KK zero mode:
\beq
              \langle \phi_0\rangle = \theta \hat f_\phi~,~~~~~~
              \langle \phi_k\rangle =0~~{\rm for~all}~~k>0~
\label{KKinit}
\eeq
where 
\beq
            \hat f_\phi~\equiv~ \sqrt{2\pi R}\, f_\Phi^{3/2}~.
\label{fdef}
\eeq
Note that all of the higher KK modes $\phi_k$ with $k>0$ must have vanishing vevs as a result
of the five-dimensional shift symmetry.

This is the situation that exists prior to (\ie, at energies higher than those associated with) 
the brane dynamics that establishes the brane mass $m$.
However, once this brane mass is established, we must shift to the {\it mass-eigenstate}\/ basis,
whereupon we see from Eq.~(\ref{zeromodematrixelement})
that Eq.~(\ref{KKinit}) now takes the form
\beq
              \langle \phi_\lambda \rangle = \theta A_\lambda \hat f_\phi~~~~~{\rm for~all}~~\lambda~.
\label{massinit}
\eeq
Thus, we see that {\it all}\/ of the mass eigenstates will generally have non-zero values.
Of course, the fact that that these vevs are all related through the coefficients $A_\lambda$
is a reflection of our original five-dimensional shift symmetry in the bulk.

The dynamics that establishes the brane mass $m$ also leads to a non-zero energy density $\rho$
associated with the configuration in Eq.~(\ref{massinit}).  In general,
the four-dimensional energy density $\rho_\lambda$ associated with 
each mode $\phi_\lambda$ 
is given by $\rho_\lambda =\half \lambda^2 \langle \phi_\lambda\rangle^2$. 
Given Eq.~(\ref{massinit}), we thus have
\beq
          \rho_\lambda~=~ \half \theta^2 \lambda^2 A_\lambda^2 \hat f_\phi^2~.
\eeq 
Of course, at any moment in the evolution of the universe, the {\it critical}\/ energy 
density is given by
\beq
      \rho_{\rm crit}~=~ 3 M_P^2 H^2
\eeq
where $M_P\equiv (8\pi G_N)^{-1/2}$ is the {\it reduced}\/ Planck scale
and where $H$ is the Hubble parameter.  
The initial abundance $\Omega_\lambda\equiv \rho_\lambda/\rho_{\rm crit}$ 
associated with the $\phi_\lambda$ mode  
is thus given by
\beq
     \Omega_\lambda^{(0)}~=~ {\theta^2 \over 6} \,\tilde \lambda^2 A_\lambda^2 
                     \,\left( {m \hat f_\phi\over M_P H}\right)^2~.
\eeq
This is, in fact, a completely general result.

We shall let 
$t_0$ denote the time at which this initial abundance is established
by the brane dynamics.
Thus $\Omega_\lambda(t_0)=\Omega_\lambda^{(0)}$.
The next question, however, is to determine the corresponding 
value of $\Omega_\lambda(t_{\rm now})$.
In order to do this, we see from Fig.~\ref{sketch}
that we must make some assumptions about whether $t_0$ is situated
during the reheating, radiation-dominated, or matter-dominated eras,
and whether the $\phi_\lambda$ mode experiences an instantaneous turn-on
at $t_0$ or a staggered turn-on at a time $t_\lambda >t_0$.
There are therefore six different cases to consider.

For simplicity we shall assume that for the modes $\phi_\lambda$ which are part of 
a staggered turn-on,
the corresponding turn-on time $t_\lambda$ occurs at the threshold $3H(t_\lambda)=2\lambda$.
(Of course, if $t_\lambda\leq t_0$, then such modes turn on only at $t_0$.)
We shall also assume that all modes in a given tower actually turn on  
during the same era as $t_0$, so that our turn-on ``cascade'' down the tower does 
not cross a boundary between two different eras. 
Finally, we shall assume that within each era,
$H(t)$ takes the approximate form $H(t)=\kappa/3t$ where 
$\kappa=2$ for the reheating and matter-dominated eras and
$\kappa=3/2$ during the radiation-dominated era.
This implies that $t_\lambda = \kappa/2\lambda$.
Note that this approximate form for $H(t)$ is generally valid at relatively late times within each era,
and we shall disregard all ${\cal O}(1)$ ``threshold'' effects associated with the boundaries between
different eras.  Thus, we shall implicitly take $\Omega_\lambda$ to be a continuous function of $t$,
as sketched in Fig.~\ref{sketch}, and we shall therefore disregard all $\tilde\lambda$-independent 
${\cal O}(1)$ numerical coefficients in those expressions for $\Omega_\lambda(t_{\rm now})$ which follow.
   
Clearly, if $t_0$ occurs during the final matter-dominated era (\ie, if $t_0>t_{\rm MRE}$), then modes
which turn on instantaneously (\ie, modes with $t_\lambda\leq t_0$) 
will have abundances
\beq
      \Omega_\lambda(t_{\rm now})~\sim~ 
     X_\lambda \Omega_\lambda^{(0)} ~\sim~ 
           \tilde \lambda^2 A_\lambda^2\, X_\lambda\, \left(\hat f_\phi\over M_P\right)^2 \, (mt_0)^2  
\label{case1}
\eeq  
where $X_\lambda$ denotes the expected damping factor due to dark-matter decays:
\beq
           X_\lambda ~\equiv~ e^{-\Gamma_\lambda (t_{\rm now}-t_0)}~.  
\label{Xdef}
\eeq
By contrast, for those modes which experience a staggered turn-on (\ie, modes with $t_\lambda>t_0$), this result becomes 
\beq
      \Omega_\lambda(t_{\rm now})~\sim~  X_\lambda\, \Omega_\lambda^{(0)} \, \left( {t_\lambda\over t_0}\right)^2
         ~\sim~  A_\lambda^2~ X_\lambda\, \left(\hat f_\phi\over M_P\right)^2 ~ 
\label{case2}
\eeq
where we have substituted the result $t_\lambda\sim 1/\lambda$ in passing to the final expression. 

By contrast, for $t_0$ within the radiation-dominated era (\ie, $t_{\rm RH}\lsim t_0\lsim t_{\rm MRE}$), 
these two cases are instead given by
\beqn
      \Omega_\lambda(t_{\rm now})
      &\sim& X_\lambda\, \Omega_\lambda^{(0)} \, \left({t_{\rm MRE}\over t_0}\right)^{1/2} \nonumber\\
      &\sim&  \tilde \lambda^2 A_\lambda^2~ X_\lambda\, 
                 \left(\hat f_\phi\over M_P\right)^2 \, (mt_0)^{3/2}\, (m t_{\rm MRE})^{1/2} 
        \nonumber\\
\label{case3}
\eeqn
and
\beqn
      \Omega_\lambda(t_{\rm now})
      &\sim& X_\lambda\, \Omega_\lambda^{(0)} \, \left( t_\lambda\over t_0\right)^2 
       \, \left({t_{\rm MRE}\over t_\lambda}\right)^{1/2} \nonumber\\
      &\sim&  \tilde \lambda^{1/2} \, A_\lambda^2~ X_\lambda\, 
                 \left(\hat f_\phi\over M_P\right)^2 \, (mt_{\rm MRE})^{1/2}~.\nonumber\\
\label{case4}
\eeqn

Finally, for $t_0$ within the reheating era (\ie, $t_0\lsim t_{\rm RH}$),
these two cases are instead given by
\beqn
      \Omega_\lambda(t_{\rm now})
      &\sim& X_\lambda\, \Omega_\lambda^{(0)} \, 
                \left( {t_{\rm MRE}\over t_{\rm RH}}\right)^{1/2} \nonumber\\
      &\sim&  \tilde \lambda^2 A_\lambda^2~ X_\lambda\, 
                 \left(\hat f_\phi\over M_P\right)^2 \, (mt_0)^{2}\, 
                \left( {t_{\rm MRE}\over t_{\rm RH}}\right)^{1/2}
        \nonumber\\
\label{case5}
\eeqn
and
\beqn
      \Omega_\lambda(t_{\rm now})
      &\sim& X_\lambda\, \Omega_\lambda^{(0)} \,
         \left( {t_\lambda\over t_0}\right)^2\, 
                \left( {t_{\rm MRE}\over t_{\rm RH}}\right)^{1/2} \nonumber\\
      &\sim&  A_\lambda^2~ X_\lambda\, 
                 \left(\hat f_\phi\over M_P\right)^2 \, 
                \left( {t_{\rm MRE}\over t_{\rm RH}}\right)^{1/2}~.
        \nonumber\\
\label{case6}
\eeqn
Interestingly, of all six cases, this is the only one which yields a
result for $\Omega_\lambda(t_{\rm now})$ which is 
parametrically independent of the scale $m$.

It is also instructive to examine the manner in which these results scale with $\tilde \lambda$.
Surveying Eqs.~(\ref{case1}) through (\ref{case6}),
we see that
the dependence of $\Omega_\lambda$ on $\tilde\lambda$ follows only three different patterns,
depending on the specific turn-on behavior experienced by the KK mode
in question
and the era during which it takes place:
\beq
       \Omega_\lambda ~\sim~ \cases{
        \tilde \lambda^2 A_\lambda^2   &   instantaneous \cr
        \tilde\lambda^{1/2}  A_\lambda^2   &   staggered (RD era)\cr
                 A_\lambda^2   &   staggered (RH/MD eras)~.\cr}
\label{lambdascalings}
\eeq
Under the assumption of misalignment production,
this result is exact and completely general.
However, given the definition in Eq.~(\ref{Alamdef}),
we may approximate
\beq
       A_\lambda~\sim~\cases{  1/\tilde\lambda & for $\tilde\lambda \ll \sqrt{1+\pi^2/y^2}$\cr
                            1/\tilde\lambda^2 & for $\tilde\lambda \gg \sqrt{1+\pi^2/y^2}$~.\cr}
\label{Alamapprox}
\eeq
In the future, we shall refer to these two approximation regimes as 
the small-$\tilde\lambda$ and large-$\tilde\lambda$ regimes;
note that while there always exists a large-$\tilde \lambda$ regime, the
existence of a small-$\tilde\lambda$ regime depends on the value of $y$.
We then find that the results in Eq.~(\ref{lambdascalings}) lead to
the
large-$\tilde\lambda$ scaling behaviors
\beq
       \Omega_\lambda ~\sim~ \cases{
        \tilde \lambda^{-2} &   instantaneous \cr
        \tilde\lambda^{-7/2} &   staggered (RD era)\cr
                 \tilde\lambda^{-4}   &   staggered (RH/MD eras)~\cr}
\label{lambdascalings2}
\eeq
as well as the small-$\tilde\lambda$ behaviors
\beq
       \Omega_\lambda ~\sim~ \cases{
        {\rm constant}~ &   instantaneous \cr
        \tilde\lambda^{-3/2} &   staggered (RD era)\cr
                 \tilde\lambda^{-2}   &   staggered (RH/MD eras)~.\cr}
\label{lambdascalings3}
\eeq
Indeed, the only $\lambda$-dependence which is not included in these results
is that which appears through the decay widths in the $X_\lambda$-factors
in Eqs.~(\ref{case1}) through (\ref{case6}).  However, these $X_\lambda$ factors only express
the physics of the eventual dark-matter {\it decay}\/ processes;  they play no role in determining the
mode abundances that exist {\it prior}\/ to decay, which is our main interest in this discussion.   
As a result, we shall disregard these $X_\lambda$ factors in what follows, understanding
that our analysis is primarily appropriate for the period that exists {\it prior}\/ 
to the onset of decays of the KK states with significant abundances. 
(Indeed, the final period of KK decays will be discussed in Sect.~III.B.3 after we have
analyzed the behavior of the decay widths $\Gamma_\lambda$ in Sect.~III.B.2.)

Given the individual abundances $\Omega_\lambda$ 
in Eqs.~(\ref{case1}) through (\ref{case6}),
we can now calculate the values of both $\Omega_{\rm tot}$ and $\eta$ that exist
prior to the onset of significant KK decays. 
Recall that $\Omega_{\rm tot}$ is nothing but the sum over all of the individual
abundances $\Omega_\lambda$, while $\eta$ describes how that total
abundance is distributed across the different modes.
Indeed, we see from Eq.~(\ref{lambdascalings2}) that
the lightest of the KK modes will always carry the greatest abundance.
It then follows from its definition in Eq.~(\ref{etadef}) that $\eta$
indicates what fraction of the total abundance is carried by the {\it excited}\/
states in the KK tower.  For this reason we shall occasionally refer to $\eta$ as the ``tower fraction''.

Let us first consider $\Omega_{\rm tot}$.  
At first glance, it might seem algebraically
cumbersome to tally these individual mode abundances $\Omega_\lambda$, since they each have 
a different $\lambda$-dependence given in Eq.~(\ref{lambdascalings}) 
and we would need to sum over all of the eigenvalue solutions to the transcendental equation
in Eq.~(\ref{coteq}).
However, it turns out that 
the resulting spectrum of $\lambda$-eigenvalues
satisfies two critical identities~\cite{DDGAxions}: 
\beq
          \sum_\lambda A_\lambda^2 ~=~1~,~~~~~~~
          \sum_\lambda \tilde\lambda^2 A_\lambda^2 ~=~1~.
\label{identities}
\eeq
These identities ultimately stem from the unitary nature of the mapping
between the KK-momentum basis and the mass-eigenstate basis.
Thus, unless our KK tower experiences a staggered turn-on during the radiation-dominated era,
summing over the individual abundances in Eq.~(\ref{lambdascalings})
is particularly simple.

What is truly remarkable about the identities in Eq.~(\ref{identities}) is that they
hold for all values of $y$.  They are thus independent of the degree of non-diagonality 
exhibited by the KK mass matrix, and independent of the degree to 
which the corresponding KK modes are mixed.  
As a result, if we assume that $\hat f_\phi$, $t_0$, and $m$ are all $y$-independent, 
it then follows that $\Omega_{\rm tot}$ will be $y$-independent as well!
Although dialing the value of $y$ (\ie, adjusting the radius of the extra spacetime dimension)
might change the {\it distribution}\/ of abundances across the infinite tower of KK states, the
 {\it total}\/ abundance remains essentially fixed.
 
\begin{figure}[t!]
\centerline{
     \epsfxsize 3.5 truein \epsfbox{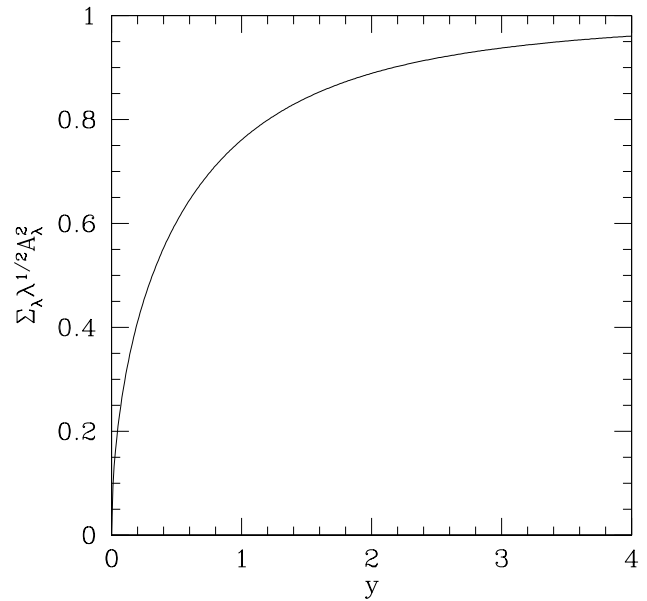} 
 }
\vskip -0.25 truein
\caption{The quantity $\sum_\lambda \tilde\lambda^{1/2} A_\lambda^2$,
plotted as a function of $y$.
Note that the total dark-matter abundance $\Omega_{\rm tot}$ prior to the onset of significant
KK decays 
is proportional to 
this quantity if the individual KK abundances are initially established through a 
staggered turn-on during the radiation-dominated era. 
As a result, this curve also illustrates the 
$y$-dependence of $\Omega_{\rm tot}$ in this case.
By contrast, for all other cases, the total abundance $\Omega_{\rm tot}$ is
$y$-independent.}
\label{yplot}
%  USE THIS ONLY IF WE WANT FOLLOWING FIGURE TO BE "JOINED" WITH THIS ONE
%  \vskip 0.2 truein
%  \centerline{
   %  \epsfxsize 3.4 truein \epsfbox{plot3.eps} 
 %  }
%  \vskip -0.25 truein
%  \caption{The tower fraction $\eta$
%  after all dark-matter modes have ``turned on'' and
%  entered the present matter-dominated epoch,
%  plotted as a function of $y$ for three different regimes of misalignment production:
%  (a) instantaneous turn-on, in which case
%  $\Omega_\lambda\sim {\tilde\lambda}^2 A_\lambda^2$;
%  (b) staggered turn-on during a radiation-dominated era, in which case  
%  $\Omega_\lambda\sim {\tilde\lambda}^{1/2} A_\lambda^2$;
%  and
%  (c) staggered turn-on during a reheating or matter-dominated era, in which
%  case 
%  $\Omega_\lambda\sim  A_\lambda^2$.
%  In each case we see that $\eta\to 0$ as $y\to \infty$, while $\eta$ approaches
%  a fixed maximum value $\eta_{\rm max}$ as $y\to 0$.}
%  \label{plot2}
\end{figure}

We stress that this result holds only for those cases in which the KK abundances 
are established either instantaneously, or through a staggered turn-on which
takes place during the reheating or matter-dominated eras.
By contrast, if these KK abundances are established through a staggered turn-on
during the {\it radiation}\/-dominated era,
the total abundance $\Omega_{\rm tot}$ today will be proportional to
$\sum_\lambda \tilde \lambda^{1/2} A_\lambda^2$.
This quantity is not $y$-independent, but rather has the $y$-dependence
shown in Fig.~\ref{yplot}.
As a result, our preferred choices for parametric quantities such as
$\hat f_\phi$, $t_0$, and $m$ would need to be altered in order
to compensate for this effect.

%================== FIGURE ============================================
%========================================================================

%================== FIGURE ============================================
\begin{figure}[t!]
\centerline{
   \epsfxsize 3.5 truein \epsfbox{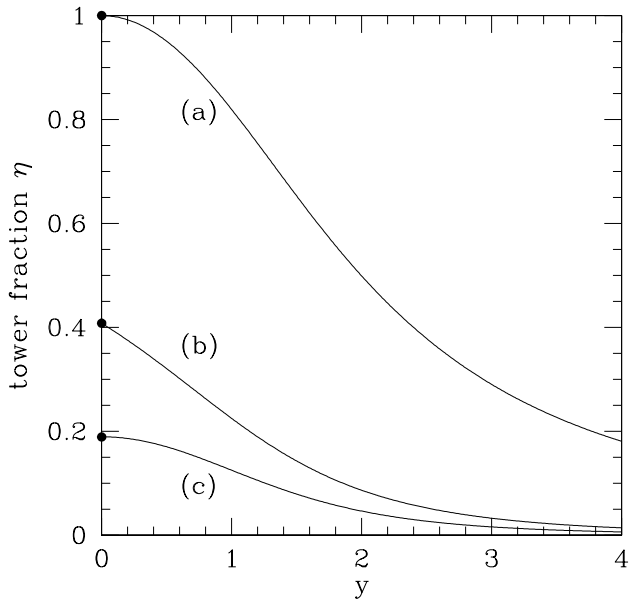} 
 }
\vskip -0.2 truein
\caption{The tower fraction $\eta$
after all dark-matter modes have ``turned on'' and
entered the present matter-dominated epoch,
plotted as a function of $y$ for three different regimes of misalignment production:
(a) instantaneous turn-on, in which case
$\Omega_\lambda\sim {\tilde\lambda}^2 A_\lambda^2$;
(b) staggered turn-on during a radiation-dominated era, in which case  
$\Omega_\lambda\sim {\tilde\lambda}^{1/2} A_\lambda^2$;
and
(c) staggered turn-on during a reheating or matter-dominated era, in which
case 
$\Omega_\lambda\sim  A_\lambda^2$.
In each case we see that $\eta\to 0$ as $y\to \infty$, while $\eta$ approaches
a fixed maximum value $\eta_{\rm max}$ as $y\to 0$.}
\label{plot2}
\end{figure}  
%========================================================================

The results in Eq.~(\ref{lambdascalings}) also allow us
to calculate the tower fraction $\eta$ prior to the onset of significant KK decays.
Indeed, the three different patterns for $\Omega_\lambda$ in Eq.~(\ref{lambdascalings}) 
imply three different distributions 
for the abundances across the different mass eigenstates in the KK tower.
Therefore, if we additionally assume that all of the states in a given tower 
simultaneously fall
into one of these three cases,
there will be three corresponding possible behaviors 
for the ``tower fraction'' $\eta$ defined in Eq.~(\ref{etadef}), viewed as a function of
the non-diagonality parameter $y$.
These results are shown in Fig.~\ref{plot2}, 
and we see that $\eta$ indeed spans a range of ${\cal O}(1)$ values, as desired.
These results are also highly $y$-dependent, illustrating that while adjusting $y$ changes
the total abundance only in certain restricted circumstances,
it changes the {\it distribution}\/ of these abundances quite substantially
in all cases.

It is easy to understand the overall features exhibited in Fig.~\ref{plot2}.
As $y\to \infty$, we enter the four-dimensional limit in which virtually
no abundance is carried by the excited KK modes.  As a result,  these modes
become completely irrelevant to the dark-matter problem,
and $\eta\to 0$.
By contrast, as $y\to 0$, 
our KK states experience maximal mixing, as a result of which the corresponding
value of $\eta$ is maximized.
Since $\lambda_n\approx (n+\half)/R$ for all $n$ in this limit, we can easily calculate
these maximum values of $\eta$, obtaining the result
that $\eta_{\rm max}=1$ in the case of instantaneous turn-on, 
while
for the case of staggered turn-on during a reheating or matter-dominated era
we have
\beqn
    \eta_{\rm max} &=&  1-4 \left[ \sum_{n=0}^\infty {1\over (n+1/2)^2}\right]^{-1} \nonumber\\
         &=&  1- {8\over \pi^2} ~\approx~ 0.189~,
\eeqn
and for the case of staggered turn-on during a radiation-dominated era
we have 
\beqn
    \eta_{\rm max} &=&  1- 2\sqrt{2}\left[\sum_{n=0}^\infty {1\over (n+1/2)^{3/2}} \right]^{-1} \nonumber\\
            &=& 1-{2\sqrt{2}\over
                (2\sqrt{2}-1) \zeta(3/2)}~\approx~ 0.408~
\eeqn
where $\zeta$ denotes the Riemann zeta-function.
All three of these limiting values are evident in Fig.~\ref{plot2}.

\subsubsection{Decay widths $\Gamma_\lambda$}

Next, we turn to the decay widths $\Gamma_\lambda$ which can be expected in such
a scenario.

Up to this point, we have 
assumed nothing more than that the bulk field in 
our setup has a vanishing bulk mass $M=0$
and a non-vanishing brane mass $m\not =0$.  As we have seen, this has proven sufficient
to allow us to determine not only the Kaluza-Klein ``spectroscopy'' of our dark towers
but also the corresponding cosmological mode abundances that emerge
from misalignment production. 
In some sense, 
it is remarkable that these results
rely on such minimal assumptions;  indeed, this happy fact
explains why our results 
thus far are extremely general and can be expected to hold for all
bulk fields for which $M=0$ and $m\not=0$. 

However,
in order to discuss the decay widths of these KK modes into Standard-Model
states, we shall require further information concerning the couplings between
the five-dimensional bulk field $\Phi$ and the four-dimensional Standard-Model
states.  In other words, we shall require further information concerning the
interaction terms that might appear within ${\cal L}_{\rm int}$
in Eq.~(\ref{branelag}) and thereby become part of our four-dimensional effective
Lagrangian.   
Of course,
it is most dangerous for the consistency and phenomenological viability of our dynamical
dark-matter scenario if these decay widths are too large.  
Our conservative approach to this problem will therefore be
to consider the worst possible scenario and determine 
how large these decay widths might be.  Since the largest decay widths will generally
arise from the 
operators of lowest possible dimension
within ${\cal L}_{\rm int}$,
the first step in our analysis is therefore to determine what forms such operators might take.

Imposing Lorentz invariance and invariance under all Standard-Model
gauge symmetries, we find that our options are fairly limited. 
The separate brane/bulk structure of this dark-matter setup requires that our
bulk dark-matter field $\Phi$ be a singlet under all Standard-Model gauge symmetries.
This implies that any combinations of brane fields to which $\Phi$ couples
must be gauge invariant by themselves.  
Moreover, in order to restrict ourselves to operators of lowest possible dimensionality,
we shall consider operators which are at most linear in $\Phi$.
We shall also assume for this discussion that $\Phi$ is real.
Letting $\psi$ denote a generic Standard-Model fermion and 
$F_{\mu\nu}$ denote a generic Standard-Model field strength for any gauge group,
we then find that the operators of lowest possible dimensionality come in two groups.
If $\Phi$ is CP-even, the lowest-dimension operators which may appear in ${\cal L}_{\rm int}$
take the form
\beq
           {1\over f_\Phi^{3/2}} \, \Phi \, \overline{\psi} \gamma^\mu \partial_\mu \psi~,~~~~~~
           {1\over f_\Phi^{3/2}} \, \Phi \, F_{\mu\nu} F^{\mu\nu}~
\label{group1}
\eeq
where $f_\Phi$ is the five-dimensional mass scale associated with $\Phi$ 
which originally appeared in Eq.~(\ref{prefdef}).
By contrast, if $\Phi$ is CP-odd, the lowest-dimension operators which may appear in 
${\cal L}_{\rm int}$
take the form
\beq
           {1\over f_\Phi^{3/2}} \, (\partial_\mu \Phi) \, \overline{\psi} \gamma^\mu \gamma^5 \psi~,~~~~~~
           {1\over f_\Phi^{3/2}} \, \Phi \, F_{\mu\nu} \tilde F^{\mu\nu}~
\label{group2}
\eeq
where $\tilde F^{\mu\nu}\sim \epsilon^{\mu\nu\rho\sigma}F_{\rho\sigma}$.
These groups of operators then respectively give rise to the four-dimensional couplings
\beq
           {1\over \hat f_\phi} \, \phi' \, \overline{\psi} \gamma^\mu \partial_\mu \psi~,~~~~~~
           {1\over \hat f_\phi} \, \phi' \, F_{\mu\nu} F^{\mu\nu}~
\label{group3}
\eeq
and
\beq
           {1\over \hat f_\phi} \, (\partial_\mu \phi') \, \overline{\psi} \gamma^\mu \gamma^5 \psi~,~~~~~~
           {1\over \hat f_\phi} \, \phi' \, F_{\mu\nu} \tilde F^{\mu\nu}~
\label{group4}
\eeq
where $\phi'$, as defined in Eq.~(\ref{phiprime}),
is the projection of $\Phi$ onto the Standard-Model brane,
and where $\hat f_\phi$ is defined in Eq.~(\ref{fdef}).  

This list exhausts the possible dimension-five operators. 
It is encouraging that we see among this list of possible operators 
the standard moduli and axion couplings --- indeed,
in the CP-even case we can even regard the linear prefactor $\phi'/\hat f_\phi$ as
the leading term of 
an exponential prefactor
$\exp( \phi'/\hat f_\phi)$, and thereby recognize the standard dilaton coupling 
in string theory.
Thus, this list of operators includes most of our cases of phenomenological interest.
At first glance, it might seem that operators of even lower dimension could be constructed ---
\eg, $\Phi \overline{\psi} \psi$ and 
 $\Phi \overline{\psi} \gamma^5 \psi$.
However, such operators are not gauge invariant because all of the 
fermions $\psi$ in the Standard Model are chiral.
Likewise, dimension-four operators such as $|\Phi|^2 |H|^2$ are also forbidden
as they would violate the shift symmetry under which $\Phi\to\Phi+{\rm constant}$.

Although kinematic effects favor the di-photon decay mode $\phi'\to \gamma\gamma$,
the four-dimensional couplings in Eqs.~(\ref{group3}) and (\ref{group4}) 
all lead to SM decay rates of the same parametric order.  
Standard calculations then lead to
an overall decay width
\beq
   \Gamma_\lambda ~\sim~ {\lambda^3\over \hat f_\phi^2} \langle \phi_\lambda | \phi'\rangle^2~ 
   ~=~ {\lambda^3\over \hat f_\phi^2} \left(\tilde \lambda^2 A_\lambda \right)^2~ 
\label{lifetimeresults}
\eeq
where we have substituted Eq.~(\ref{phiprimematrixelement}) in the final step.
Use of Eqs.~(\ref{Alamapprox}) then leads to the large-$\tilde\lambda$ behavior
$\Gamma_\lambda\sim \tilde \lambda^3$ as well as the small-$\tilde\lambda$ behavior
$\Gamma_\lambda\sim \tilde \lambda^5$.

Before concluding our discussion of the decay widths, it is important
to note that there will generally exist
many competing decay modes for our KK states which
do not exclusively involve Standard-Model particles as end-products.
One example includes intra-ensemble decays (\ie, decays {\it within}\/ the
KK tower, from heavier KK states to lighter KK states);
indeed, this possibility will be discussed in general terms in the Appendix.
Also, in cases involving multiple fields in the bulk,
it is possible for bulk KK states of one species to decay
to bulk states of another species.

While such decays can be important on a number of cosmological
and phenomenological levels, 
they generally do not significantly diminish the abundance of 
what might be termed ``dark matter'' or increase the corresponding
abundance of what might be termed ``visible matter''.
Moreover, it is often the case that such decays are significantly suppressed relative
to the KK decays that proceed directly to Standard-Model brane states.
Such suppressions can occur for a variety of reasons, some
of which depend on the fact that physics in the bulk is often governed
directly by (and therefore suppressed by) the gravitational Planck scale,
and some of which 
are the consequences of 
extra restrictive symmetries 
which exist purely in the bulk and which therefore do not apply to decays  
of bulk fields into brane fields.
Of course, a detailed analysis of this question requires specifying a
particular bulk field, along with a complete Lagrangian for the theory
including its gravitational interactions.  
While such an analysis is beyond the
scope of this theoretical overview, 
an analysis of this sort does appear in Refs.~\cite{paper2,paper3}
where it is shown that such decays are indeed greatly suppressed
in a specific realistic model of dynamical dark matter.
This result therefore confirms our general expectations in one specific
example.

We have therefore assumed in this paper that the primary decay mode
for each KK bulk mode is directly into a Standard-Model brane state.
However, as discussed in the Appendix,
our dynamical dark-matter scenario can easily be generalized to
accommodate more complex decay channels 
if this should ultimately prove appropriate in a given situation.

\subsubsection{Balancing lifetimes against abundances}

Having calculated the spectrum of abundances $\Omega_\lambda$ and the spectrum of
decay widths $\Gamma_\lambda$
across our KK tower, we can now see exactly how KK towers manage to balance lifetimes
against abundances.
Combining the results in Eqs.~(\ref{lambdascalings2})
and (\ref{lifetimeresults}),
we find that for large $\lambda$ our KK towers indeed always obey a balancing equation
of the form anticipated in Eq.~(\ref{ab}):
\beqn
 {\rm instantaneous}\/:&&   \Omega_\lambda \Gamma_\lambda^{2/3}\sim{\rm constant}\nonumber\\ 
 {\rm staggered~(RD~era)}:&& \Omega_\lambda \Gamma_\lambda^{7/6}\sim{\rm constant} \nonumber\\
 {\rm staggered~(RH/MD~eras)}:&& \Omega_\lambda \Gamma_\lambda^{4/3}\sim{\rm constant}~.\nonumber\\ 
\label{bal1}
\eeqn
Indeed, this asymptotic behavior holds for $\tilde\lambda\gg \sqrt{1+\pi^2/y^2}$.
Thus, we see that our KK towers succeed in balancing lifetimes against abundances in a robust
fashion, regardless of the particular type of ``turn-on'' they experience and
regardless of the cosmological era during which this ``turn-on'' takes place.

While Eq.~(\ref{bal1}) describes the crucial asymptotic behavior at the ``top'' of the
KK tower, a similar set of relations describes the ``bottom'' of each KK tower.
Combining the small-$\tilde \lambda$ behavior in Eq.~(\ref{lambdascalings3}) 
with the small-$\tilde\lambda$ result $\Gamma_\lambda\sim \tilde\lambda^5$ leads to the
relations
\beqn
 {\rm instantaneous}\/:&&   \Omega_\lambda \sim{\rm constant}\nonumber\\ 
 {\rm staggered~(RD~era)}:&& \Omega_\lambda \Gamma_\lambda^{3/10}\sim{\rm constant} \nonumber\\
 {\rm staggered~(RH/MD~eras)}:&& \Omega_\lambda \Gamma_\lambda^{2/5}\sim{\rm constant}~\nonumber\\ 
\label{bal0}
\eeqn
for $\tilde\lambda\ll \sqrt{1+\pi^2/y^2}$.
We stress, however, that this behavior is relevant only at the bottom of a KK tower,
and only for relatively small $y$. 
Indeed, regardless of the value of $y$, the behavior of the abundances
and decay widths always eventually shifts to satisfy the relations in Eq.~(\ref{bal1}) 
as we pass to higher and higher modes in a given KK tower.

Given these results for the abundances $\Omega_\lambda$ and decay widths $\Gamma_\lambda$,
we can now calculate the general $(\alpha,\beta)$ scaling coefficients that
appear in Eq.~(\ref{step2}).  These results also enable us to deduce an ``effective'' equation of state
for our ensemble of decaying dark-matter KK components.
The values of $\alpha$, of course, are directly evident from Eq.~(\ref{bal1}) for large $\tilde\lambda$
and from Eq.~(\ref{bal0}) for small $\tilde\lambda$.
Likewise, since the states in our KK tower are nearly equally spaced throughout
the tower, 
we know that the density of states per unit $\lambda$ is essentially $\lambda$-independent:
$n_\lambda\sim \lambda^0$.  Per unit of $\Gamma$, this translates into  
$n_\Gamma\sim n_\lambda |{d\Gamma/ d\lambda}|^{-1}\sim \Gamma^{(1-x)/x}$ for $\Gamma\sim \lambda^x$.
We thus have $\beta= -2/3$ for large $\lambda$, and $\beta= -4/5$ for small $\lambda$.

We therefore conclude that for large $\lambda$, a general KK tower has the scaling coefficients
\beq
       (\alpha,\beta)~=~\cases{ (-2/3,-2/3)   &   instantaneous \cr
                        (-7/6,-2/3)   &   staggered (RD era)\cr
                        (-4/3,-2/3)   &   staggered (RH/MD eras)~.\cr} 
\label{straw1}
\eeq 
By contrast, for small $\tilde\lambda$, these results are modified to become
\beq
       (\alpha,\beta)~=~\cases{ (0,-4/5)   &   instantaneous \cr
                        (-3/10,-4/5)   &   staggered (RD era)\cr
                        (-2/5,-4/5)   &   staggered (RH/MD eras)~.\cr} 
\label{straw2}
\eeq

Given these $(\alpha,\beta)$ scaling coefficients, we can also calculate the effective equation-of-state
function $w_{\rm eff}(t)$ which describes the collective effects of the 
decays of the individual modes along the KK tower.
Indeed, as we have seen in Sect.~II, the behavior of this function $w_{\rm eff}(t)$  
depends critically on the value of the sum $x\equiv \alpha+\beta$.
However, given the results in Eqs.~(\ref{straw1}) and (\ref{straw2}), 
we can easily tabulate the values of $x$ for the different cases under study,
obtaining the results shown in Table~\ref{alphabetatable}.
As we see from Table~\ref{alphabetatable}, most of the $x$-values for a general
KK tower tend to cluster near $x\lsim -1$.
 {\it This is remarkable, since we have already shown in Sect.~II that this is precisely
the range for $x$ which is preferred phenomenologically.}
We thus see that a KK tower indeed serves as an excellent realization
of dynamical dark matter.

%========================== Table ==========================================
\begin{table}[t!]
\begin{tabular}{||c||c|c||}
\hline
\hline
 ~ & ~large $\tilde \lambda$~ & ~small $\tilde\lambda$~ \cr
\hline
instantaneous          & $-4/3$   &   $-4/5$  \cr
staggered (RD era)     & ~$-11/6$~    & ~$-11/10$~          \cr
~staggered (RH/MD eras)~ & $-2$    &  $-6/5$          \cr
\hline
\hline
\end{tabular}
\caption{Values of the equation-of-state parameter $x\equiv \alpha+\beta$ for different 
portions of a general KK tower with different ``turn-on'' phenomenologies.  We observe that
KK towers naturally give rise to values $x\lsim -1$, which is precisely the range
favored phenomenologically.  } 
\label{alphabetatable}
\end{table}
%========================== end of Table ===================================

One feature which emerges from Table~\ref{alphabetatable} is that 
regardless of the turn-on behavior of the individual modes,
the value of $x$ generally decreases as we pass from the large-$\tilde\lambda$  regime
to the small-$\tilde\lambda$ regime.
This generally corresponds to passing from 
early times (during which the decays of the heavier KK modes dominate the physics)
to later times (during which only the lighter KK modes are still present).
Indeed, this transition typically occurs for values of $\lambda\sim \sqrt{1+\pi^2/y^2}$
which decrease as a function of $y$. 
Thus, we do not expect to see the small-$\tilde\lambda$ behavior 
emerge strongly except for later times in small-$y$ scenarios.

\begin{figure*}[tbh]
\centerline{
   \epsfxsize 2.33 truein \epsfbox{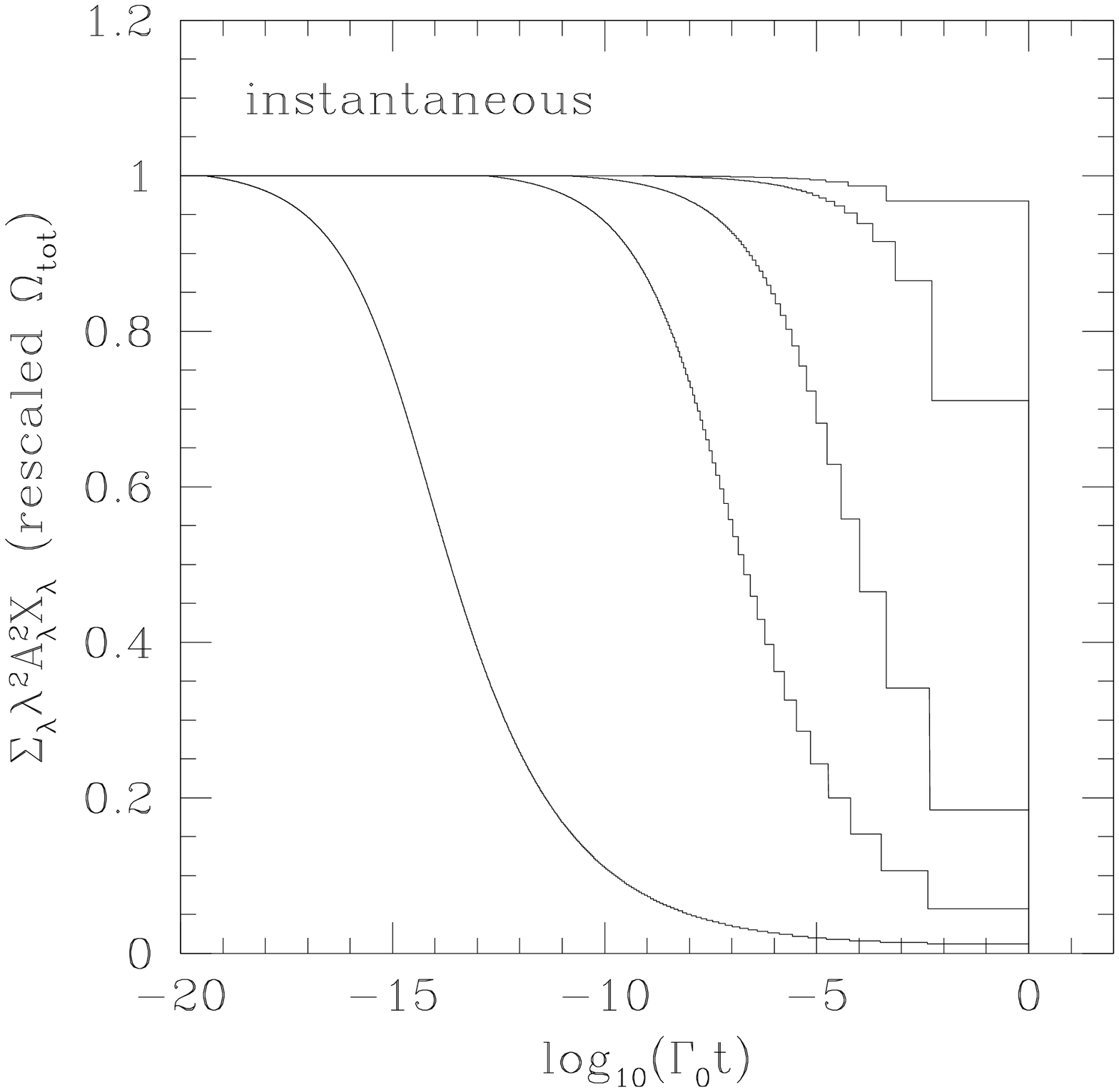} 
   \epsfxsize 2.33 truein \epsfbox{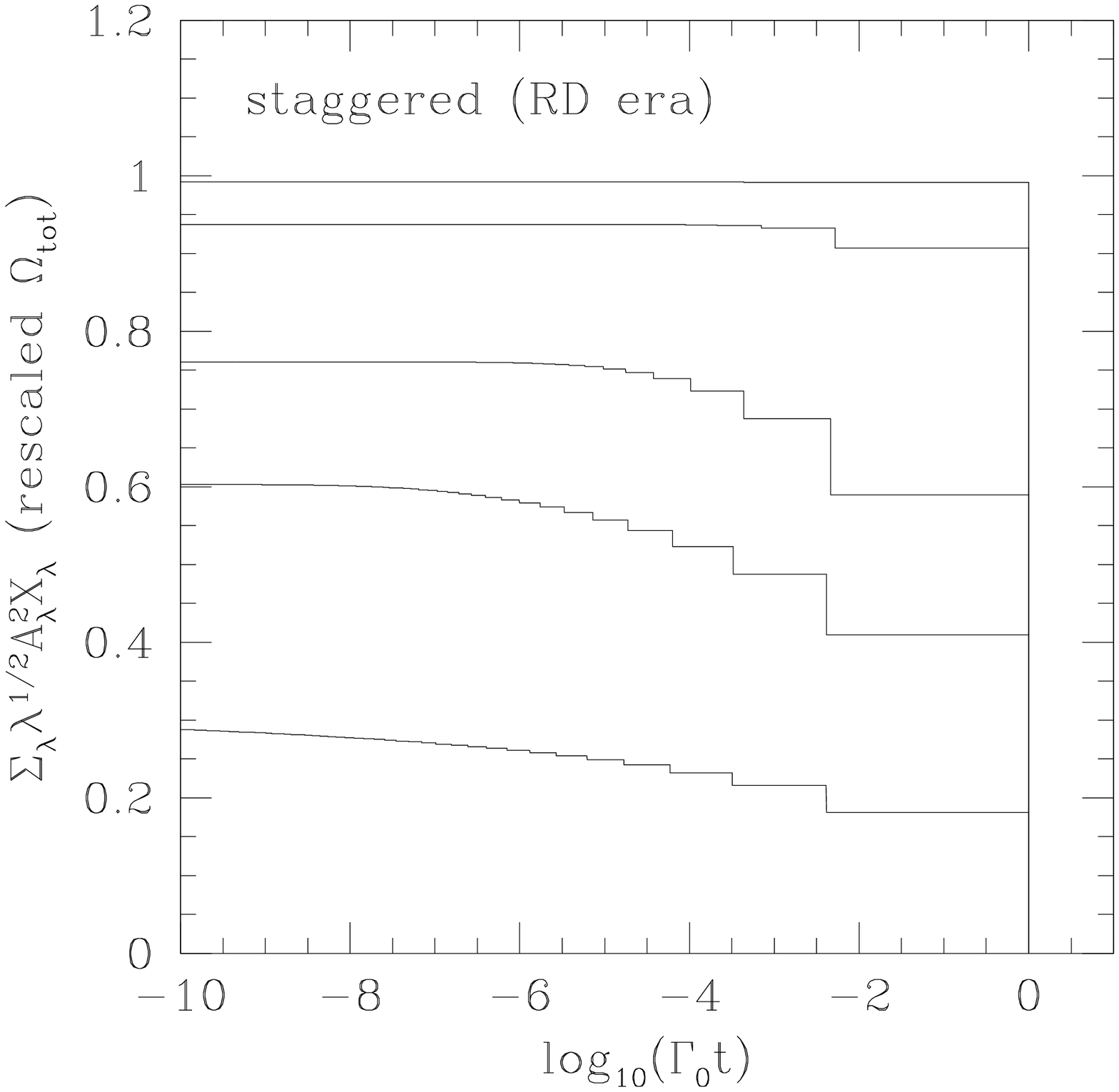} 
   \epsfxsize 2.33 truein \epsfbox{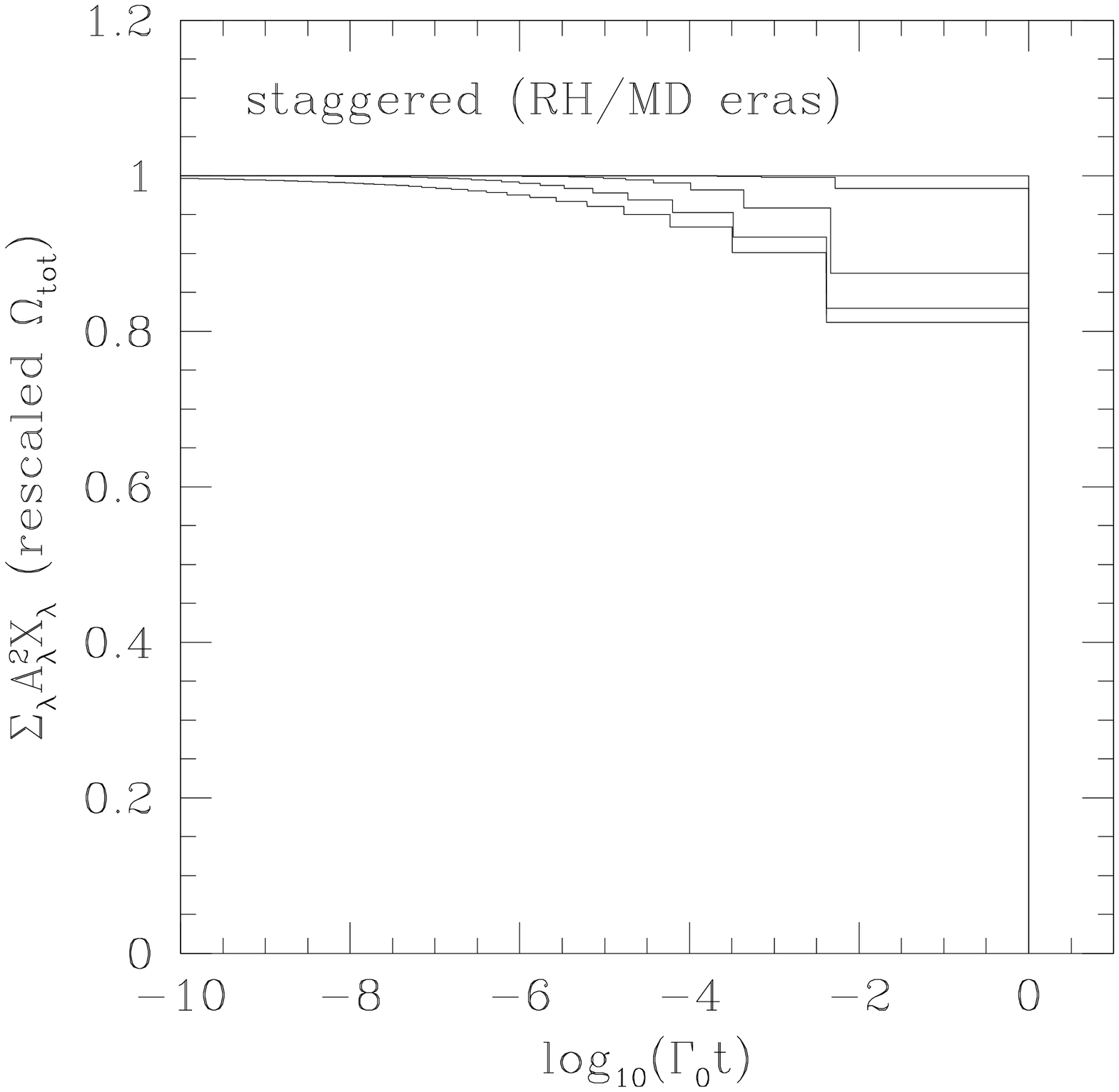} 
 }
\vskip -0.2 truein
\caption{The (rescaled) total dark-matter abundance $\Omega_{\rm tot}$, plotted as a function
of time for each of the three cases relevant for a general KK tower.  
In each panel, the uppermost curve corresponds to $y=10$ and the successively
lower curves correspond to $y=3$, $y=1$, $y=0.5$, and $y=0.1$.
In order to compare curves with different values of $y$,
we have plotted $\log_{10}(\Gamma_0 t)$ on the 
horizontal time axis, where $\Gamma_0$ is the 
decay width of the lightest KK mass eigenstate associated with each curve.
This ensures that the curves
share a common horizontal location at which
$\Omega_{\rm tot}$ vanishes in each case,
signifying the decay
of the final, lightest state in the KK tower.  
We see from these results that
overall shape of the time-dependence of 
$\Omega_{\rm tot}$ is highly $y$-dependent: 
the $y\to\infty$ limit corresponds to the usual scenario
of a single dark-matter particle
(with $\Omega_{\rm tot}$ remaining essentially constant until this particle decays),
while smaller values of $y$ correspond to situations in which 
$\Omega_{\rm tot}$ is distributed across multiple KK modes with different decay widths.
It is this property which leads to
a time-dependent $\Omega_{\rm tot}$ and thus a non-trivial ``effective'' equation of state
for the dark KK tower.} 
\label{omtotplots}
\end{figure*}

Needless to say, all of the above conclusions are predicated on approximations
which model the KK tower according to certain power-law scaling behaviors.
It is therefore natural to wonder how robust these conclusions 
actually are when compared with the results of a 
complete numerical calculation which uses the exact numerical values for 
the eigenvalues $\tilde \lambda$ across
the entire KK tower and which avoids any approximations for the coefficients
$A_\lambda$ which appear in the KK mode abundances and decay widths.
However, it is straightforward to perform such a calculation.
In Fig.~\ref{omtotplots},
we plot a rescaled version of
the total dark-matter abundance $\Omega_{\rm tot}$ as
a function of time during its final decay-dominated period,
assuming (as in Sect.~II) that these decays occur 
during the present matter-dominated cosmological era. 
Each panel in Fig.~\ref{omtotplots} corresponds to one of the three different 
cases that describe how the individual abundances in the KK tower
might have been established;  indeed,
following the results in Eq.~(\ref{lambdascalings}),
this ``rescaled'' $\Omega_{\rm tot}$ 
is defined in each case as $\sum_\lambda \tilde \lambda^k A_\lambda^2 X_\lambda$ 
with $k=2$ (first panel of
Fig.~\ref{omtotplots}),
$k=1/2$ (second panel), and $k=0$ (third panel). 
Moreover, in making these plots, we have assumed that each KK state decays instantaneously
at $t=\tau_\lambda\equiv \Gamma_\lambda^{-1}$ so that the contributions from individual
states will be readily discernible.
This is tantamount to approximating $X_\lambda$ in Eq.~(\ref{Xdef})
as $X_\lambda(t)\approx \Theta(\Gamma_\lambda^{-1}-t)$.
Finally,
in order to compare curves with different values of $y$, 
an overall normalization for the time axis for each curve has been chosen 
such that the time $t$ is expressed in units of $\Gamma_0^{-1}$, where $\Gamma_0$ is the 
decay width of the lightest KK mass eigenstate.
As a result the curves in Fig.~\ref{omtotplots} share a common location at which
$\Omega_{\rm tot}$ ultimately vanishes in each case,
signifying the eventual decay
of the final, lightest state in the KK tower.

%================== FIGURE ============================================
%========================================================================

%================== FIGURE ============================================
\begin{figure*}[tbh]
\centerline{
   \epsfxsize 2.33 truein \epsfbox{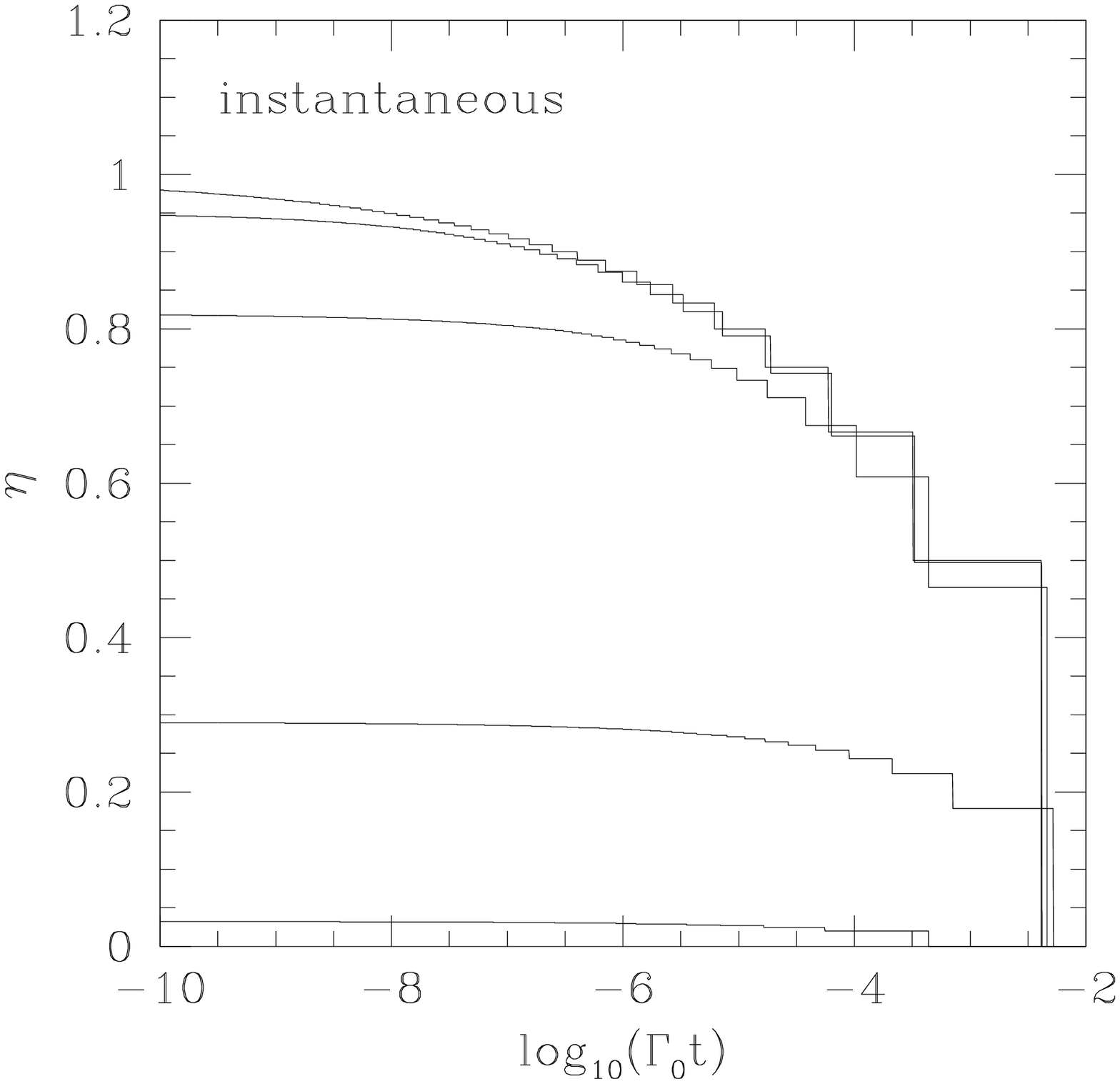} 
   \epsfxsize 2.33 truein \epsfbox{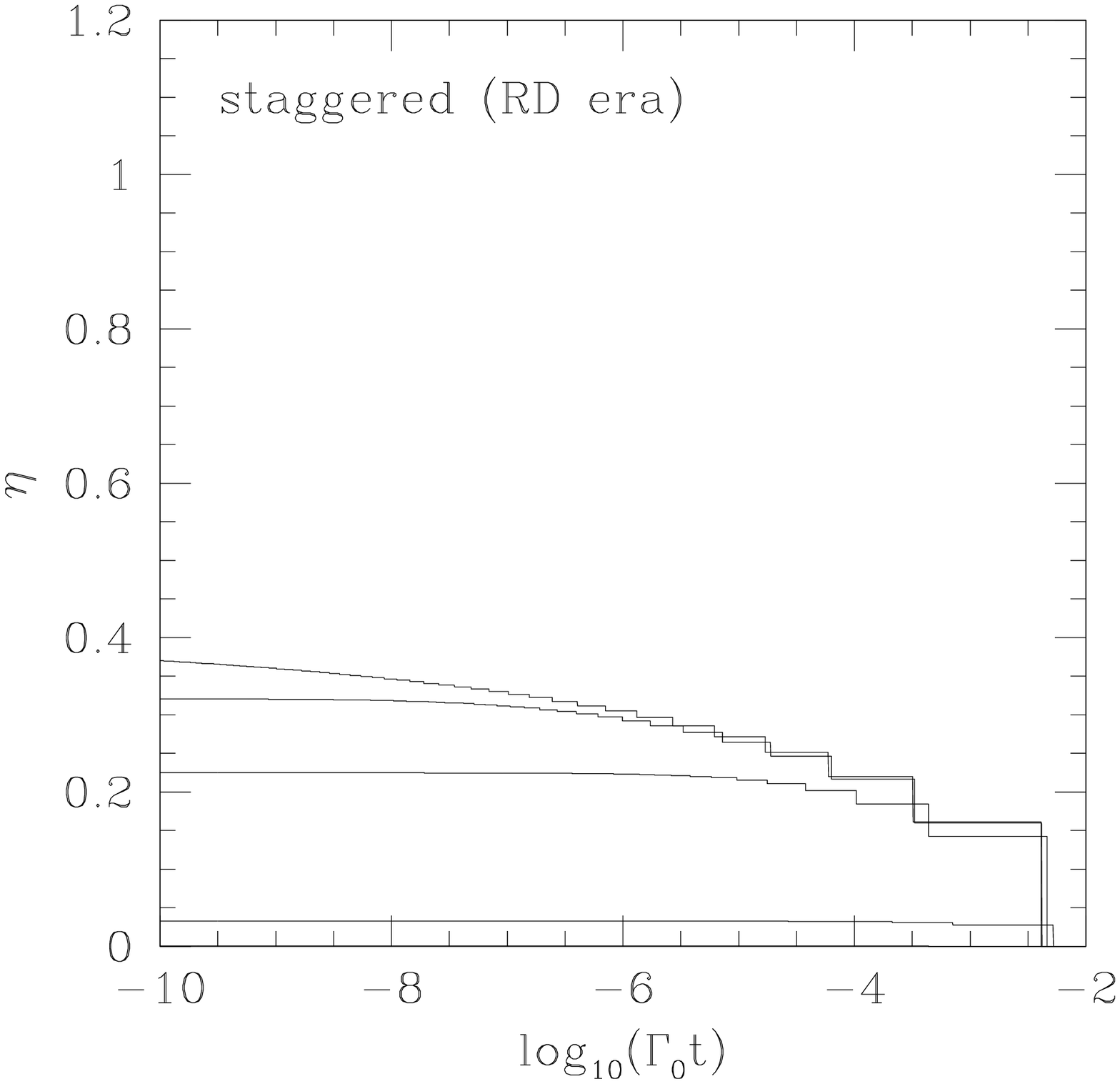} 
   \epsfxsize 2.33 truein \epsfbox{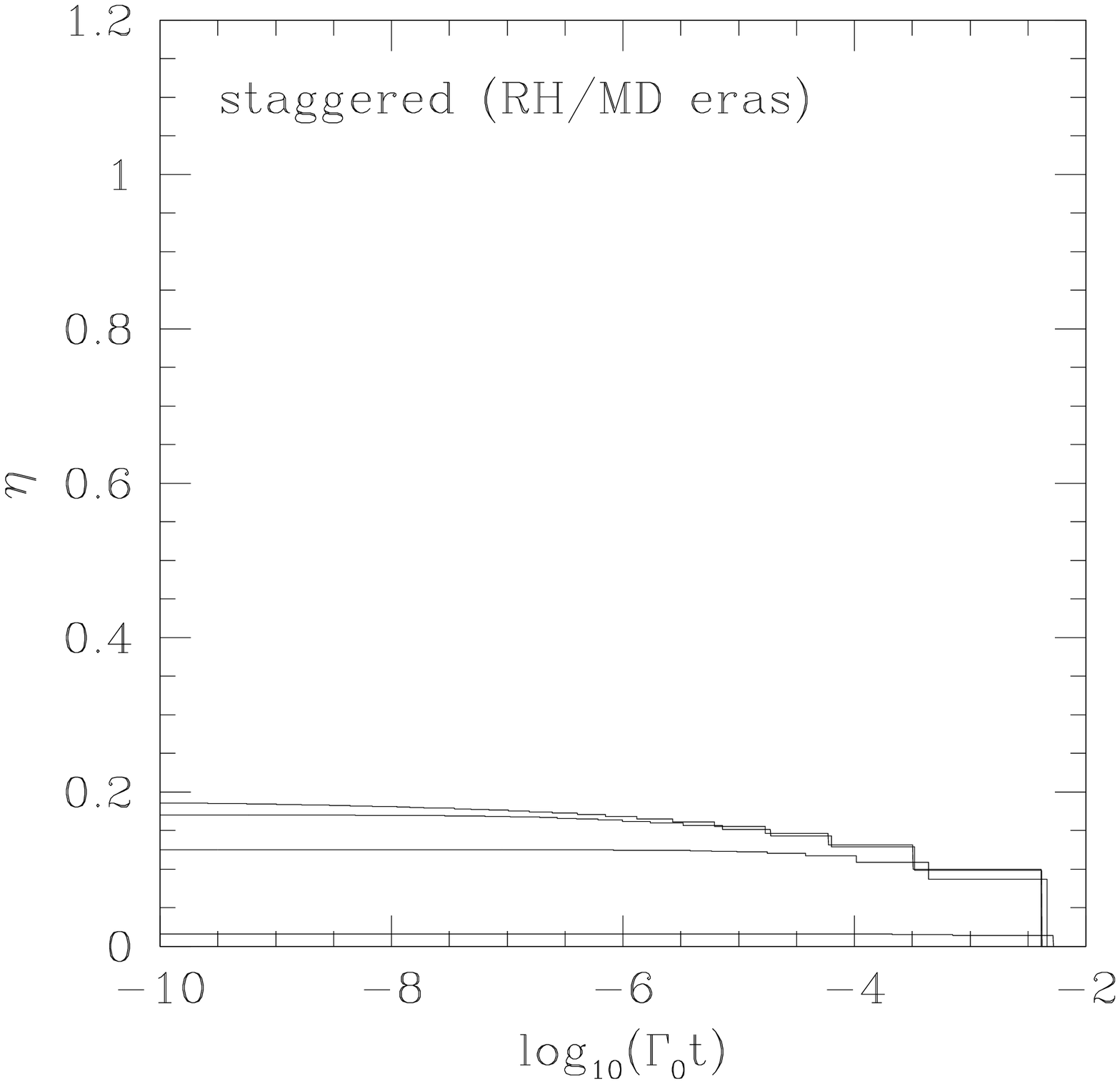} 
 }
\vskip -0.2 truein
\caption{The tower fraction $\eta$, plotted as a function
of time for each of the three cases relevant for a general KK tower.  
In each panel, the lowest curve corresponds to $y=10$ (not visible in the center
and right panels) and the successively
higher curves correspond to $y=3$, $y=1$, $y=0.5$, and $y=0.1$.
Thus, these values for $\eta$ directly 
correspond to the values of $\Omega_{\rm tot}$ plotted in Fig.~\protect\ref{omtotplots}.
In general, we observe that $\eta$ increases with decreasing $y$,
and that the maximum value of $\eta$ shown for each
curve is consistent with the results of Fig.~\protect\ref{plot2}.}  
\label{etaplots}
\end{figure*}  
%========================================================================

Although 
the total dark-matter abundance
$\Omega_{\rm tot}$ ultimately vanishes at $\log(\Gamma_0 t)=0$ for all curves, 
we see that the 
overall time-evolution of 
$\Omega_{\rm tot}$ as we approach this vanishing point is highly $y$-dependent. 
As $y\to\infty$,
we see that $\Omega_{\rm tot}$ remains constant until it experiences 
a single, sudden, complete decay;
this of course corresponds to the traditional scenario of a single dark-matter particle.
By contrast, for smaller values of $y$, we see that multiple modes with different decay widths
carry the total dark-matter
abundance $\Omega_{\rm tot}$;  as a result, the resulting transition of $\Omega_{\rm tot}$ from
its maximum value to zero is more gentle.
In all cases the quantity $1-\eta$ indicates the relative size 
of this final ``last step'' down to $\Omega_{\rm tot}=0$;  note that the
results for $\eta$ implicitly shown here in terms of the relative final step size 
are consistent with those shown in Fig.~\protect\ref{plot2}.
Likewise, the {\it initial}\/ values of $\Omega_{\rm tot}$ also confirm
our expectations discussed earlier:  the instantaneous 
and staggered (RH/MD) cases have initial values at $\Omega_{\rm tot}=1$,
in accordance with Eq.~(\ref{identities}) for all $y$,
while the initial values shown on the second panel
are $y$-dependent and 
correspond to the values shown in Fig.~\ref{yplot}.

Using the results for $\Omega_{\rm tot}(t)$ shown in Fig.~\ref{omtotplots},
we can now proceed to calculate 
the corresponding tower fractions $\eta(t)$.
The results are shown in Fig.~\ref{etaplots}. 
As expected, we see that $\eta$ increases with decreasing $y$.
Moreover, we can now see directly that the maximum value of $\eta$ shown for each
curve is consistent with the results of Fig.~\protect\ref{plot2}.

%================== FIGURE ============================================
\begin{figure*}[tbh]
\centerline{
   \epsfxsize 2.33 truein \epsfbox{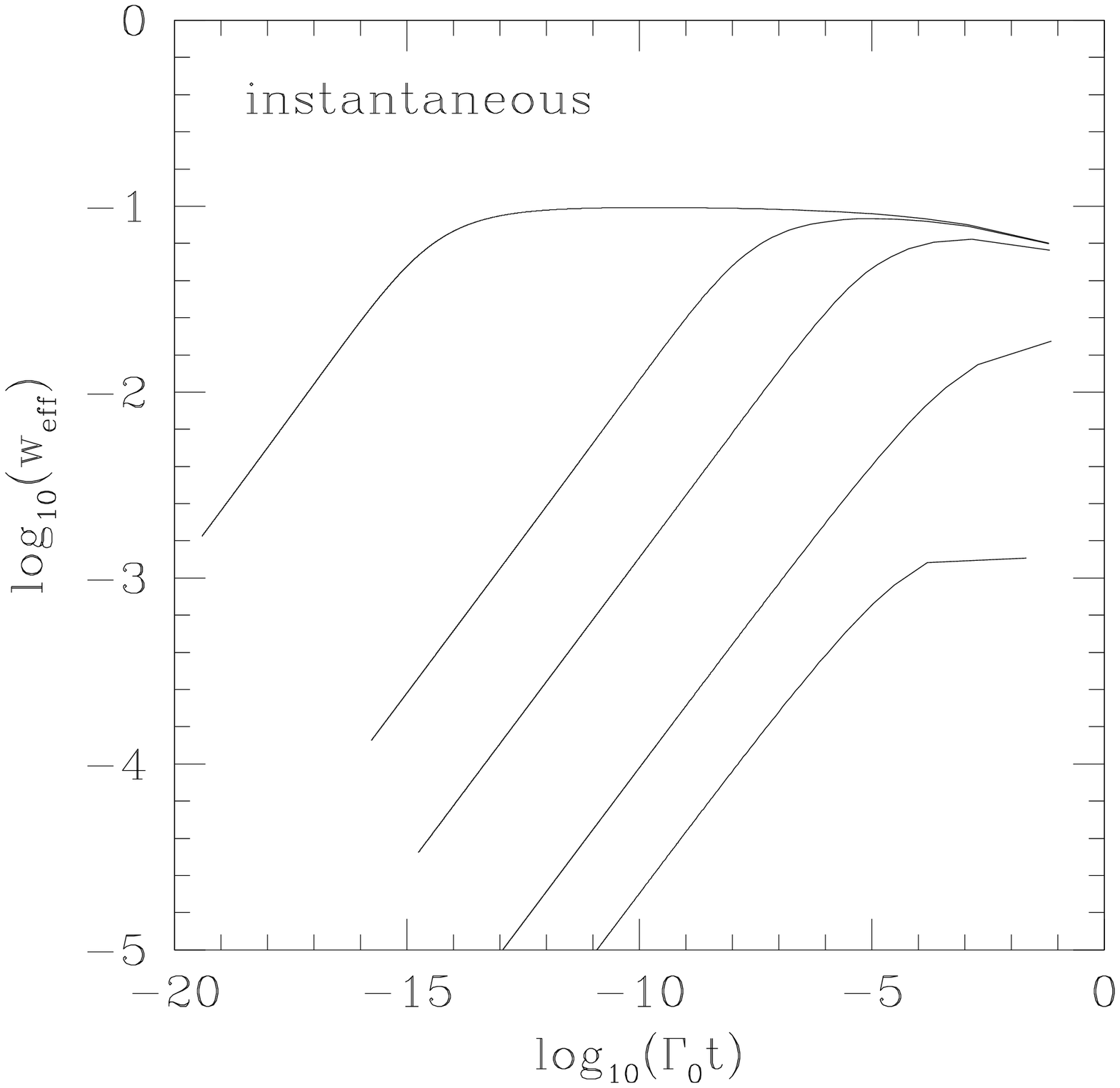} 
   \epsfxsize 2.33 truein \epsfbox{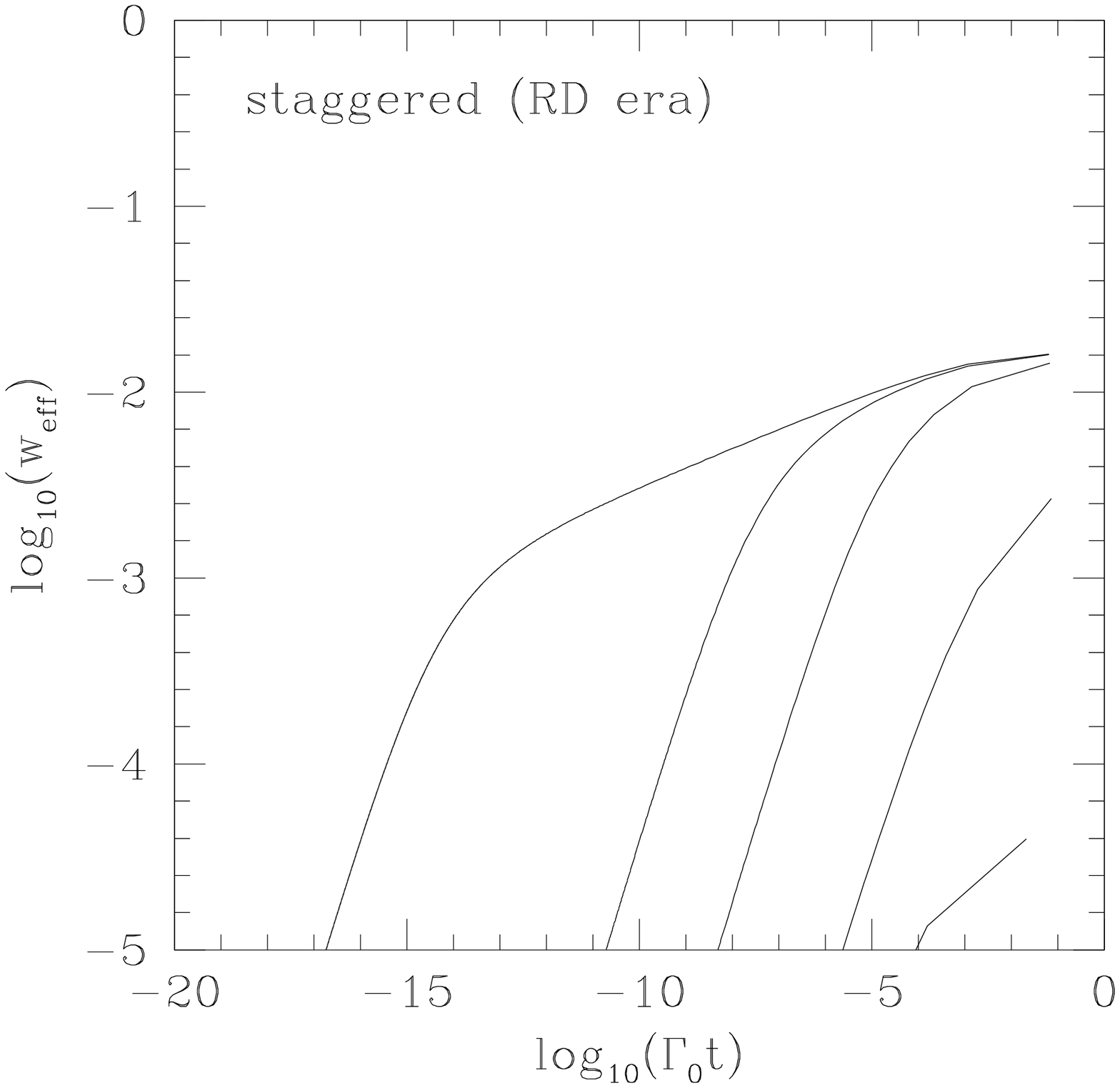} 
   \epsfxsize 2.33 truein \epsfbox{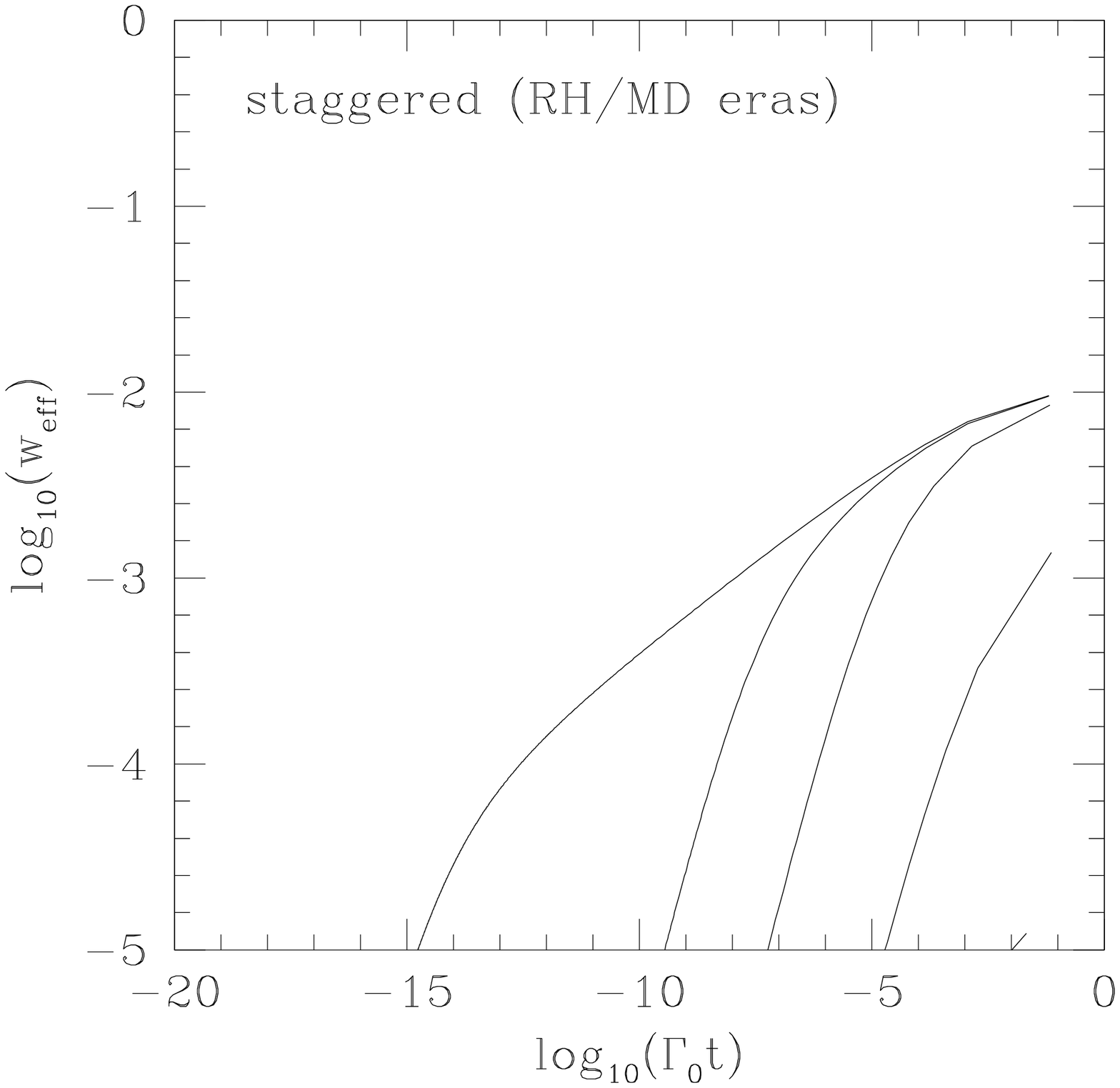} 
 }
\vskip -0.2 truein
\caption{
The effective dark-matter equation-of-state parameter $w_{\rm eff}$, plotted as a function
of time for each of the three cases relevant for a general KK tower.  
In each panel, the lowest curve corresponds to $y=10$ and the successively
higher curves correspond to $y=3$, $y=1$, $y=0.5$, and $y=0.1$.
Thus, these values for $w_{\rm eff}$ directly 
correspond to the values of $\Omega_{\rm tot}$ plotted in Fig.~\protect\ref{omtotplots},
or equivalently the values of $\eta$ plotted in Fig.~\protect\ref{etaplots}. 
In all cases, we see that $w_{\rm eff}\to 0$ as $t\to 0$.
Note that although the values of $\Omega_{\rm tot}$ were plotted in Fig.~\protect\ref{omtotplots}
only up to an overall rescaling factor, the values of $w_{\rm eff}$ plotted here
are insensitive to this rescaling and are thus meaningful on an absolute scale.
We thus see that $w_{\rm eff}$ never exceeds $0.1$ for a general KK tower, and is
generally much smaller.}
\label{wplots}
\end{figure*}  
%========================================================================

Using the results for $\Omega_{\rm tot}(t)$ shown in Fig.~\ref{omtotplots},
we can also proceed to calculate 
the corresponding equation-of-state function $w_{\rm eff}(t)$ 
which follows from the definition in Eq.~(\ref{weffdef}).
Equivalently, we can use the 
results for $\eta(t)$ shown in Fig.~\ref{etaplots} along
with the definition in Eq.~(\ref{weffdefeta}).
In either case, the results are shown in Fig.~\ref{wplots}. 
In passing from Figs.~\ref{omtotplots} and \ref{etaplots} to Fig.~\ref{wplots},
we have calculated logarithmic slopes numerically for each successive
KK decay event and then plotted a continuous function.
It is clear that the results in Fig.~\ref{wplots} are in complete
agreement with our general expectations for $w_{\rm eff}$ from Sect.~II:
in each case we observe the general tendency that $w_{\rm eff}\to 0$ as $t\to 0$, 
and likewise in all but one case
$w_{\rm eff}$ approaches a pole at $t=\Gamma_0^{-1}$ (corresponding to the fact
that $\Omega_{\rm tot}$ hits zero upon the decay of the final, lightest
dark-matter mode in the KK tower).

These results are also in agreement with our expectations
based on the KK scaling coefficients 
in Table~\ref{alphabetatable}.
As $y$ decreases, we see the emergence of a definite shift in the behavior 
of $w_{\rm eff}(t)$ as we transition from early times to later times.
This corresponds to the shift from large-$\tilde \lambda$ behavior
to small-$\tilde\lambda$ behavior in Table~\ref{alphabetatable}.
Indeed, in the case of an ``instantaneous'' turn-on, we
even observe that our function $w_{\rm eff}(t)$ develops a slight {\it maximum}\/
for smaller values of $y$, 
shifting from increasing behavior to slightly decreasing behavior.
This change in the slope of $w_{\rm eff}(t)$ for this particular situation
is directly correlated with the fact that the value of $x$ in Table~\ref{alphabetatable}
shifts from $x< -1$ to $x> -1$.  We see from both Table~\ref{alphabetatable} and
Fig.~\ref{wplots} that this is the only case in which such interesting
behavior occurs.   

At first glance, it might seem surprising that we are able to 
obtain effective equation-of-state functions $w_{\rm eff}(t)$ which depend on $x$, but which are
otherwise universal when plotted versus $\Gamma_0 t$.
Indeed, $w_{\rm eff}(t)$  depends
on a number of parameters:  not just
the dimensionless exponents $\alpha$ and $\beta$ in Eq.~(\ref{step2}),
but also dimensionful quantities such as 
the leading coefficients $A$ and $B$ in Eq.~(\ref{step2})
as well as physical parameters such as
$\Omega_{\rm CDM}$
which are involved in setting a boundary condition for
$\Omega_{\rm tot}$.
Indeed, all of these parameters 
appear in our approximate results for $\Omega_{\rm tot}(t)$ 
in Eqs.~(\ref{result}) and (\ref{logresult}).
However, the important point is that while
$\Omega_{\rm tot}$ depends on all of these dimensionful quantities somewhat
independently, $w_{\rm eff}(t)$ depends on them in only one particular
combination.
This was already apparent in Eqs.~(\ref{weffresult}) and 
(\ref{weffresultlog}), where the combination   
in question was nothing but $w_\ast \equiv w_{\rm eff}(t_{\rm now})$.

Of course, the results for $w_{\rm eff}(t)$ in Eqs.~(\ref{weffresult}) and (\ref{weffresultlog})
were respectively derived from the results for $\Omega_{\rm tot}(t)$ in Eqs.~(\ref{result}) and
(\ref{logresult}), and these in turn were realized by taking our boundary condition
to be $\Omega_{\rm tot}(t_{\rm now})=\Omega_{\rm CDM}$.
However, we can equivalently write our boundary condition in the form
$\Omega_{\rm tot}(1/\Gamma_0)= 0$,
where we are assuming that each KK mode with mass $\lambda$
decays promptly at $t=\tau_\lambda\equiv \Gamma_\lambda^{-1}$,
with $\Gamma_0$ denoting the width of the lightest mass eigenmode.
Following the same algebraic manipulations as in Sect.~II
then leads to equations of state 
which are written in terms of $\Gamma_0$ rather than $w_\ast$:
\beq
   w_{\rm eff}(t) ~=~ \cases{
     \displaystyle  \half \displaystyle (x+1) \left\lbrack 1-(\Gamma_0 t)^{x+1}\right\rbrack^{-1} 
                   & for $x\not= -1$\cr
     ~&~\cr
     \half (\Gamma_0 t)^{-1} & for $x= -1$~\cr}
\eeq
\vskip 0.3 truein
\noindent where $x\equiv \alpha+\beta$.
Thus, when expressed in terms of the dimensionless time variable $\Gamma_0 t$ as in Fig.~\ref{wplots},
our $w_{\rm eff}(t)$-functions are indeed universal, depending only on $x$.

It is important to bear in mind that
in this section
we have made only minimal assumptions concerning the precise
nature of this KK tower or the identity of the fields it represents.
We therefore expect that all of the features we have discussed in this section
will hold quite generally,
regardless of the identity of the particular field(s) which happen to populate
the bulk and constitute the KK tower.

%========================================================================
\section{Dynamical dark matter:  ~Novel signatures and phenomenological constraints}

Having described the general theoretical structure of our dynamical dark-matter framework,
we now present several additional features of this framework which are likely to be 
of importance in enabling this framework to satisfy phenomenological constraints.
As in other sections, our discussion here will be restricted to broad, model-independent 
themes, and we shall present
a detailed phenomenological analysis of 
one specific dynamical dark-matter scenario in Refs.~\cite{paper2,paper3}.

It turns out that there are three phenomenological features which are unique to dynamical dark matter,
and which under certain circumstances 
might be taken as signatures (or even ``smoking guns'') of the overall framework.

\begin{itemize}
\item  {\sl \underbar{No well-defined dark-matter mass or cross-section:}}
      ~~First, since the dark-matter ``candidate'' within the dynamical dark-matter
       framework is not a single particle, but rather an ensemble of particles,    
       it does not have a specific
        mass or cross-section associated with it.
       This represents a marked difference relative to most other dark-matter proposals,
       and implies that the kinematics associated with the production and decay of dynamical
       dark matter is likely to be quite different from that associated with more traditional
       single-component dark matter.  This could have dramatic consequences for 
       collider phenomenology (and potentially for direct detection), 
       and may also have a number of cosmological implications.

\item  {\sl \underbar{Coupling suppression for light modes:}}
     ~~Second, it is almost inevitable that the eventual phenomenological success or failure
     of specific dynamical dark-matter scenarios involving
      large extra dimensions will ultimately rest in part on the couplings between the 
      dynamical dark matter in the bulk and the Standard-Model states on the brane.  
      Such couplings are of critical importance because they govern the degree to which
      this dark matter might be ``visible'' to the Standard Model.
      From intuitions based on studies of KK graviton dynamics, one might suspect 
     that all mass eigenstates in the bulk 
     would couple to the brane with equal strength.  However, it turns out 
     that the opposite is true for theories with a brane mass:
     while the most of the couplings between the fields on the brane and the mass eigenstates 
     in the bulk are indeed uniform,
     the couplings between the brane and the {\it lightest}\/ mass eigenstates in the bulk 
     are significantly suppressed.  
     We have already seen this behavior in Fig.~\ref{plot1}, where we plot the coupling matrix
     element $\langle \phi_\lambda|\phi'\rangle= \tilde \lambda^2 A_\lambda^2$ as 
     a function of $\lambda$:
     although this coupling always reaches an asymptotic value for sufficiently large $\lambda$,
     this coupling is significantly suppressed for small $\lambda$.
     Indeed,
     the magnitude of this suppression can be controlled by varying the 
     non-diagonality parameter $y$.

     This suppression feature is of immense phenomenological importance, since the couplings
     of the lightest dark-matter eigenstates to the brane are precisely those which 
     are the most dangerous
     for the viability of our dynamical dark-matter framework.
     Thus, this suppression feature can be very important in relaxing
     numerous phenomenological bounds on dynamical dark matter, and thereby 
     constitutes an unexpected additional
     effect which can help dynamical dark matter stay dark despite its multitude of states.

\item  {\sl \underbar{Decoherence:}} ~~Finally, 
     there is an additional feature associated with dynamical dark matter 
     in large extra dimensions which can play an important role in its phenomenology:
     this is the phenomenon of 
     ``decoherence''~\cite{DDGAxions}.  As we have seen in Sect.~III, only one 
     particular linear combination of bulk dark-matter modes $\phi_n$ can 
     couple to the brane:  this is the linear combination $\phi'$.  However, 
     once $\phi'$ is created through an interaction with the brane, it rapidly 
     {\it decoheres}\/ as it propagates because it is not a mass eigenstate.  

     One way to understand this decoherence involves simple quantum mechanics:  
     because $\phi'$ consists of a huge number of different mass eigenstates, 
     and because the masses of these eigenstates are generally not related 
     to each other through rational multiplicative factors, the different 
     mass eigenstates fall out of phase with each other under time-evolution and will not 
     reconstitute $\phi'$ within finite time.  Thus, they cannot couple to the
     brane at later times, and essentially become ``invisible'' as far as physics 
     on the brane is concerned.  
     Another (quantum field-theoretic) way to describe the same phenomenon is 
     simply 
     that the amplitude associated with any process that involves 
     the production and subsequent detection of dark matter on the brane will have 
     multiple individual contributions, each associated with the propagation of an 
     intermediate state consisting of an individual dark-matter component. 
     However, because these individual components have different masses, their 
     corresponding amplitudes accrue different phases.  These amplitudes therefore 
     destructively interfere within the calculation of any cross-section sum.

     This decoherence phenomenon can have important phenomenological consequences.  Indeed,
     decoherence generically induces a suppression of the cross-section for any process involving
     virtual dark-matter particles   
     by a factor of $N$, where $N$ is the number of such particles being exchanged.
     This, then, is yet another mechanism which helps dynamical dark matter stay dark.
     We emphasize that this feature is not specifically extra-dimensional;  it applies to any
     dark-matter framework in which the dark matter has many components of different masses,
     and in which only a specific linear combination of those components can couple to Standard-Model states.
\end{itemize}

Needless to say, dynamical dark matter must ultimately be subjected to all of the phenomenological
bounds and constraints that apply to more traditional dark-matter candidates.
However, because dynamical dark matter consists of a vast ensemble of individual states which
are not necessarily stable on cosmological time scales, 
many of these constraints take unusual forms in this context.
We shall therefore now provide a quick overview of the different classes of 
laboratory, astrophysical,
and cosmological constraints
which apply to dark matter in general, and then indicate the forms they can be expected to take
within the context of dynamical dark matter.
Once again, we emphasize that our goal here is merely to provide a model-independent
theoretical overview in which we restrict ourselves to addressing a single question:  
 {\it for each class of constraints that apply to theories of dark matter,
what combinations of parameters are bounded in the traditional framework and
how do these combinations translate into our dynamical dark-matter framework?}  
Explicit details concerning a particular dynamical dark-matter scenario can be found
in Refs.~\cite{paper2,paper3}.

Broadly speaking, there are four classes of constraints which apply to any candidate
theory of dark matter.
 
\begin{itemize}

\item  First, there are general constraints on the relic dark-matter abundance
and on the dark-matter equation of state.  
The constraints on the dark-matter abundance $\Omega_{\rm tot}(t)$ 
are similar to those which apply in
traditional dark-matter scenarios:  
 $\Omega_{\rm tot}(t_{\rm now})$ must match the observed
dark-matter relic abundance $\Omega_{\rm CDM}$;
the dark-matter ensemble must not cause the universe
to become matter-dominated too early;  {\it etc}\/.  
However, our scenario differs from traditional dark-matter scenarios in that
it gives rise to an equation-of state-parameter $w_{\rm eff}$ which can be different from zero
and which generally exhibits a non-trivial time dependence.
Astrophysical and
cosmological considerations therefore imply
additional constraints on our equation-of-state function $w_{\rm eff}(t)$,
or equivalently on the scaling coefficients $(\alpha,\beta)$.

\item  Second, there are constraints on dark matter which 
    derive from physical processes
    in which dark matter is produced through its interactions with Standard-Model
    particles but not subsequently detected.
    For example, there are collider constraints on processes in which 
    dark-matter particles manifest themselves as missing energy.
    Furthermore, if the dark-matter candidates in question are sufficiently light,
    additional constraints can be derived from limits on dark-matter production 
    by astrophysical sources.  For example, dark-matter particles produced 
    in stars and supernovae can carry away energy from these sources 
    very efficiently.  This can lead to
    observable effects on stellar lifetimes, energy-loss rates from supernovae, {\it etc}\/. 
    Indeed, observational limits on the magnitudes of these effects imply stringent 
    bounds on any light particle whose interactions with the Standard-Model fields are 
    highly suppressed.

    To see how such considerations constrain the parameters of a generic dark-matter 
    model, let us consider a traditional single-component scenario in which the dark-matter
    candidate resides in a hidden sector. 
    The dominant interaction between the dark sector and the Standard Model 
    in such scenarios occurs through non-renormalizable operators $\mathcal{O}_n$ of
    mass dimension $n$, suppressed by inverse powers of some large mass scale $\Lambda$ 
    associated with the cutoff of the theory.  For example,  
    $\Lambda$ might be an effective Planck scale
    $M_P$ in the case of dark-matter candidates associated with gravity,
    or a particle mass $M_R$ in the case of candidates such as a right-handed neutrino,
    or a dynamical scale such as the Peccei-Quinn scale $f_{\rm PQ}$ in the case of axions.
    The cross-section for dark-matter 
    production will therefore be suppressed by a factor of $\Lambda^{2(4-n)}$, and 
    constraints in this class thus ultimately become constraints on $\Lambda$.

     For example, in the specific dynamical dark-matter scenario presented in 
     Sect.~III, the leading
     operators have mass-dimension five, and $\Lambda$ is equated with the 
     suppression scale $\hat f_\phi$.
     Thus, constraints in this class ultimately become 
     bounds on $1/\hat f_\phi^2$.
     Or, phrased directly in terms of the decay widths and abundances which are the bedrock
     of our scenario, these constraints yield bounds on 
     $\sum_\lambda \lambda^{-3} \Gamma_\lambda$, where the sum over mass eigenstates
     includes only those states which are kinematically relevant for the 
     process in question. 
    
    We conclude, then, that constraints of this type tend to place bounds on the particular
    combination  $\sum_\lambda \lambda^{-3} \Gamma_\lambda$.
    However, this quantity is significantly affected by the coupling-suppression effect
    discussed above. 
    As a result, such bounds can often turn out to be considerably weaker
    that one might imagine at first glance.   

\item  Third, there are constraints that arise from situations in which dark matter
     is produced through its interactions with Standard-Model particles and is then
    subsequently detected (either directly or indirectly) via those same interactions.
     Here we have in mind not only astrophysical production with subsequent detection
     on earth, but also any process involving virtual dark-matter particles. 
    Which physical processes of this sort serve to constrain a particular dark-matter 
    particle are extremely model-specific.  Axions and other similar particles, for 
    example, are constrained primarily by helioscope searches, microwave-cavity 
    experiments, {\it etc}\/.; other particles are more stringently constrained by collider 
    limits, and so forth.  Nevertheless, a few generic observations can be made.

     If we assume, as above, that the dark matter resides in a hidden sector, 
     it then follows that the cross-sections for processes of this 
     sort are proportional to $\Lambda^{4(4-n)}$.  Thus, once again, limits on 
     such cross-sections ultimately become bounds on $\Lambda$.  For example,  
     in the specific dynamical dark-matter scenario presented in 
     Sect.~III, they become
     bounds on $1/\hat f_\phi^4$, or equivalently    
     on the quantity $(\sum_\lambda \lambda^{-3} \Gamma_\lambda)^2$.
     In terms of overall mass scales, we might approximate this quantity as
     $\sum_\lambda \lambda^{-6} \Gamma_\lambda^2$,
    but we must also bear in mind that the cross-terms within such products can be
    significant.  Indeed, these are precisely the situations in which the decoherence
    phenomenon discussed above can play a role.  Thus, these constraints 
     might also turn out to be significantly weaker than they might at first sight appear.

\item   Finally, there are constraints on dark-matter decays and annihilations.
    As far as decays are concerned, we have in mind constraints such as 
    those from big-bang nucleosynthesis, 
    measurements of the cosmic microwave background, observations
    of the diffuse X-ray and gamma-ray backgrounds, {\it etc}\/.
    The basic idea behind all of these constraints is that the decays of
    a cosmological population of dark-matter particles can result in measurable
    deviations from the standard cosmology at times $t\gsim 1$~s, or leave
    (unobserved) imprints on these backgrounds.
    In situations in which dark-matter annihilation cross-sections are sufficiently
    large, various additional limits (such as those from typical indirect-detection 
    methods) would also apply.

    Let us focus on those constraints related to dark-matter decays, as these are generic
    to dynamical dark-matter scenarios.  (By contrast, constraints related to 
    dark-matter annihilation tend to be somewhat model-dependent,
    and indeed often do not apply.)
    Roughly speaking, in a traditional single-component dark-matter scenario, 
    such dark-matter decay constraints tend to place bounds on the
    product $\Omega_\chi \Gamma_\chi$, where $\Omega_\chi$ and $\Gamma_\chi$
    are the abundance and decay width of our dark-matter field $\chi$,
    suitably evaluated during the appropriate cosmological period.
    In a dynamical dark-matter framework, by contrast, this now becomes
    a constraint on 
\beq
        \langle \Gamma \rangle ~\equiv~ \sum_\lambda \Omega_\lambda \Gamma_\lambda
\eeq
    where again our abundances and widths are to be evaluated during the 
    cosmological epoch during which decays can contribute to the effect in question 
    (the disruption of BBN, distortions of the CMB, \etc).
    It is important to note that this dependence on time effectively truncates 
    the range of the above sum to those states whose lifetimes fall roughly
    within the time scale associated with that epoch.  
    Put another way, 
    only those states whose masses lie within certain characteristic ranges contribute
    to the sum. 
    Of course, at a mathematical level, the behavior of this sum ultimately depends 
    critically on the balancing
    relations that happen to hold across the entire dark-matter ensemble. 
    Note that these arguments will be addressed more rigorously in
    Refs.~\cite{paper2,paper3}.

\end{itemize}

Finally, for completeness, we also mention two further 
classes of constraints which must also be borne in mind. 
Unlike the previous constraints, these are substantially more model-dependent.

\begin{itemize}

\item  First, there can be phenomenological bounds that accompany
      (and are therefore specific to)  particular realizations of dynamical dark matter.
       For example, we have seen that an infinite tower of Kaluza-Klein states furnishes
       an excellent realization of dynamical dark matter.
      However, the extra-dimension brane/bulk framework brings with it a whole host of
     additional bounds and constraints, some of which come from the fact that we are now
     attempting to do standard physics within such a context 
     (\eg, the need for a late-time-reheating (LTR) cosmology~\cite{ADD}), 
      and others of which place bounds on the context itself
    (\eg, E\"otv\"os-type or Cavendish-type ``fifth-force'' experiments 
     which restrict the allowed sizes of the extra dimensions).
    Such constraints are clearly highly model-dependent, and frequently they are
    also wholly independent of the general dynamical dark-matter framework.

\item Finally, there can also be constraints that arise 
     simply for reasons of theoretical self-consistency.
     For example, if we assume (as we have done here) that our initial dark-matter 
     abundances are determined through misalignment production, then we must insist 
    that misalignment production indeed dominates over other production
    mechanisms such as thermal production.
    This, of course, yields a non-trivial constraint on the parameters of the model. 
    Likewise,  
    the assumption that dynamical dark matter in the bulk decays preferentially
    to Standard-Model states on the brane, rather than to other dynamical dark-matter states
    in the bulk, implies yet another self-consistency requirement.
    Once again, however, constraints in this class tend to be highly model-dependent
    and therefore do not represent generic constraints on the 
    dynamical dark-matter framework. 
 
\end{itemize}

This is clearly a fairly long list of constraints, and one must not minimize
the impact that they can have in ruling out specific dark-matter proposals.
In Refs.~\cite{paper2,paper3}, however, we shall study one particular realization
of dynamical dark matter, and we shall exhaustively work through all of the constraints
relevant for this particular realization.
We shall find that for this particular scenario, our 
dynamical dark-matter framework indeed survives all known
laboratory, astrophysical, and cosmological constraints.
This will thereby furnish us with an ``existence proof'' 
that dynamical dark-matter can indeed be a viable, alternative framework
for addressing the central questions in dark-matter physics.

Finally, we remark that for some purposes it may also be interesting 
to consider the phenomenology
associated with our overall dark-matter framework
when the individual component decay widths $\Gamma_i$ are extremely small.  
Of course, in the actual limit $\Gamma_i\to 0$ we know that
$\Omega_{\rm tot}$ approaches a constant in the final, matter-dominated era; 
likewise, $w_{\rm eff}\to 0$.  In this respect, this limit of our framework begins
to resemble a traditional {\it non}\/-dynamical dark-matter framework, thereby allowing
us to evade many of the most stringent 
phenomenological constraints coming from dark-matter decays in the early universe.
However, even in this limit, our framework nevertheless continues to retain 
those distinctive features which stem from 
its underlying multi-component nature.
For example, we can still have $\eta\not=0$.
We also still have the possibility of staggered turn-ons 
as well as the possibility of coupling suppression for light modes.
We even continue to have decoherence, even though many of the processes for
which decoherence is most phenomenologically relevant will already tend to be suppressed in 
the $\Gamma_i\to 0$ limit.
And perhaps most importantly, 
the dark matter in our framework will continue to evade simple characterization
in terms of a single well-defined mass or cross-section.

Needless to say, we are not particularly interested in the limit $\Gamma_i\to 0$.
Indeed, we regard the dynamical aspects of our
dark-matter framework to be among its most intriguing features and key signatures.
However, the freedom to tune the values of $\Gamma_i$ 
relative to the other dimensional parameters in our framework
is important from a theoretical standpoint because 
it illustrates that our overall dark-matter framework possesses
a means of ``dialing'' the scale associated with its dynamical aspects 
relative to those associated with its multi-component aspects.  
This is particularly relevant because 
the dynamical aspects of our framework are
often ultimately subject to an entirely different set of phenomenological bounds and constraints 
than those governing its multi-component aspects.
Thus, the freedom to independently adjust the scales associated with 
these different aspects of our dark-matter framework 
gives this framework an added flexibility 
when it comes to satisfying many of the phenomenological bounds discussed above.

Of course, within the particular higher-dimensional brane/bulk context discussed 
in Sect.~III, this freedom
may initially appear to be lacking:
a single five-dimensional mass scale $f_\Phi$ 
governs not only the magnitudes
of the abundances of individual dark-matter components but also the magnitudes of their 
corresponding decay widths.  Indeed, in particular realizations of this framework,
even the brane mass $m$ can be tied to $f_\Phi$, and we shall see an explicit example
of this in Ref.~\cite{paper2}.  However, there is in principle no reason why the 
mass scale $f_\Phi$ which appears in Eq.~(\ref{prefdef}) and which ultimately sets the scale for dark-matter 
abundances needs to be the same
as the mass scale $f_\Phi$ which appears in Eqs.~(\ref{group1}) and (\ref{group2}) and which ultimately
sets the scale for decay widths.
Indeed, identifying these two quantities is merely a 
minimal assumption about the energy scales in our higher-dimensional theory,
and we can easily envision more complex scenarios in which these two mass scales are significantly
different.

%===============================================================================
\section{Conclusions and Discussion}

In this paper, we have introduced a new framework for dark-matter physics 
which we call ``dynamical dark matter''.
Unlike most approaches to the dark-matter problem
which hypothesize the existence of a single, stable, dark-matter particle,
our dynamical dark-matter framework may be characterized 
by the fact that the requirement of stability is replaced by a delicate 
balancing between cosmological abundances and lifetimes
across a vast ensemble of individual dark-matter
components.  This setup therefore collectively produces
a time-varying cosmological dark-matter abundance, and
decays of the different dark-matter components 
can occur 
continually throughout the evolution of the universe.

Although our framework is quite general and need not be tied to a specific set
of particles or theoretical models,
we have shown that one natural realization of this scenario consists of 
a tower of KK states corresponding to a single higher-dimensional
field propagating in the bulk of large extra spacetime dimensions.
Indeed, as we have shown, the states in such a ``dark tower''
naturally obey inverse ``balancing'' equations that  
relate their abundances and decay widths in just the right manner. 
Remarkably, this remains true even if the stability 
of the KK tower itself is entirely unprotected.
Our dynamical dark-matter scenario is therefore well-motivated
both in field theory and string theory, and can even
be used to constrain the cosmological viability of
certain limits of string theory.
We have also seen that within this context,
dynamical dark matter generically gives rise to certain
phenomena such as coupling suppression and decoherence which may help  
to explain why dark matter is dark and thus far unobserved.
Such phenomena transcend those usually associated with 
traditional single-component dark matter,
and may in some sense be viewed as unique signatures for a dark-matter 
framwork such as ours which rests on the existence of a large multitude 
of individual dark-matter components.

Needless to say, there are many possible generalizations of our basic dynamical
dark-matter framework.  Some of these apply
to dynamical dark matter in general, while others are more specific
to realizations involving extra dimensions.
For example, insofar as our general dark-matter ensemble is concerned,
there are several natural extensions which can be contemplated.

\begin{itemize}

\item  Not all components within the ensemble need be scalars.  Higher-spin
    fields may also be considered.
     We may even demand that our ensemble be supersymmetric, although
     there would be no obvious need for $R$-parity within such supersymmetric
     extensions as far as dark-matter considerations are concerned.

\item  In this paper, we have examined the case of relatively simple 
      dimension-five couplings between the components in the ensemble 
      and the fields of the Standard Model.  
    However, different scenarios may involve different coupling structures,
   and thus different models will lead to their 
    own distinctive phenomenologies.  

\item Continuing along these lines, we have assumed in this paper that all of the
   components of our dynamical dark-matter ensemble are neutral under 
    the Standard-Model gauge symmetries.  While this choice is particularly 
   convenient, allowing the possibility of specific realizations of our
   scenario in higher-dimensional    
   brane/bulk Kaluza-Klein theories, there is nothing
   intrinsic to the dynamical dark-matter framework that requires this
   to be the case.  In particular, some or all of the components of our 
   dark-matter ensemble could have ${\cal O}(1)$ $SU(2)$ weak interactions with 
   the Standard Model.  This would, of course, undoubtedly tighten many of the 
   phenomenological constraints on such scenarios;  
    likewise, scattering processes involving such dark-matter components are 
    also generally likely to play
    an important role and would need to be included in the analysis along the
   lines discussed in the Appendix.  However, as long as the lifetimes of the ensemble
   components are sufficiently balanced against their abundances,
   the basic features of our framework will remain intact.

\item  In this paper, we have considered all decays of our ensemble
   components to be essentially instantaneous.  However, such decays
    really have an exponential time-dependence.  The fact that these decays have
    different widths can thus lead to further non-trivial effects
   on the time-dependence of the total dark-matter abundance
    associated with the ensemble.

\item  The primary decay mode
      for a given dark-matter component within our ensemble 
    need not always be directly into 
     Standard-Model states.  In particular, it is also possible to consider
     decays from heavier ensemble components into lighter ensemble
    components.  Note that in this sense,
    we are viewing the ensemble as consisting of all states which
    are neutral under Standard-Model symmetries, including fields which
    reside in what might
    in more traditional contexts be considered a hidden sector.
    Such intra-ensemble decays could significantly alter
    the sorts of balancing equations which might arise across
    our dynamical ensemble, and as we shall discuss in the Appendix,
    they can thereby modify the time-dependence
    associated with $\Omega_{\rm tot}$, $\eta$, and $w_{\rm eff}$.

\item  Misalignment production need not be the only mechanism through
    which the abundances of our individual components are initially
    established.  Many other mechanisms (\eg, thermal production,
    decays arising from topological defects, {\it etc}\/.) also provide
    ways of populating the different components, and can likewise
    lead to different resulting phenomenologies for dynamical
    dark matter.  In particular, it would be very interesting (and
    relevant for our overall framework) to see whether the correct
    sorts of balancing relations might arise for situations in which
    our different ensemble components are populated in the manner
    of a standard WIMP --- \ie, by thermal freeze-out.

\item  At many points in this paper, we have made assumptions that
     simplify our analysis.  For example, in Sect.~II we have assumed
    that $t_1$, the time by which our staggered turn-on has ended,
     is less than $t_2$, the time at which significant dark-matter decays
     commence.  
      Likewise, we have assumed for much of our analysis of abundances
     in Sect.~III that a staggered turn-on, if it occurs, takes place
     entirely within a single epoch (either RH, RD, or MD).  
     While such assumptions prove useful
      for analyzing the effects of different features of our
     framework individually,
    there is nothing intrinsic to the dynamical dark-matter framework
    which requires that these features be separated in this way, 
    and numerous extensions and combinations of these features are possible.

\item  Although we have discussed several different signatures 
     which are unique to dynamical dark matter,
     it is likely that our discussion has only begun to scratch the surface. 
     It would be interesting to investigate what other kinds of
     signatures are also possible within this framework.

\item  Finally, our discussion in this paper has assumed a standard
    FRW cosmology.  However, it would be interesting to repeat this
   analysis for a $\Lambda {\rm CDM}$ cosmology (and also for versions thereof
   with low reheat temperatures, as appropriate for theories with
    large extra dimensions).
    Indeed, within such cosmologies, quantities such as
   $\Omega_{\rm tot}$ will experience additional types of 
   time-dependence beyond those
    considered here.

\end{itemize}

Likewise, within the specific framework of large extra dimensions in which
our dynamical ensemble of dark-matter components is represented by an 
infinite tower of Kaluza-Klein states,
there are also numerous generalizations and extensions 
which may be contemplated.

\begin{itemize}

\item  We may consider situations involving multiple species of bulk fields.
    For example, the bulk can be a crowded place, consisting of a whole
   plethora of particles which are neutral under all Standard-Model gauge symmetries:
     these include gravitons and gravitini, axions and other axion-like particles, 
     string-theory moduli, right-handed neutrinos, and so forth.
   From the point of view of physics on the brane, all of these states
    can be considered ``dark matter'', and their contributions to quantities
    such as $\Omega_{\rm tot}$ must all be considered within the overall dynamical 
    dark-matter framework.

\item  We also need not restrict ourselves to a single extra spacetime dimension.
   Multiple extra dimensions are also possible.

\item  Likewise, our extra dimensions need not necessarily be flat.  Warped extra
   dimensions will give rise to an entirely different KK spectroscopy, and as a result
    the phenomenology of dynamical dark matter within such contexts is likely
    to be significantly different from what has been presented here.

\end{itemize}

The above represent ideas for generalizing our overall dynamical dark-matter
scenario.  However, it may also be possible to use this kind of dynamical framework in order
to address questions that go beyond dark matter {\it per se}.
While some of these are relatively straightforward, others are indeed quite speculative.

\begin{itemize}

\item One of the key features of dynamical dark matter is that quantities such as 
   $\Omega_{\rm tot}$ are {\it dynamical}\/ (\ie, time-dependent) in this framework --- 
   even during the current matter-dominated epoch during which the dark-matter abundance
   is normally thought to be roughly constant.
   It is therefore possible  that such a dynamic approach could serve as a starting point towards 
   addressing the cosmic coincidence problem.

\item Further along these lines, it might also be possible to address the
   cosmological constant within a similar framework.
   For example, the energy density associated with each scalar 
    field $\phi_i$ prior to its ``turn-on''
   behaves as dark energy rather than dark matter.
  Thus in this respect the cosmological constant in our framework is time-dependent as
   well, and this sort of dynamic cosmological constant might even persist into
    the current epoch if there continue to exist light scalar modes which have
    not yet turned on.
    In this case, the process of dark-matter decays would necessarily {\it overlap}\/  
    with the process of a staggered turn-on, and one might be able to develop a consistent
    theory in which a vast ensemble of states gives rise to both dynamical dark matter 
    and dynamical dark energy.

\item  The framework of dynamical dark matter might also provide a new means of
    placing phenomenological bounds on string theory (and in particular, 
    on candidate string models).
    After all, string models are typically rife with 
    ``bulk'' fields ---  even if their extra dimensions are compactified
    at or near the traditional Planck scale.  
     Some of these fields (such as moduli) are model-dependent:  they depend
    on the particular kind of compactification geometry employed in the construction
   of the candidate string model and their masses depend on the particular stabilization
    mechanisms, if any, which have been employed.
    By contrast, some of these fields are generically model-{\it independent}\/:
    these include  all fields associated with the 
   (super)gravity multiplet, such as the
   graviton, dilaton, other higher-form fields, and their possible superpartners.
     Indeed, if the Standard Model is restricted to a stack of D-branes within
    a given string construction,
     the corresponding ``bulk fields'' 
     include {\it all}\/ string states which
    do not couple to those branes.

    While these fields are typically required for the self-consistency of the string,
    those that are massless require stabilization.
    Indeed, this is nothing but the standard moduli problem.
    However, depending on the specific cosmological properties of these fields, 
    it is also true that their abundances 
    must necessarily be considered as contributing either to the total
    dark energy or the total dark matter of the universe --- even after they are stabilized. 
    An analysis of their cosmological effects is then likely to run along the lines we have
    presented here, and the cosmological viability of the underlying candidate string model 
   thus necessarily becomes an issue to be studied within a dynamical dark-matter
   (or dynamical dark-energy) framework.
    Indeed, in such cases our dark-matter ensemble could potentially include
    not only string KK modes, but also (a subset of) string oscillator modes and string winding modes.

\item  One central feature of our dynamical dark-matter scenario is the
   phenomenon in which an ensemble of decaying ``stuff'' with
   one equation of state collectively simulates ``stuff'' with a different equation of state.
   For example, in the specific dynamical dark-matter scenario presented here,
   an ensemble of decaying dark-{\it matter}\/ states (\ie, each with $w=0$) 
   collectively simulates an effective equation of state with $w_{\rm eff}>0$.  
   This notion of using a vast ensemble of states with one equation of state
    to simulate another is, we believe, worthy of exploration in its own right,
    independently of the specific uses for dark-matter physics that we have presented here.
   
\item  Further along the above lines, it is natural to ask whether we could construct
     an ensemble of individual components of ``stuff'' with {\it negative}\/ $w$.
     The decays of these components within the ensemble would therefore act to increase the
    effective value of $w$, and perhaps even simulate $w=0$.  
     In other words, it is possible that dark matter might not even need to be comprised
    of  matter!  In some sense, this is the converse of the scenario we have presented
   here, in which individual {\it matter}\/ components collectively produce a value for
    $w_{\rm eff}$ which, though not too different from zero, is still non-zero.
    Indeed, both scenarios may represent equally legitimate approaches to the dark-matter problem.

\item  Pursuing these lines still further, one can even speculate as to whether
     an ensemble of decaying dark-{\it energy}\/ components (each with $w= -1$) 
    could simulate dark {\it matter}\/ (with $w=0$).
      Indeed, such individual dark-energy components need be nothing
    more complicated than a set of scalar fields $\phi_i$ which decay (potentially
   into a hidden sector) {\it prior}\/ to turning on.
   Such an approach might then ``unify'' dark energy and dark matter
   as simultaneously stemming from a primordial ensemble of scalar fields.

\item  Along entirely different lines, there is another phenomenon
   inherent in our dynamical dark-matter framework which is potentially interesting
   in its own right:
   this is the phenomenon (discussed in Sect.~III)
   in which a KK tower appears to have periodic modings for its heavier
    modes, but anti-periodic modings for its lighter modes.
   As we have seen in Sect.~III, this result emerges rather generically,
   requiring only a bulk field that somehow accrues a non-zero brane mass.
   This phenomenon is extremely interesting, because one normally
    associates the modings of a given field with its boundary conditions 
    around non-contractible loops in a topologically non-trivial space,
    or equivalently with the magnitudes of the fluxes which might penetrate
    those loops.
    This phenomenon therefore seems to suggest a mechanism by which such
    modings or fluxes might become effectively energy-dependent.
    
\item  Finally, the general phenomenon of decoherence is interesting
   in its own right.  This might be extremely relevant for axion invisibility
    (see, \eg, the discussion in Ref.~\cite{DDGAxions}),
     and is also likely to be of more general applicability.  Indeed,
    this might provide an interesting approach to the moduli problem in string
    theory, and explain why moduli such as the dilaton are not observed.

\end{itemize}

We see, then, that our dynamical dark-matter framework appears to be
pregnant with numerous possibilities for extension and generalization.
However, even as a framework for dark-matter physics,
we caution that our presentation here has been limited to only
the broadest model-independent theoretical aspects and features.
  In particular, it still remains to 
    choose a specific realization of this scenario in terms of
   a particular species of bulk field, and examine the phenomenological
    consequences of such a choice in complete detail.  
     In other words, it still remains to build an actual {\it model}\/ of dark
  matter within this framework.
    However, this is precisely what we shall do in Refs.~\cite{paper2,paper3},
    and we shall verify there that 
    our specific models satisfy all known collider,
    astrophysical, and cosmological constraints.
We thus conclude that the dynamical dark-matter framework 
can indeed serve as a viable alternative to the
standard paradigm of a single, stable, dark-matter
particle, and that dynamical dark matter therefore has a legitimate place
alongside other approaches to dark-matter physics.

%============================================================================= 

\section*{Acknowledgments}

\smallskip
We are happy to thank 
Z.~Chacko, T.~Tait, and N.~Weiner for discussions.
%  We also acknowledge the US taxpayers whose dollars 
%  help to support fundamental science research in the US.~
This work was supported in part
by the Department of Energy under Grants~DE-FG02-04ER-41291 
and DE-FG02-04ER-41298.
The opinions and conclusions expressed here are those of the authors,
and do not represent either the Department of Energy or the National Science Foundation.

%=============================================================================

%  \appendix  % indicates that all subsequent sections are really appendices
\appendix*  % indicates that only one subsequent section exists, and is really an appendices

\section{~Intra-ensemble decays}

Throughout this paper, we have 
implicitly assumed that
each component of our dark-matter ensemble preferentially decays directly into one or more Standard-Model
states rather than into another, lighter component {\it within}\/ the ensemble. 
In other words, we have been assuming that the decays associated with 
the widths $\Gamma_i$ take the direct {\it extra-ensemble}\/ form $\phi_i\to {\rm SM}$,
and that such direct decays dominate over all possible {\it intra-ensemble}\/ decays
which might produce other dark-matter components among their end-products.
However, it is easy to generalize our overall framework to include 
cases in which this assumption is relaxed.

Towards this end, 
it proves useful to start by rewriting Eq.~(\ref{diffeq1}) in a more useful form.
Recall that $\Omega_i\equiv \rho_i/\rho_{\rm crit}$, where $\rho_i$ is the energy density
associated with our oscillating $\phi_i$ field and where $\rho_{\rm crit}\equiv 3M_P^2 H^2$, 
where $M_P$ is the reduced Planck mass.  
Eq.~(\ref{diffeq1}) can therefore be rewritten as
\beq
   \dot\rho_i + \left( 3 H + \Gamma_i\right) \rho_i~=~0~.
\label{diffe}
\eeq
However, because this energy density $\rho_i$ is entirely associated with the coherent
oscillations of the (zero-momentum modes of the) scalar $\phi_i$ field, it is possible
to repackage this energy density in terms of an effective number density 
$n_i\equiv \rho_i/m_i$.
We then find
\beq
   \dot n_i + \left( 3 H + \Gamma_i\right) n_i~=~0~.
\label{diffstep1}
\eeq
Indeed, Eq.~(\ref{diffstep1}) describes the evolution of the number densities
associated with each of the oscillating components in our ensemble under the assumption
that the only decays available for these components are direct decays into Standard-Model
states, \ie, $\phi_i \to {\rm SM}$, with widths $\Gamma_i$.

Let us now consider what happens if we introduce an additional
set of intra-ensemble decays of the form
\beq
      \phi_i~\to~ \sum_j N_{ij}^{(\alpha)} \phi_j +X^{(\alpha)}~.
\label{decaychannel}
\eeq
Here the $\alpha$-index labels the specific decay channel,
and $N_{ij}^{(\alpha)}$
are non-negative integers describing the multiplicities of the $\phi_j$ particles
produced in this decay channel (each of which may be assumed to have $m_j<m_i$). 
Likewise, $X^{(\alpha)}$ collectively represents 
any fields 
 {\it outside}\/ our dark-matter ensemble (potentially including Standard-Model fields) 
which may also happen to be produced in this decay process. 
We shall let $\Gamma_i^{(\alpha)}$ denote the width associated with the decay in Eq.~(\ref{decaychannel}).

The inclusion of such additional decay channels into our discussion leads to 
two additional effects on the time-evolution of the number densities $n_i$.
First, there will be an additional decline in $n_i$ due to the new decay channels
for $\phi_i$ which are now available.
However, there is also the possibility of
an {\it increase}\/ in $n_i$ due to the {\it production}\/ of $\phi_i$
particles from the decays of presumably heavier components $\phi_j$ within the ensemble.
Indeed, 
we find that Eq.~(\ref{diffstep1}) is now replaced with
the {\it coupled}\/ system of differential equations
\beq
   \dot n_i + \left( 3 H + \Gamma_i + \sum_\alpha \Gamma_i^{(\alpha)} \right) n_i~=~
             \sum_j \left(\sum_\alpha N_{ji}^{(\alpha)} \Gamma_j^{(\alpha)}\right) n_j~.
\label{diffeq2}
\eeq
 
In general, the solutions to Eq.~(\ref{diffeq2}) can exhibit a number of striking
behaviors.
Not only can there be direct decays into Standard-Model states, as before,
but there can also be ``cascade'' decays that take place entirely 
within the dark-matter ensemble,
from heavier states down to lighter states.
Indeed, a given state can also decay directly into 
Standard-Model states at any point along the cascade.
As a result,
a particularly rich and subtle phenomenology can easily ensue 
depending on the relations between $\Gamma_i$ and $\Gamma_i^{(\alpha)}$,
with different portions of the ensemble exhibiting different particle-decay patterns
in a manner reminiscent of the vacuum-decay patterns studied in Ref.~\cite{DTvacua}.
The corresponding values of $n_i$ can then alternatively rise and fall as time evolves.

The possibility of such dark-to-dark intra-ensemble decays also allows even more striking
features to emerge.  For example, once a given heavy state $\phi_i$ ``turns on'', it can potentially decay 
to lighter states $\phi_k$ which, because of their relative lightness, have 
not yet turned on.  Thus, in this way, we see that 
a given component $\phi_k$ within our ensemble 
can simultaneously contribute
to dark matter (in the form of daughter particles from the decays of heavier 
dark-matter components $\phi_i$);  to ``dark radiation''
(if the momenta of these $\phi_k$ daughter particles are large compared to their
masses);  and to dark energy
(in the form of the energy still trapped in the overdamped field $\phi_k$).

Given the result in Eq.~(\ref{diffeq2}), it might
seem at first glance to be a straightforward exercise to obtain
a corresponding set of coupled differential equations 
for the energy densities $\rho_i$ and the abundances $\Omega_i$.
Indeed, all that would seem to be necessary would be
to start with Eq.~(\ref{diffeq2}) in place of Eq.~(\ref{diffstep1})
and essentially
reverse the process that led from Eq.~(\ref{diffe}) to Eq.~(\ref{diffstep1}).
However, in going from Eq.~(\ref{diffe}) to
Eq.~(\ref{diffstep1}) we needed to assume that all of the energy density $\rho_i$ 
was in the form of coherent zero-momentum mode oscillations of the field $\phi_i$,
and this will no longer be true when intra-ensemble decays are possible.
Indeed, the daughters $\phi_i$ which are produced through such intra-ensemble decays
are literal particles --- they have their own momenta and  energies which are
governed by the kinematics of the specific intra-ensemble decays which produced them.
As a result, while it is still valid to discuss the time-evolution of a total number
density $n_i$ as in Eq.~(\ref{diffeq2}), 
we cannot simply identify $\rho_i= m_i n_i$ in order
to obtain a corresponding set of equations for the energy densities $\rho_i$ or abundances $\Omega_i$.

In order to handle this calculation correctly,
it is first necessary to express the relations in Eq.~(\ref{diffeq2}) in terms of the 
phase-space distributions $f_i(|\vec p_i|,t)$ [or equivalently $f_i(E_i,t)$, simply denoted $f_i$] 
associated with each field $\phi_i$.
This will essentially yield a Boltzmann equation which describes the 
time-evolution of these distributions.
To do this, we observe that in general we may write
\beqn
      n_i &=& \int {d^3 \vec p_i\over (2\pi)^3} f_i\nonumber\\
 \Gamma_i &=&  {1\over 2E_i} \int \left[d\pi_a (1\pm f_a)\right] 
             \left[d\pi_b (1\pm f_b)\right]  \cdots |{\cal M}|^2~,\nonumber\\
\eeqn
where $\Gamma_i$ is the width for a generic decay of the form $\phi_i\to a+b+...$,
where ${\cal M}$ 
is the corresponding matrix element 
(including an implicit Dirac
$\delta$-function to enforce momentum conservation),
where the momentum-integration measures are given by 
$d\pi\equiv g d^3 \vec p/[(2\pi)^3 2E]$ with $g$ signifying the number
of associated degrees of freedom, and where the $\pm$ signs are chosen
positive for bosons and negative for fermions.
The left side of Eq.~(\ref{diffeq2}) then takes the form
\beq
      \int {d^3 \vec p_i\over (2\pi)^3}
      \left[    \dot f_i + 
     \left( 3H + \Gamma_i + \sum_\alpha \Gamma_i^{(\alpha)}\right) f_i \right]
\label{LHS}
\eeq
and the right side of Eq.~(\ref{diffeq2}) takes the form
\beqn
   &&\sum_{j,\alpha} N_{ji}^{(\alpha)} \int
            [d\pi_j f_j]   
            \left\lbrace \prod_{k\not=i} [d\pi_k (1+ f_k)]^{N_{jk}^{(\alpha)}} \right\rbrace 
          \nonumber\\
            && ~~~~~ [d\pi_i (1+ f_i)]^{N_{ji}^{(\alpha)}} [d\pi_X (1\pm f_X)] |{\cal M}|^2~~~~~~~~~
\label{RHS}
\eeqn
where ${\cal M}$ is the matrix element for the 
decay $\phi_j\to \sum_k N_{jk}^{(\alpha)} \phi_k +X^{(\alpha)}$.
Equating the $d^3 \vec p_i$ integrands in Eqs.~(\ref{LHS}) and (\ref{RHS}) then yields the result
\beqn
   &&   \dot f_i + 
    \left( 3H + \Gamma_i + \sum_\alpha \Gamma_i^{(\alpha)}\right) f_i  ~=~\nonumber\\ 
   && {1+ f_i\over 2E_i} 
       \sum_{j,\alpha} N_{ji}^{(\alpha)}
          \int [d\pi_j f_j]   
            \left\lbrace \prod_{k\not=i} [d\pi_k (1+ f_k)]^{N_{jk}^{(\alpha)}} \right\rbrace 
          \nonumber\\
            && ~~~~~~~~ [d\pi_i (1+ f_i)]^{N_{ji}^{(\alpha)}-1} [d\pi_X (1\pm f_X)] |{\cal M}|^2~~~
            \nonumber\\
\label{urdiffeq}
\eeqn
where we have recognized that although
there are in principle $N_{ji}^{(\alpha)}$ different ways of 
identifying an integrand with respect to $d^3\vec p_i$ within Eq.~(\ref{RHS}),
each yields the same result and therefore any choice will suffice.
               
Eq.~(\ref{urdiffeq}) is a set of coupled differential equations for the phase-space distributions $f_i$.
Indeed, while $\Gamma_i$ is independent of the $f_i$, the quantity $\Gamma_i^{(\alpha)}$ has a hidden
dependence on all $f_j$ for which $N_{ij}^{(\alpha)}\not=0$.
However, despite the complexity of this system of equations,  
our remaining task is conceptually easy:
we simply begin with the distributions
\beq
           f_i(|\vec p_i|, t_0)~=~ 4\pi^3 m_i (\phi_i^{(0)})^2\, \delta^3 (\vec p_i)  
\label{initialdists}
\eeq
at the time $t_0$ when our abundances are initially established,
and then use the results in Eq.~(\ref{urdiffeq}) in order to evolve these distributions
forward in time. 
Indeed, the initial distributions in Eq.~(\ref{initialdists})  
reflect nothing more than the assertion that the original state of our ensemble
consists of fields $\phi_i$ whose zero-momentum modes are displaced by some amount $\phi_i^{(0)}$ from
their minima in field space, with a resulting energy density given by $\rho_i \equiv \half m_i^2 (\phi_i^{(0)})^2$.
While the time-evolution of the effective number densities $n_i$  
can then be obtained from Eq.~(\ref{urdiffeq}) by integrating this equation with respect to
the measure $\int {d^3 \vec p_i /(2\pi)^2}$
[thereby reproducing Eq.~(\ref{diffeq2})],
the time-evolution of the corresponding energy densities $\rho_i$ can be 
obtained from Eq.~(\ref{urdiffeq}) by integrating Eq.~(\ref{urdiffeq}) with respect to
the alternative measure 
$\int [{d^3 \vec p_i /(2\pi)^3}]  E_i = 
 \int [{d^3 \vec p_i /(2\pi)^3}]  \sqrt{|\vec p_i|^2 + m_i^2}$. 
This, then, provides us with the desired coupled differential equations for the energy densities $\rho_i$,
from which it is then trivial to obtain 
the corresponding equations for the abundances $\Omega_i$.

Needless to say, there are many caveats which must be 
borne in mind when applying this formalism.
First, in general we must require that such intra-ensemble decays 
not produce daughter particles with great momenta, for then our dark matter would
not be sufficiently cold.  
Likewise, although we have included the possibility of intra-ensemble decays in the
above analysis, we have 
disregarded the possible contributions from
 {\it scattering}\/ processes which also potentially involve our ensemble components.
Indeed, this is generally an excellent approximation for gravitons, moduli, axions, and other fields
which are very weakly coupled.
We have also disregarded the effects of inverse decays.

%========================================================================
%========================================================================
%========================================================================

\end{document}